\documentclass[10pt,a4paper]{article}
\usepackage{graphicx,amsfonts,amssymb,amsmath, hyperref}
\usepackage{a4wide}

\newif\ifhyper
\hypertrue
\ifhyper
\hypersetup{
   citecolor = {green},
   colorlinks = {true}, 
   urlcolor = {blue} 
}
\fi
\newcommand{\beq}{\begin{equation}}
\newcommand{\eeq}{\end{equation}}
\newcommand{\beqa}{\begin{eqnarray}}
\newcommand{\eeqa}{\end{eqnarray}}
\newcommand{\ket} [1] {\vert #1 \rangle}
\newcommand{\bra} [1] {\langle #1 \vert}
\newcommand{\braket}[2]{\langle #1 | #2 \rangle}

\begin{document}

\title{A Practical Introduction to Tensor Networks: \\ Matrix Product States and Projected Entangled Pair States}

\author{Rom\'an Or\'us \footnote{\emph{E-mail address:} roman.orus@uni-mainz.de} \\ 
\multicolumn{1}{p{0.9\textwidth}}{\centering\emph{Institute of Physics, Johannes Gutenberg University, 55099 Mainz, Germany}}}
\maketitle

\begin{abstract}

This is a partly non-technical introduction to selected topics on tensor network methods, based on several lectures and introductory seminars given on the subject. It should be a good place for newcomers to get familiarized with some of the key ideas in the field, specially regarding the numerics. After a very general introduction we motivate the concept of tensor network and provide several examples. We then move on to explain some basics about Matrix Product States (MPS) and Projected Entangled Pair States (PEPS). Selected details on some of the associated numerical methods for 1$d$ and $2d$ quantum lattice systems are also discussed.  

\end{abstract}

\newpage 
\enlargethispage{1cm}
\tableofcontents
\clearpage

\section{Introduction}

During the last years, the field of Tensor Networks has lived an explosion of results in several directions. This is specially true in the study of quantum many-body systems, both theoretically and numerically. But also in directions which could not be envisaged some time ago, such as its relation to the holographic principle and the AdS/CFT correspondence in quantum gravity \cite{holo, holographic}. Nowadays, Tensor Networks is  rapidly evolving as a field and is embracing an interdisciplinary and motivated community of researchers. 

This paper intends to be an introduction to selected topics on the ever-expanding field of Tensor Networks, mostly focusing on some practical (i.e. algorithmic) applications of Matrix Product States and Projected Entangled Pair States. It is mainly based on several introductory seminars and lectures that the author has given on the topic, and the aim is that the unexperienced reader can start getting familiarized with some of the usual concepts in the field. Let us clarify now, though, that we do not plan to cover \emph{all} the results and techniques in the market, but rather to present some insightful information in a more or less comprehensible way, sometimes also trying to be intuitive, together with further references for the interested reader. In this sense, this paper is not intended to be a complete review on the topic, but rather a useful manual for the beginner. 

The text is divided into several sections. Sec.\ref{sec2} provides a bit of background on the topic. Sec.\ref{sec3} motivates the use of Tensor Networks, and in Sec.\ref{sec4} we introduce some basics about Tensor Network theory such as contractions, diagrammatic notation, and its relation to quantum many-body wave-functions. In Sec.\ref{sec5} we introduce some generalities about Matrix Product States (MPS) for $1d$ systems and Projected Entangled Pair States (PEPS) for $2d$ systems. Later in Sec.\ref{sec6} we explain several strategies to compute expectation values and effective environments for MPS and PEPS, both for finite systems as well as systems in the thermodynamic limit. In Sec.\ref{sec7} we explain generalities on two families of methods to find ground states, namely variational optimization and imaginary time evolution. Finally, in Sec.\ref{sec8} we provide some final remarks as well as a brief discussion on further topics for the interested reader.  

\section{A bit of background}
\label{sec2}

Understanding quantum many-body systems is probably the most challenging problem in condensed matter physics. For instance, the mechanisms behind high-$T_c$ superconductivity are still a mystery to a great extent despite many efforts \cite{highTcCup}. Other important condensed matter phenomena beyond Landau's paradigm of phase transitions have also proven very difficult to understand, in turn combining with an increasing interest in new and exotic phases of quantum matter. Examples of this are, to name a few, topologically ordered phases (where a pattern of long-range entanglement prevades over the whole system) \cite{topo}, quantum spin liquids (phases of matter that do not break any symmetry) \cite{spinliquid}, and deconfined quantum criticality (quantum critical points between phases of fundamentally-different symmetries) \cite{deconf1}.

The standard approach to understand these systems is based on proposing simplified models that are believed to reproduce the relevant interactions responsible for the observed physics, e.g. the Hubbard and $t-J$  models in the case of high-$T_c$ superconductors \cite{hubbard}. Once a model is proposed, and with the exception of some lucky cases where these models are exactly solvable, one needs to rely on faithful numerical methods to determine their properties.  

\begin{figure}[h]
\begin{centering}
\includegraphics[width=12cm]{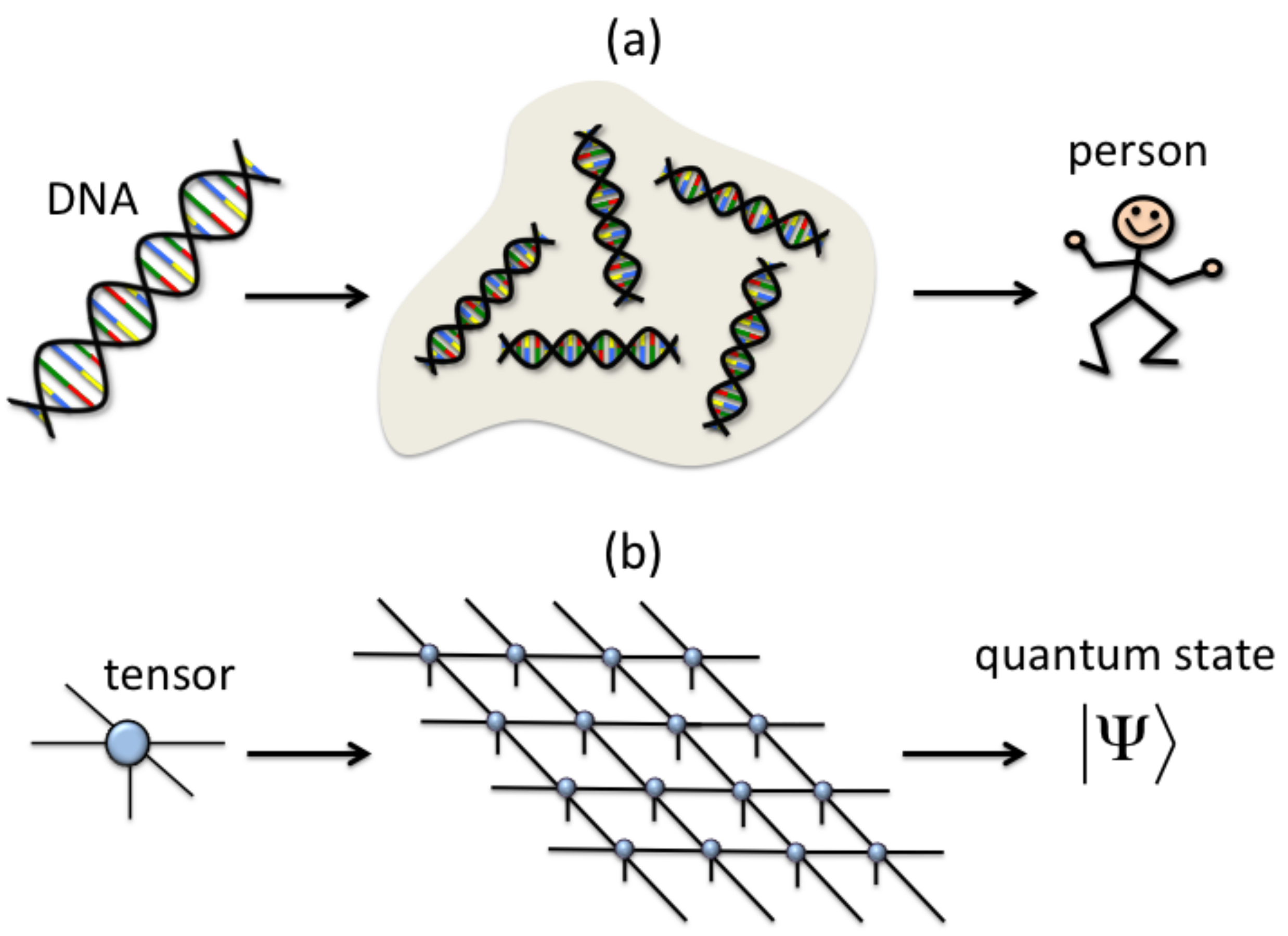} 
\par\end{centering}
\caption{(color online) (a) The DNA is the fundamental building block of a person. In the same way, (b) the tensor is the fundamental building block of a quantum state (here we use a diagrammatic notation for tensors that will be made more precise later on). Therefore, we could say that the tensor is the DNA of the wave-function, in the sense that the whole wave-function can be reconstructed from it just by following some simple rules. \label{fig1}}
\end{figure}

As far as numerical simulation algorithms are concerned, \emph{Tensor Network (TN) methods} have become increasingly popular in recent years to simulate strongly correlated systems \cite{tn}. In these methods the wave function of the system is described by a network of interconnected tensors. Intuitively, this is like a decomposition in terms of LEGO\textsuperscript{\textregistered} pieces, and where entanglement plays the role of a ``glue" amongst the pieces. To put it in another way, the tensor is the DNA of the wave-function, in the sense that the whole wave-function can be reconstructed from this fundamental piece, see Fig.(\ref{fig1}). More precisely, TN techniques offer efficient descriptions of quantum many-body states that are based on the entanglement content of the wave function. Mathematically, the amount and structure of entanglement is a consequence of the chosen network pattern and the number of parameters in the tensors. 

The most famous example of a TN method is probably the Density Matrix Renormalization Group (DMRG) \cite{dmrg1,dmrg2, pbc1, dmrg4}, introduced by Steve White in 1992. One could say that this method has been the technique of reference for the last 20 years to simulate $1d$ quantum lattice systems. However, many important breakthroughs coming from quantum information science have underpinned the emergence of many other algorithms based on TNs. It is actually quite easy to get lost in the soup of names of all these methods, e.g. Time-Evolving Block Decimation (TEBD) \cite{tebd, itebd}, Folding Algorithms \cite{folding}, Projected Entangled Pair States (PEPS) \cite{PEPS}, Tensor Renormalization Group (TRG) \cite{TRG}, Tensor-Entanglement Renormalization Group (TERG) \cite{TERG}, Tensor Product Variational Approach \cite{TPVA}, Weighted Graph States \cite{weight}, Entanglement Renormalization (ER) \cite{ER}, Branching MERA \cite{branching}, String-Bond States \cite{StringBond1}, Entangled-Plaquette States \cite{EntangPlaq}, Monte Carlo Matrix Product States \cite{MCMPS}, Tree Tensor Networks \cite{TTN}, Continuous Matrix Product States and Continuous Tensor Networks \cite{CMPS}, Time-Dependent Variational Principle (TDVP) \cite{TDVP}, Second Renormalization Group (SRG)\cite{SRG}, Higher Order Tensor Renormalization Group (HOTRG) \cite{HOTRG}... and these are just some examples. Each one of these methods has its own advantages and disadvantages, as well as optimal range of applicability.  

A nice property of TN methods is their flexibility. For instance, one can study a variety of systems in different dimensions, of finite or infinite size \cite{iDMRG, itebd, iPEPS,dctm,TRG,SRG,HOTRG,vdma}, with different boundary conditions \cite{pbc1, pbc2}, symmetries \cite{sym}, as well as systems of bosons \cite{boson1}, fermions \cite{ferm} and frustrated spins \cite{frus}.  Different types of phase transitions \cite{phase} have also been studied in this context. Moreover, these methods are also now finding important applications in the context of quantum chemistry \cite{qc} and lattice gauge theories \cite{lgt}, as well as interesting connections to quantum gravity, string theory and the holographic principle \cite{holographic}. The possibility of developing algorithms for infinite-size systems is quite relevant, because it allows to estimate the properties of a system directly in the thermodynamic limit and without the burden of finite-size scaling effects\footnote{We shall see that translational invariance plays a key role in this case.}. Examples of methods using this approach are iDMRG \cite{iDMRG} and iTEBD \cite{itebd} in $1d$ (the ``i" means \emph{infinite}), as well as iPEPS \cite{iPEPS}, TRG/SRG \cite{TRG, SRG}, and HOTRG \cite{HOTRG} in $2d$. From a mathematical perspective, a number of developments in tensor network theory have also come from the field of low-rank tensor approximations in numerical analysis \cite{lowrank}.

\section{Why Tensor Networks?}
\label{sec3}

Considering the wide variety of numerical methods for strongly correlated systems that are available, one may wonder about the necessity of TN methods at all. This is a good question, for which there is no unique answer. In what follows we give some of the reasons why these methods are important and, above all, necessary. 

\subsection{New boundaries for classical simulations}

All the existing numerical techniques have their own limitations. To name a few: the exact diagonalization of the quantum Hamiltonian (e.g. Lanczos methods \cite{lanczos}) is restricted to systems of small size, thus far away from the thermodynamic limit where quantum phase transitions appear. Series expansion techniques \cite{series} rely on perturbation theory calculations. Mean field theory \cite{meanfield} fails to incorporate faithfully the effect of quantum correlations in the system. Quantum Monte Carlo algorithms \cite{qmc} suffer from the sign problem, which restricts their application to e.g. fermionic and frustrated quantum spin systems. Methods based on Continuous Unitary Transformations \cite{cut} rely on the approximate solution of a system of infinitely-many coupled differential equations. Coupled Cluster Methods \cite{coupled} are restricted to small and medium-sized molecules. And Density Functional Theory \cite{dft} depends strongly on the modeling of the exchange and correlation interactions amongst electrons. Of course, these are just some examples. 

TN methods are not free from limitations either. But as we shall see, their main limitation is very different: the amount and structure of the entanglement in quantum many-body states. This new limitation in a computational method extends the range of models that can be simulated with a classical computer in new and unprecedented directions. 

\subsection{New language for (condensed matter) physics}

\begin{figure}[h]
\begin{centering}
\includegraphics[width=10cm]{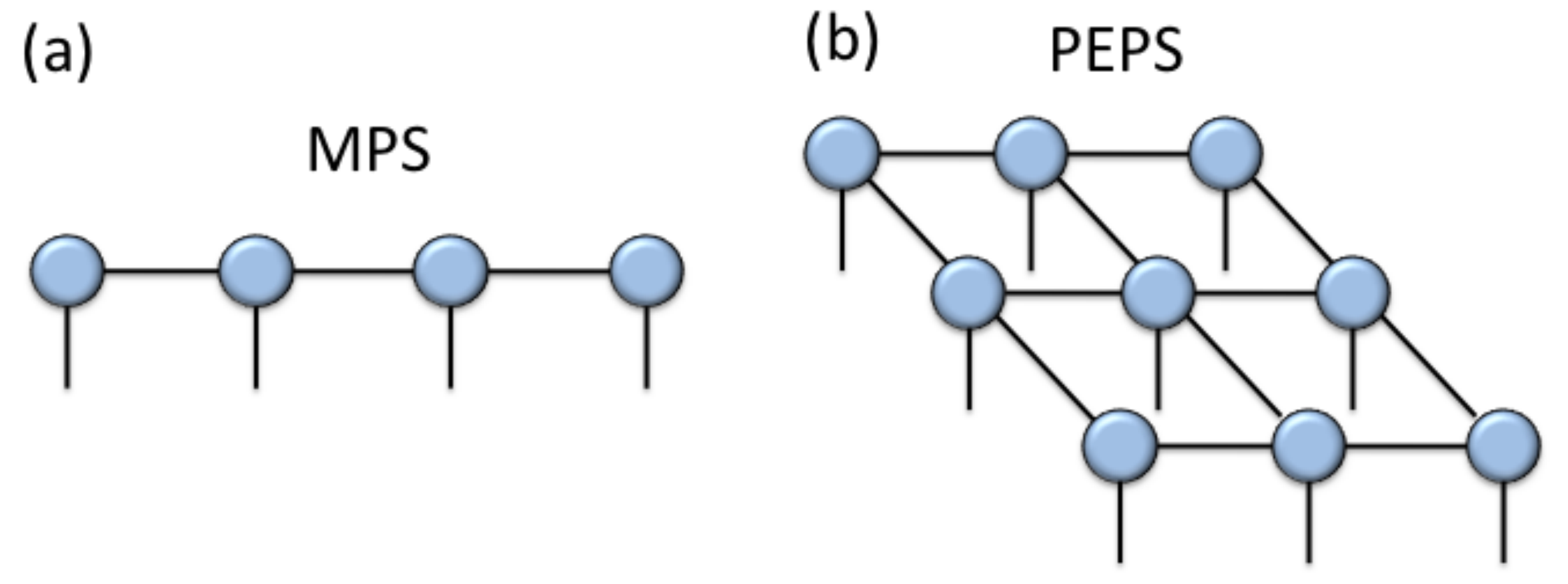} 
\par\end{centering}
\caption{(color online) Two examples of tensor network diagrams: (a) Matrix Product State (MPS) for 4 sites with open boundary conditions; (b) Projected Entangled Pair State (PEPS) for a $3 \times 3$ lattice with open boundary conditions.  \label{fig2}}
\end{figure}

TN methods represent quantum states in terms of networks of interconnected tensors, which in turn capture the relevant entanglement properties of a system. This way of describing quantum states is radically different from the usual approach, where one just gives the coefficients of a wave-function in some given basis. When dealing with a TN state we will see that, instead of thinking about complicated equations, we will be drawing \emph{tensor network diagrams}, see Fig.(\ref{fig2}). As such, it has been recognized that this tensor description offers \emph{the natural language} to describe quantum states of matter, including those beyond the traditional Landau's picture such as quantum spin liquids and topologically-ordered states. This is a new language for condensed matter physics (and in fact, for \emph{all} quantum physics) that makes everything much more visual and which brings new intuitions, ideas and results. 

\subsection{Entanglement induces geometry}

Imagine that you are given a quantum many-body wave-function. Specifying its coefficients in a given local basis does not give any intuition about the structure of the entanglement between its constituents. It is expected that this structure is different depending on the dimensionality of the system: this should be different for $1d$ systems, $2d$ systems, and so on. But it should also depend on more subtle issues like the criticality of the state and its correlation length. Yet, naive representations of quantum states do not possess any explicit information about these properties. It is desirable, thus, to find a way of representing quantum sates where this information is explicit and easily accessible. 

As we shall see, a TN has this information directly available in its description in terms of a network of quantum correlations. In a way, we can think of TN states as quantum states given in some \emph{entanglement representation}. Different representations are better suited for different types of states ($1d$, $2d$, critical...), and the network of correlations makes explicit the \emph{effective lattice geometry} in which the state actually lives. We will be more precise with this in Sec.\ref{sec42}. At this level this is just a nice property. But in fact, by pushing this idea to the limit and turning it around, a number of works have proposed that geometry and curvature (and hence gravity) could emerge naturally from the pattern of entanglement present in quantum states \cite{geo}. Here we will not discuss further this fascinating idea, but let us simply mention that it becomes apparent that the language of TN is, precisely, the correct one to pursue this kind of connection.   

\subsection{Hilbert space is far too large}

This is, probably, \emph{the} main reason why TNs are a key description of quantum many-body states of Nature. For a system of e.g. $N$ spins $1/2$, the dimension of the Hilbert space is $2^N$, which is exponentially large in the number of particles. Therefore, representing a quantum state of the many-body system just by giving the coefficients of the wave function in some local basis is an inefficient representation. The Hilbert space of a quantum many-body system is a really big place with an incredibly large number of quantum states. In order to give a quantitative idea, let us put some numbers: if $N \sim 10^{23}$ (of the order of the Avogadro number) then the number of basis states in the Hilbert space is $\sim O(10^{10^{23}})$, which is much larger (in fact \emph{exponentially} larger) than the number of atoms in the observable universe, estimated to be around $10^{80}$! \cite{sizeuniverse}  

\begin{figure}
\begin{centering}
\includegraphics[width=7cm]{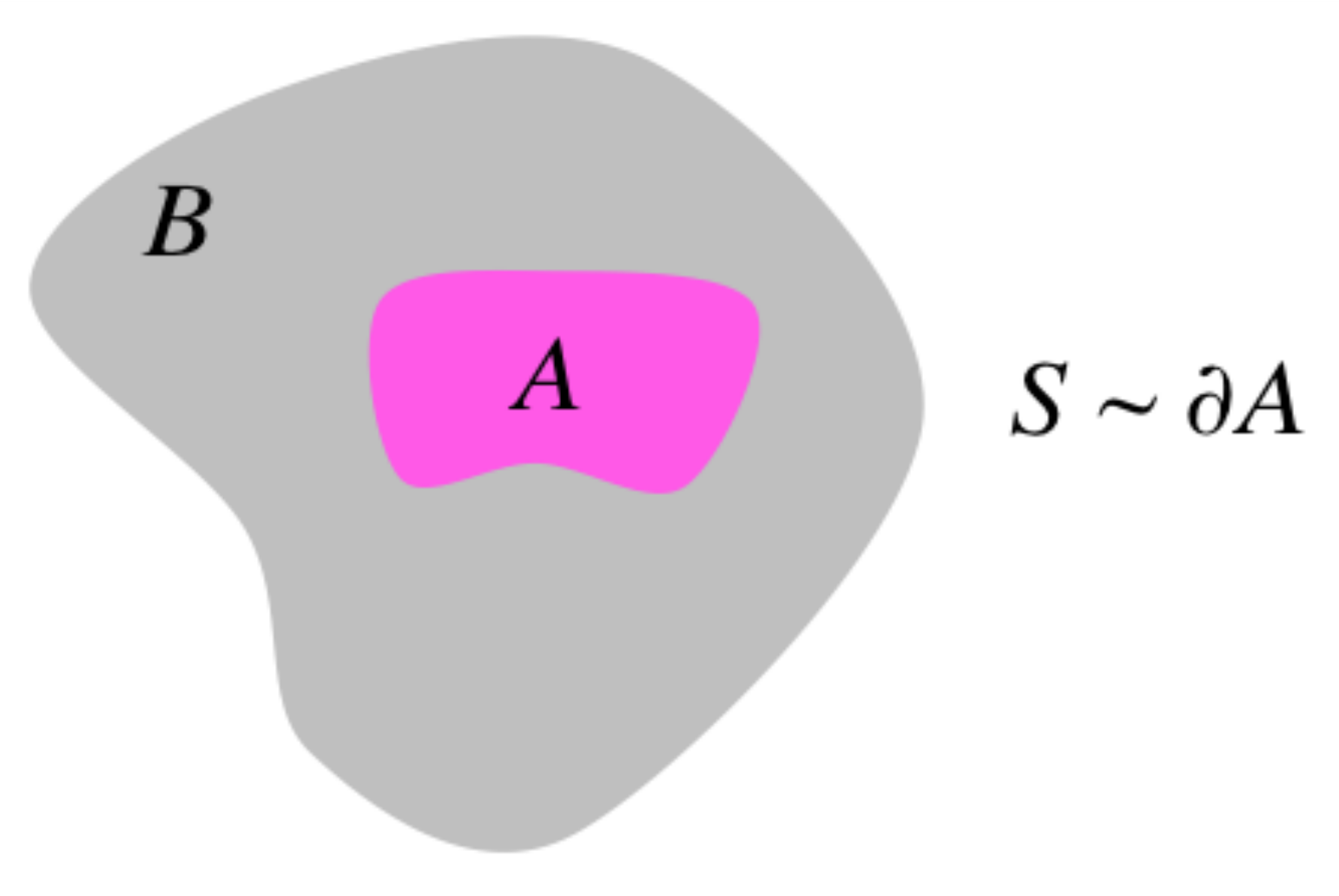} 
\par\end{centering}
\caption{(color online) The entanglement entropy between $A$ and $B$ scales like the size of the boundary $\partial A$ between the two regions, hence $S \sim \partial A$. \label{fig3}}
\end{figure}

Luckily enough for us, not all quantum states in the Hilbert space of a many-body system are equal: some are more relevant than others. To be specific, many important Hamiltonians in Nature are such that the interactions between the different particles tend to be local (e.g. nearest or next-to-nearest neighbors)\footnote{Here we will not enter into deep reasons behind the locality of interactions, and will simply take it for granted. Notice, though, that there are also many important condensed-matter Hamiltonians with non-local interactions, e.g., with a long-range Coulomb repulsion.}. And locality of interactions turns out to have important consequences. In particular, one can prove that low-energy eigenstates of gapped Hamiltonians with local interactions obey the so-called \emph{area-law} for the entanglement entropy \cite{arealaw}, see Fig.(\ref{fig3}). This means that the entanglement entropy of a region of space tends to scale, for large enough regions, as the size of the boundary of the region and not as the volume\footnote{For gapless Hamiltonians there may be multiplicative and/or additive corrections to this behavior, see e.g. Refs. \cite{fermion, spinbose,heisenberg}.}. And this is a very remarkable property, because a quantum state picked at random from a many-body Hilbert space will most likely have a entanglement entropy between subregions that will scale like the volume, and not like the area. In other words, \emph{low-energy states of realistic Hamiltonians are not just ``any" state in the Hilbert space: they are heavily constrained by locality so that they must obey the entanglement area-law}. 

\begin{figure}[h]
\begin{centering}
\includegraphics[width=8cm]{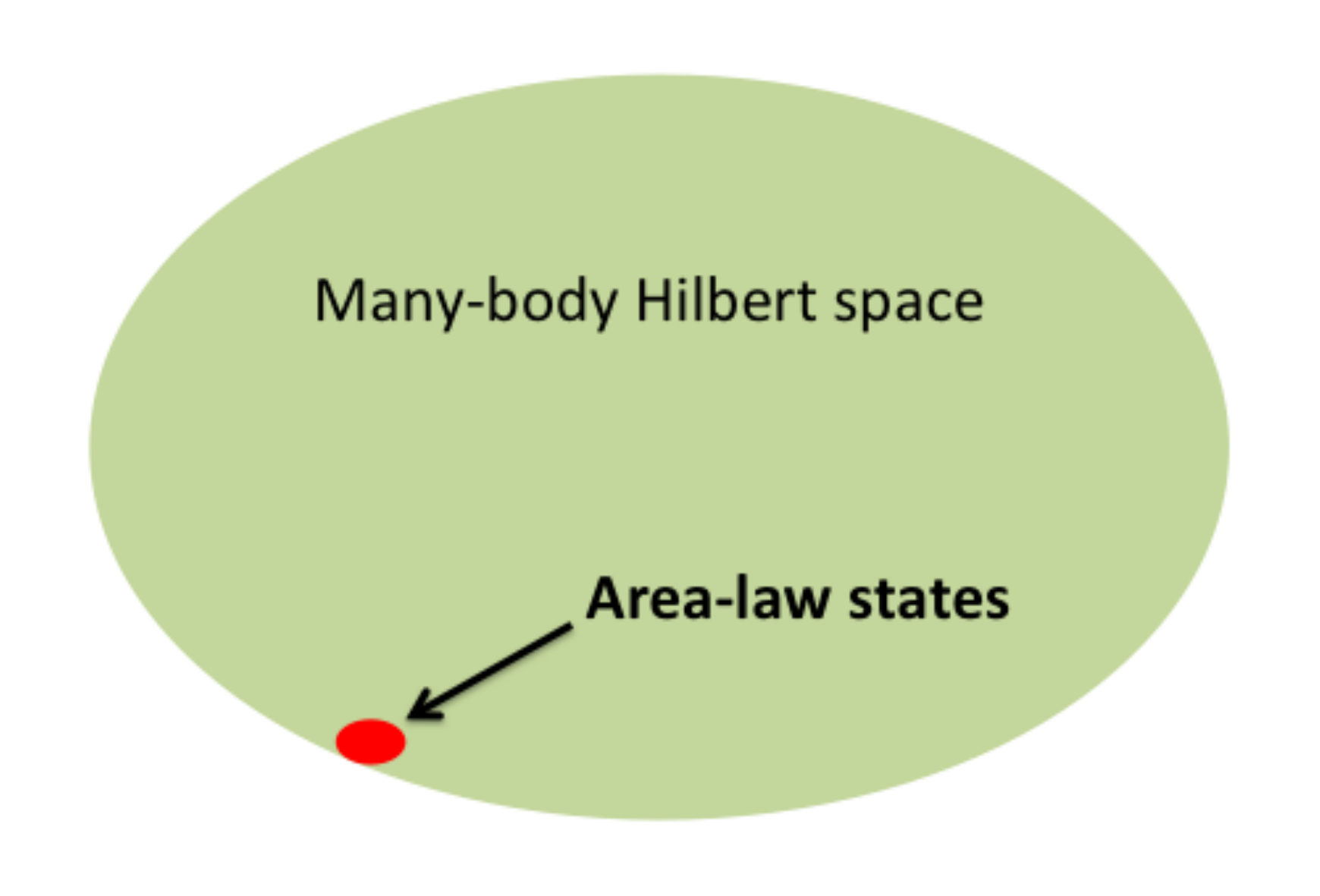} 
\par\end{centering}
\caption{(color online) The manifold of quantum states in the Hilbert space that obeys the area-law scaling for the entanglement entropy corresponds to a tiny corner in the overall huge space. \label{fig4}}
\end{figure}

By turning around the above consideration, one finds a dramatic consequence: it means that not ``any" quantum state in the Hilbert space can be a low-energy state of a gapped, local Hamiltonian. Only those states satisfying the area-law are valid candidates. Yet, the manifold containing these states is just a tiny, exponentially small, corner of the gigantic Hilbert space (see Fig.(\ref{fig4})). 
This corner is, therefore, the \emph{corner of relevant states}. And if we aim to study states within this corner, then we better find a tool to target it \emph{directly} instead of messing around with the full Hilbert space. Here is where the good news come: it is the family of TN states the one that targets this most relevant corner of states \cite{Hastings}. Moreover, recall that Renormalization Group (RG) methods for many-body systems aim to, precisely, identify and keep track of the relevant degrees of freedom to describe a system. Thus, it looks just natural to devise RG methods that deal with this relevant corner of quantum states, and are therefore based on TN states.

In fact, the consequences of having such an immense Hilbert space are even more dramatic. For instance, one can also prove that by evolving a quantum many-body state a time $O({\rm poly}(N))$ with a local Hamiltonian, the manifold of states that can be reached in this time is also exponentially small \cite{poul}. In other words: the vast majority of the Hilbert space is reachable only after a time evolution that would take $O(\exp(N))$ time. This means that, given some initial quantum state (which quite probably will belong to the relevant corner that satisfies the area-law), most of the Hilbert space is \emph{unreachable in practice}. To have a better idea of what this means let us put again some numbers: for $N\sim 10^{23}$ particles, by evolving some quantum state with a local Hamiltonian, reaching most of the states in the Hilbert space would take $\sim O(10^{10^{23}})$ seconds. Considering that the best current estimate for the age of the universe is around $10^{17}$ seconds \cite{ageuniverse}, this means that we should wait around the exponential of one-million times the age of the universe to reach most of the states available in the Hilbert space. Add to this that your initial state must also be compatible with some locality constraints in your physical system (because otherwise it may not be truly physical), and what you obtain is that all the quantum states of many-body systems that you will ever be able to explore are contained in a exponentially small manifold of the full Hilbert space. This is why the Hilbert space of a quantum many-body systems is sometimes referred to as a \emph{convenient illusion} \cite{poul}: it is convenient from a mathematical perspective, but it is an illusion because no one will ever see most of it.  

\section{Tensor Network theory}
\label{sec4}

Let us now introduce some mathematical concepts. In what follows we will define what a TN state is, and how this can be described in terms of TN diagrams. We will also introduce the TN representation of quantum states, and explain the examples of Matrix Product States (MPS) for $1d$ systems \cite{MPS}, and Projected Entangled Pair States (PEPS) for $2d$ systems \cite{PEPS}. 

\subsection{Tensors, tensor networks, and tensor network diagrams}

For our purposes, a \emph{tensor} is a multidimensional array of complex numbers. The \emph{rank} of a tensor is the number of indices. Thus, a rank-$0$ tensor is scalar ($x$), a rank-$1$ tensor is a vector ($v_{\alpha}$), and a rank-$2$ tensor is a matrix ($A_{\alpha \beta}$). 

An \emph{index contraction} is the sum over all the possible values of the repeated indices of a set of tensors. For instance, the matrix product
\beq
C_{\alpha \gamma} = \sum_{\beta=1}^D A_{\alpha \beta} B_{\beta \gamma}
\label{mat}
\eeq
is the contraction of index $\beta$, which amounts to the sum over its $D$ possible values.  One can also have more complicated contractions, such as this one:  
\beq
F_{\gamma \omega \rho \sigma} = \sum_{\alpha, \beta, \delta, \nu, \mu =1}^D A_{\alpha \beta \delta \sigma} B_{\beta \gamma \mu} C_{\delta \nu \mu \omega} E_{\nu \rho \alpha} \ ,
\label{tn}
\eeq
where  for simplicity we assumed that contracted indices $\alpha, \beta, \delta, \nu$ and $\mu$ can take $D$ different values. As seen in these examples, the contraction of indices produces new tensors, in the same way that e.g. the product of two matrices produces a new matrix. Indices that are not contracted are called \emph{open indices}. 

A \emph{Tensor Network} (TN) is a set of tensors where some, or all, of its indices are contracted according to some pattern. Contracting the indices of a TN is called, for simplicity, \emph{contracting the TN}. The above two equations are examples of TN. In Eq.(\ref{mat}), the TN is equivalent to a matrix product, and produces a new matrix with two open indices. In Eq.(\ref{tn}), the TN corresponds to contracting indices $\alpha, \beta, \delta, \nu$ and $\mu$ in tensors $A, B, C$ and $E$ to produce a new rank-$4$ tensor $F$ with open indices $\gamma, \omega, \rho$ and $\sigma$. In general, the contraction of a TN with some open indices gives as a result another tensor, and in the case of not having any open indices the result is a scalar. This is the case of e.g. the scalar product of two vectors, 
\beq
C = \sum_{\alpha=1}^D A_{\alpha} B_{\alpha} \ ,
\label{scal}
\eeq
where $C$ is a complex number (rank-$0$ tensor). A more intrincate example could be
\beq
F = \sum_{\alpha, \beta, \gamma, \delta, \omega, \nu, \mu =1}^D A_{\alpha \beta \delta \gamma} B_{\beta \gamma \mu} C_{\delta \nu \mu \omega} E_{\nu \omega \alpha} \ ,
\label{tn2}
\eeq
where all indices are contracted and the result is again a complex number $F$. 

\begin{figure}[h]
\begin{centering}
\includegraphics[width=10cm]{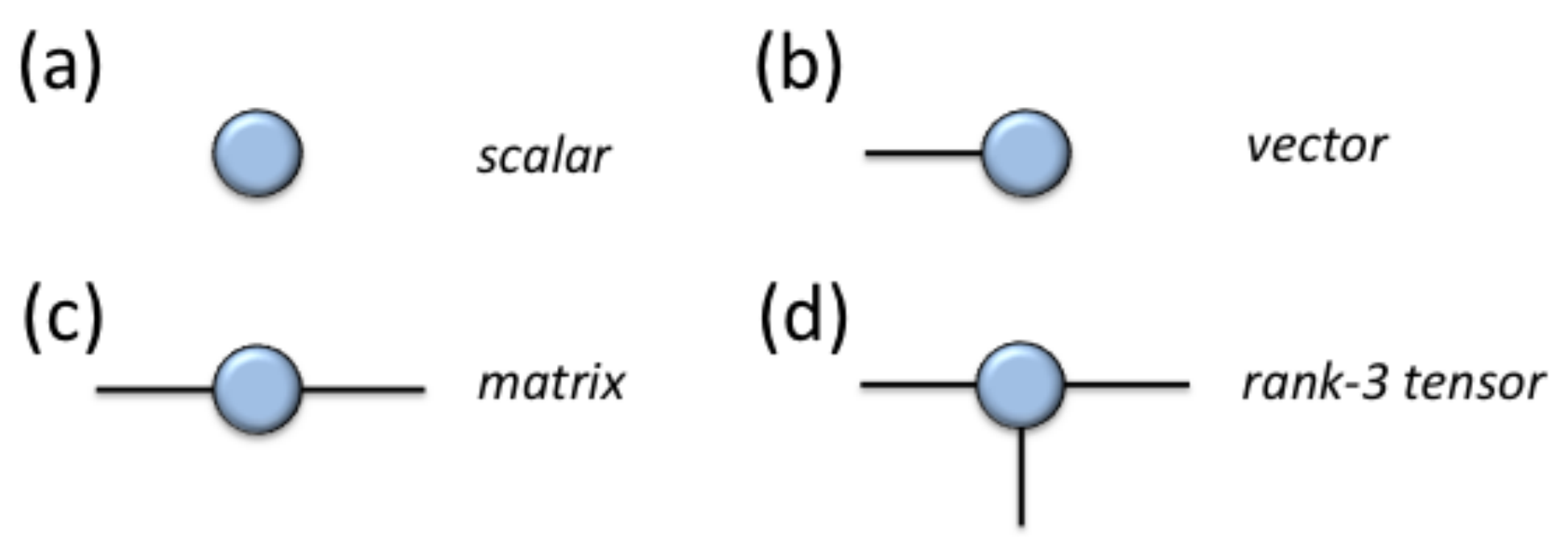} 
\par\end{centering}
\caption{(color online) Tensor network diagrams: (a) scalar, (b) vector, (c) matrix and (d) rank-3 tensor \label{fig5}}
\end{figure}

Once this point is reached, it is convenient to introduce a diagrammatic notation for tensors and TNs in terms of \emph{tensor network diagrams}, see Fig.(\ref{fig5}). In these diagrams tensors are represented by shapes, and indices in the tensors are represented by lines emerging from the shapes. A TN is thus represented by a set of shapes interconnected by lines. The lines connecting tensors between each other correspond to contracted indices, whereas lines that do not go from one tensor to another correspond to open indices in the TN. 

\begin{figure}[h]
\begin{centering}
\includegraphics[width=10cm]{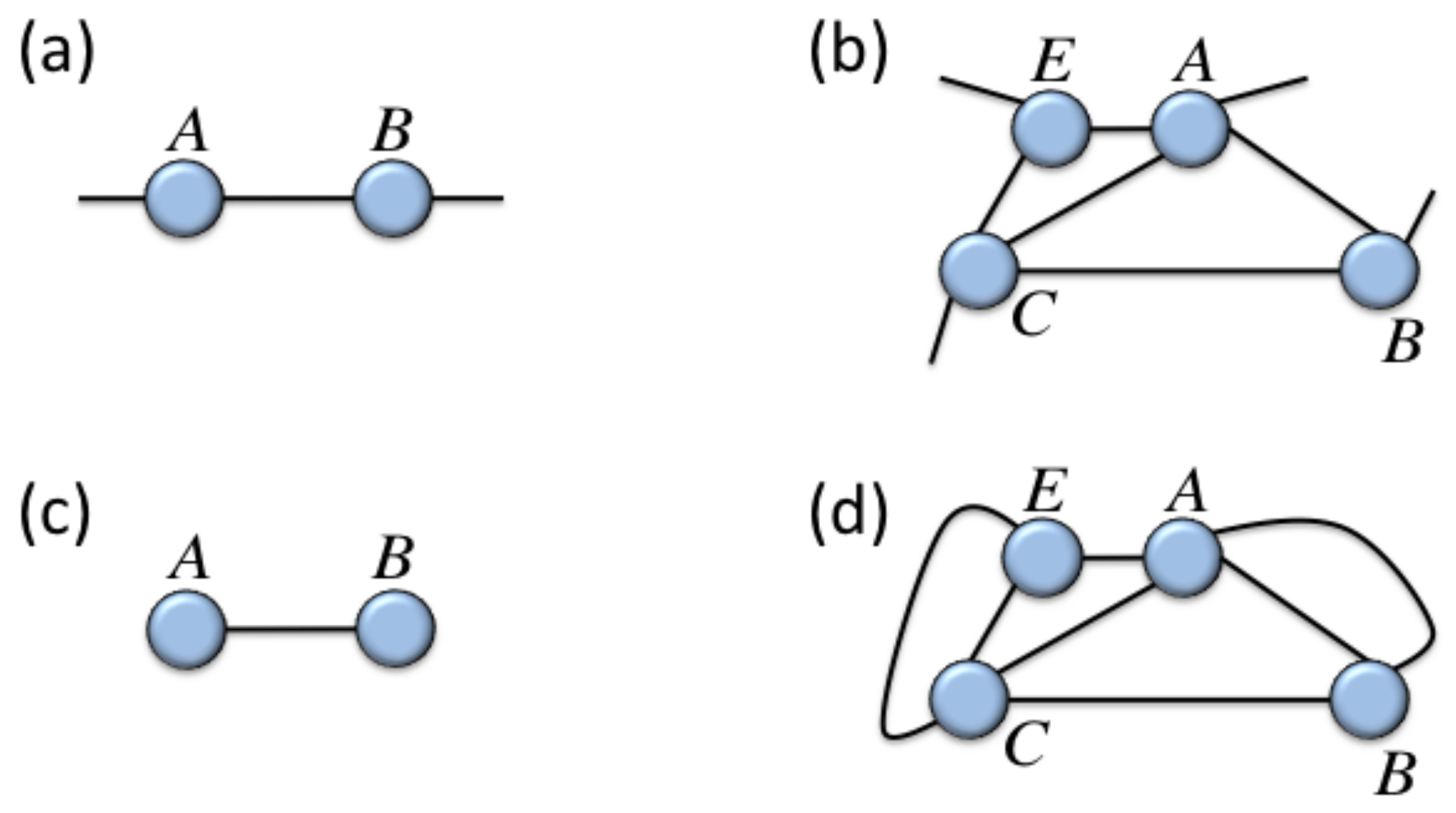} 
\par\end{centering}
\caption{(color online) Tensor network diagrams for Eqs.(\ref{mat}, \ref{tn}, \ref{scal}, \ref{tn2}): (a) matrix product, (b) contraction of 4 tensors with 4 open indices, (c) scalar product of vectors, and (d) contraction of 4 tensors without open indices. \label{fig6}}
\end{figure}

\begin{figure}[h]
\begin{centering}
\includegraphics[width=5cm]{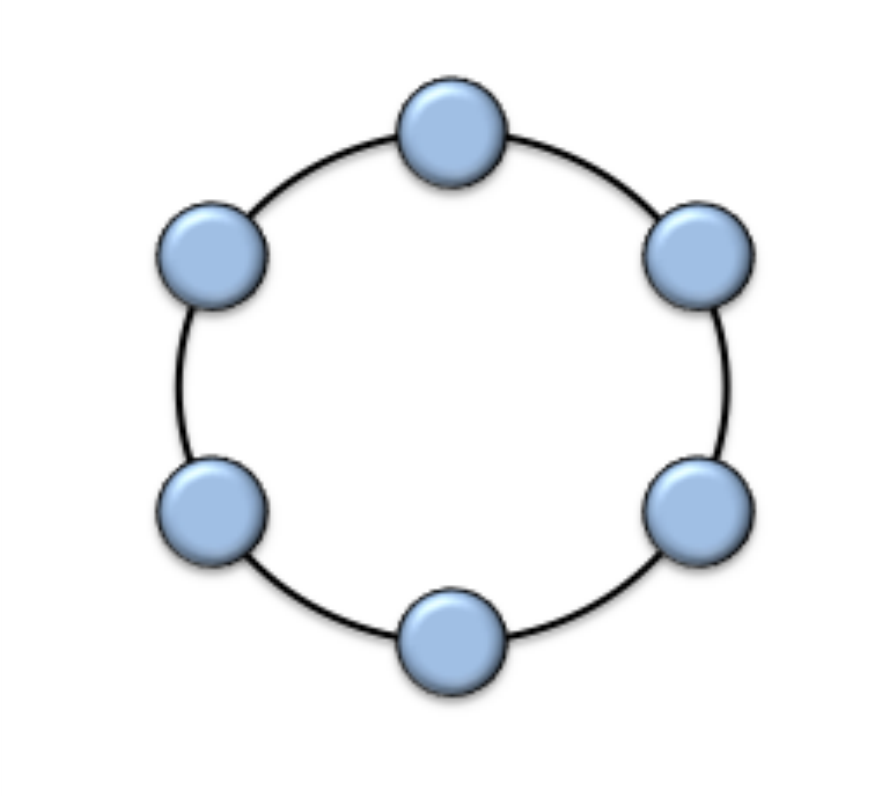} 
\par\end{centering}
\caption{(color online) Trace of the product of 6 matrices. \label{fig7}}
\end{figure}

Using TN diagrams it is much easier to handle calculations with TN. For instance, the contractions in Eqs.(\ref{mat}, \ref{tn}, \ref{scal}, \ref{tn2}) can be represented by the diagrams in Fig.(\ref{fig6}). 
Also tricky calculations, like the trace of the product of 6 matrices, can be represented by diagrams as in Fig.(\ref{fig7}). 
From the TN diagram the cyclic property of the trace becomes evident. This is a simple example of why TN diagrams are really useful: unlike plain equations, TN diagrams allow to handle with complicated expressions in a visual way. In this manner many properties become apparent, such as the cyclic property of the trace of a matrix product. In fact, you could compare the language of TN diagrams to that of Feynman diagrams in quantum field theory. Surely it is much more intuitive and visual to think in terms of drawings instead of long equations. Hence, from now on we shall only use diagrams to represent tensors and TNs. 

\begin{figure}[h]
\begin{centering}
\includegraphics[width=10cm]{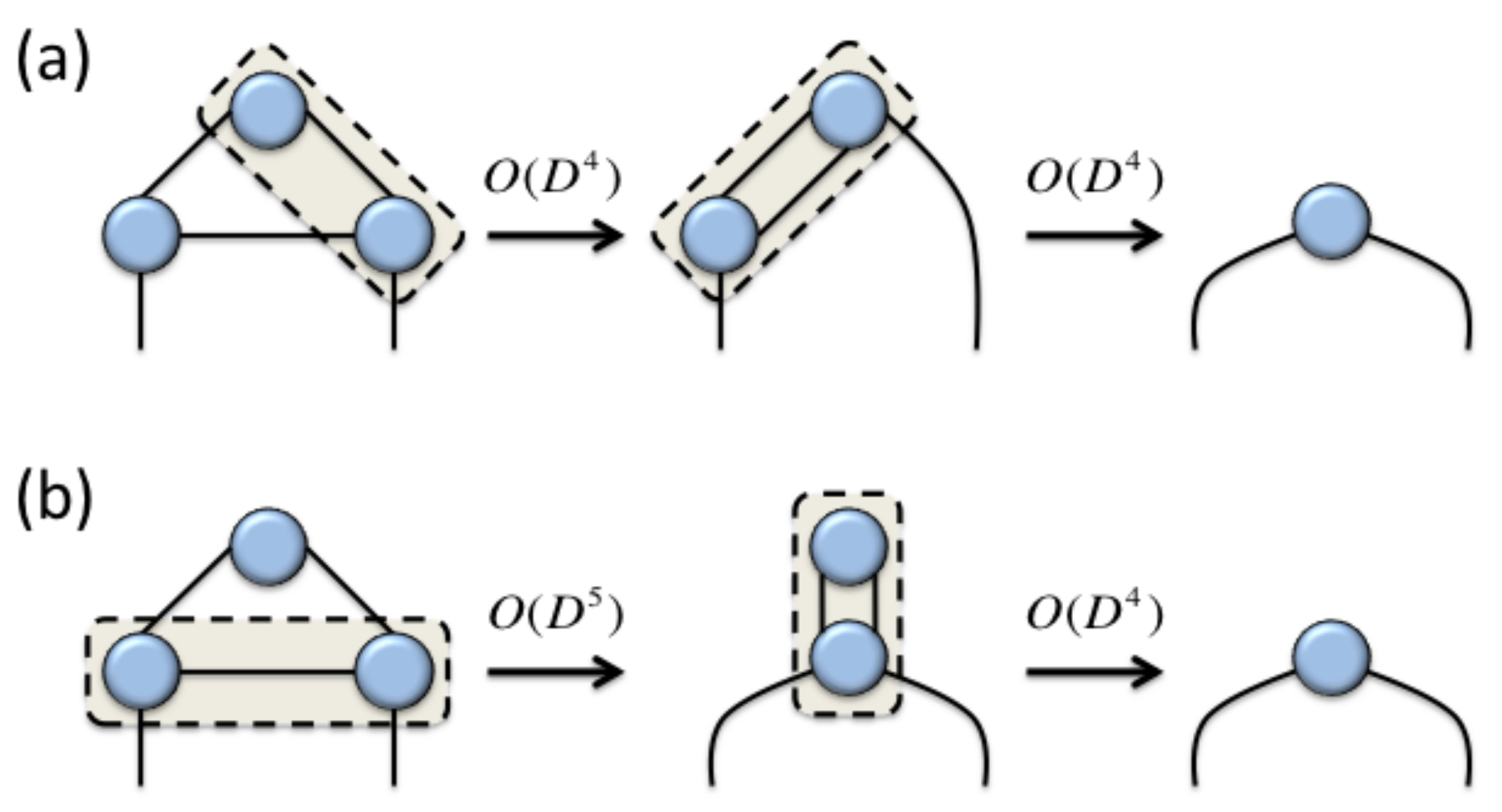} 
\par\end{centering}
\caption{(color online) (a) Contraction of 3 tensors in $O(D^4)$ time; (b) contraction of the same 3 tensors in $O(D^5)$ time.} \label{fig8}
\end{figure}

There is an important property of TN that we would like to stress now. Namely, that the total number of operations that must be done in order to obtain the final result of a TN contraction depends heavily on the order in which indices in the TN are contracted. See for instance Fig.(\ref{fig8}). Both cases correspond to the same overall TN contraction, but in one case the number of operations is $O(D^4)$ and in the other is $O(D^5)$. This is quite relevant, since in TN methods one has to deal with many contractions, and the aim is to make these \emph{as efficiently as possible}. For this, finding the optimal order of indices to be contracted in a TN will turn out to be a crucial step, specially when it comes to programming computer codes to implement the methods. To minimize the computational cost of a TN contraction one must optimize over the different possible orderings of pairwise contractions, and find the optimal case. Mathematically this is a very difficult problem, though in practical cases this can be done usually by simple inspection. 

\subsection{Breaking the wave-function into small pieces}
\label{sec42}

Let us now explain the TN representation of quantum many-body states. For this, we consider a quantum many-body system of $N$ particles. The degrees of freedom of each one of these particles can be described by $p$ different states. Hence, we are considering systems of $N$ $p$-level particles. For instance, for a quantum many-body system such as the spin-1/2 Heisenberg model we have that $p = 2$, so that each particle is a 2-level system (or qubit). For a given system of this kind, any wave function $\ket{\Psi}$  that describes its physical properties can be written as
\beq
\ket{\Psi} = \sum_{i_1 i_2 \ldots i_N} C_{i_1 i_2 \ldots i_N} \ket{i_1}\otimes \ket{i_2} \otimes \cdots \otimes \ket{i_N}
\label{st}
\eeq
once an individual basis $\ket{i_r}$ for the states of each particle $r = 1, ..., N$ has been chosen. In the above equation, $C_{i_1 i_2 \ldots i_N}$ are $p^N$ complex numbers (independent up to a normalization condition), $i_r = 1,..., p$ for each particle $r$, and the symbol $\otimes$ denotes the tensor product of individual quantum states for each one of the particles in the many-body system. 

Notice now that the $p^N$ numbers $C_{i_1 i_2 \ldots i_N}$ that describe the wave function $\ket{\Psi}$  can be understood as the coefficients of a tensor $C$ with $N$ indices $i_1 i_2 \ldots i_N$, where each of the indices can take up to $p$ different values (since we are considering $p$-level particles). Thus, this is a tensor of rank $N$, with $O(p^N)$ coefficients. This readily implies that the number of parameters that describe the wave function of Eq.(\ref{st}) is exponentially large in the system size.

\begin{figure}[h]
\begin{centering}
\includegraphics[width=11cm]{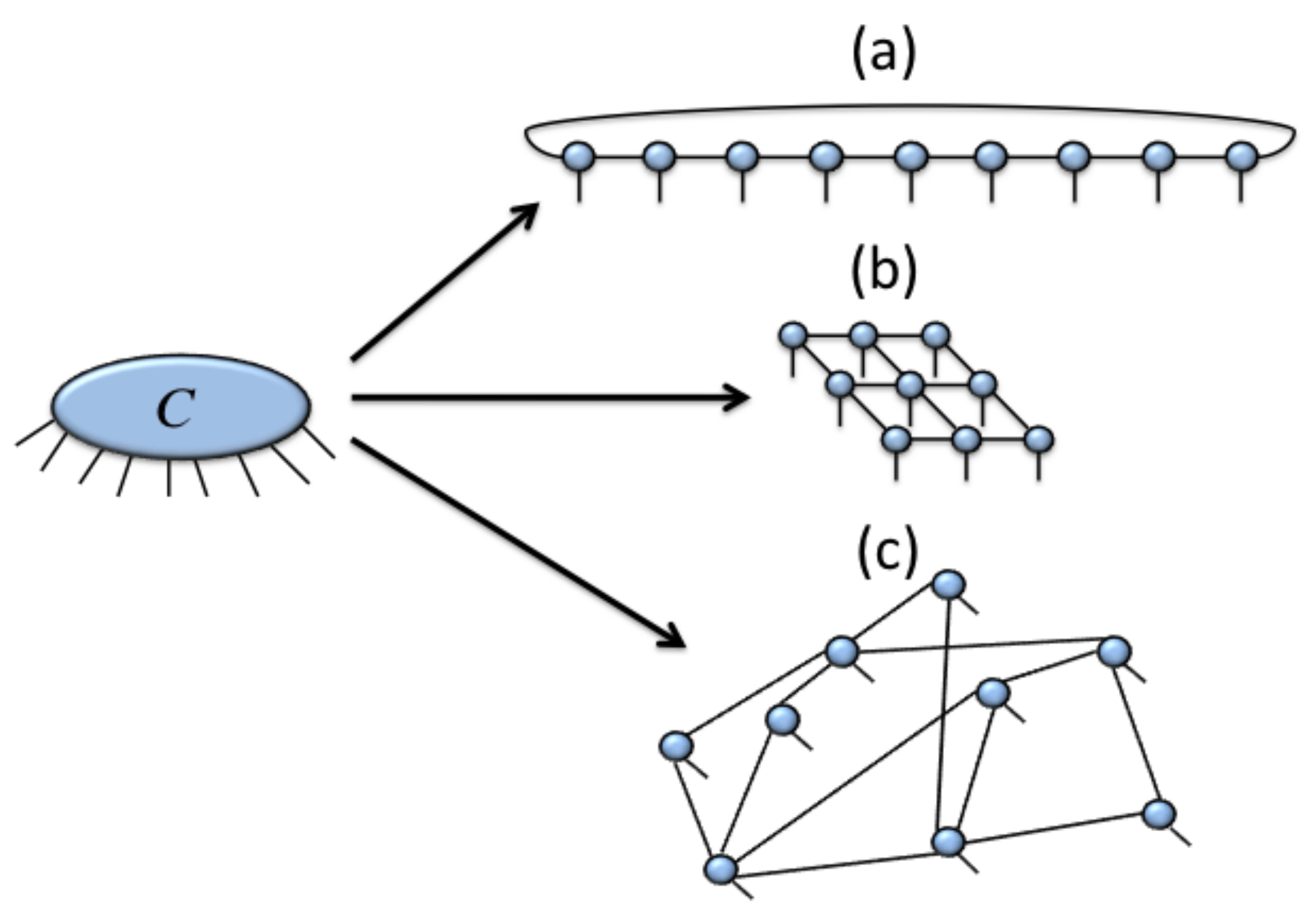} 
\par\end{centering}
\caption{(color online) Tensor network decomposition of tensor $C$ in terms of (a) an MPS with periodic boundary conditions, (b) a PEPS with open boundary condition, and (c) an arbitrary tensor network. \label{fig9}}
\end{figure}

Specifying the values of each one of the coefficients $C_{i_1 i_2 \ldots i_N}$ of tensor $C$ is, therefore, a computationally-inefficient description of the quantum state of the many-body system. One of the aims of TN states is to reduce the complexity in the representation of states like $\ket{\Psi}$ by providing an accurate description of the expected entanglement properties of the state. This is achieved by replacing the ``big" tensor $C$ by a TN of ``smaller" tensors, i.e. by a TN of tensors with smaller rank (see Fig.(\ref{fig9}) for some examples in a diagrammatic representation). This approach amounts to decomposing the ``big" tensor $C$ (and hence state $\ket{\Psi}$) into ``fundamental DNA blocks", namely, a TN made of tensors of some smaller rank, which is much easier to handle. 

Importantly, the final representation of $\ket{\Psi}$ in terms of a TN typically depends on a polynomial number of parameters, thus being a computationally efficient description of the quantum state of the many-body system. To be precise, the total number of parameters $m_{{\rm tot}}$ in the tensor network will be
\beq
m_{{\rm tot}} = \sum_{t=1}^{N_{\rm tens}} m(t) \ ,  
\eeq
where $m(t)$ is the number of parameters for tensor $t$ in the TN and $N_{{\rm tens}}$ is the number of tensors. For a TN to be practical $N_{{\rm tens}}$ must be sub-exponential in $N$, e.g. $N_{{\rm tens}} = O({\rm poly}(N))$, and sometimes even $N_{{\rm tens}} = O(1)$. Also, for each tensor $t$ the number of parameters is 
\beq
m(t) = O\left(\prod_{a_t = 1}^{{\rm rank}(t)} D(a_t)\right) \ ,
\eeq
where the product runs over the different indices $a_t = 1, 2, \ldots , {\rm rank}(t)$ of the tensor, $D(a_t)$ is the different possible values of index $a_t$, and rank$(t)$ is the number of indices of the tensor. Calling $D_t$ the maximum of all the numbers $D(a_t)$ for a given tensor, we have that 
\beq
m(t) = O\left(D_t^{{\rm rank}(t)}\right) \ .
\eeq
Putting all the pieces together, we have that the total number of parameters will be 
\beq
m_{{\rm tot}} = \sum_{t=1}^{N_{\rm tens}} O\left(D_t^{{\rm rank}(t)}\right)  = O({\rm poly}(N) {\rm poly}(D)) \ ,  
\eeq
where $D$ is the maximum of $D_t$ over all tensors, and where we assumed that the rank of each tensor is bounded by a constant. 

To give a simple example, consider the TN in Fig.(\ref{fig9}.a). This is an example of a \emph{Matrix Product State} (MPS) with periodic boundary conditions, which is a type of TN that will be discussed in detail in the next section. Here, the number of parameters is just $O(N pD^2)$, if we assume that open indices in the TN can take up to $p$ values, whereas the rest can take up to $D$ values. Yet, the contraction of the TN yields a tensor of rank $N$, and therefore $p^N$ coefficients. Part of the magic of the TN description is that it shows that these $p^N$ coefficients are not independent, but rather they are obtained from the contraction of a given TN and therefore have a structure. 

Nevertheless, this efficient representation of a quantum many-body state does not come for free. The replacement of tensor $C$ by a TN involves the appearance of extra degrees of freedom in the system, which are responsible for ``gluing the different DNA blocks" together. These new degrees of freedom are represented by the connecting indices amongst the tensors in the TN. The connecting indices turn out to have an important physical meaning: they represent the structure of the many-body entanglement in the quantum state $\ket{\Psi}$, and the number of different values that each one of these indices can take is a quantitative measure of the amount of quantum correlations in the wave function. These indices are usually called \emph{bond or ancillary indices}, and their number of possible values are referred to as \emph{bond dimensions}. The maximum of these values, which we called above $D$, is also called the \emph{bond dimension of the tensor network}. 

\begin{figure}[h]
\begin{centering}
\includegraphics[width=14cm]{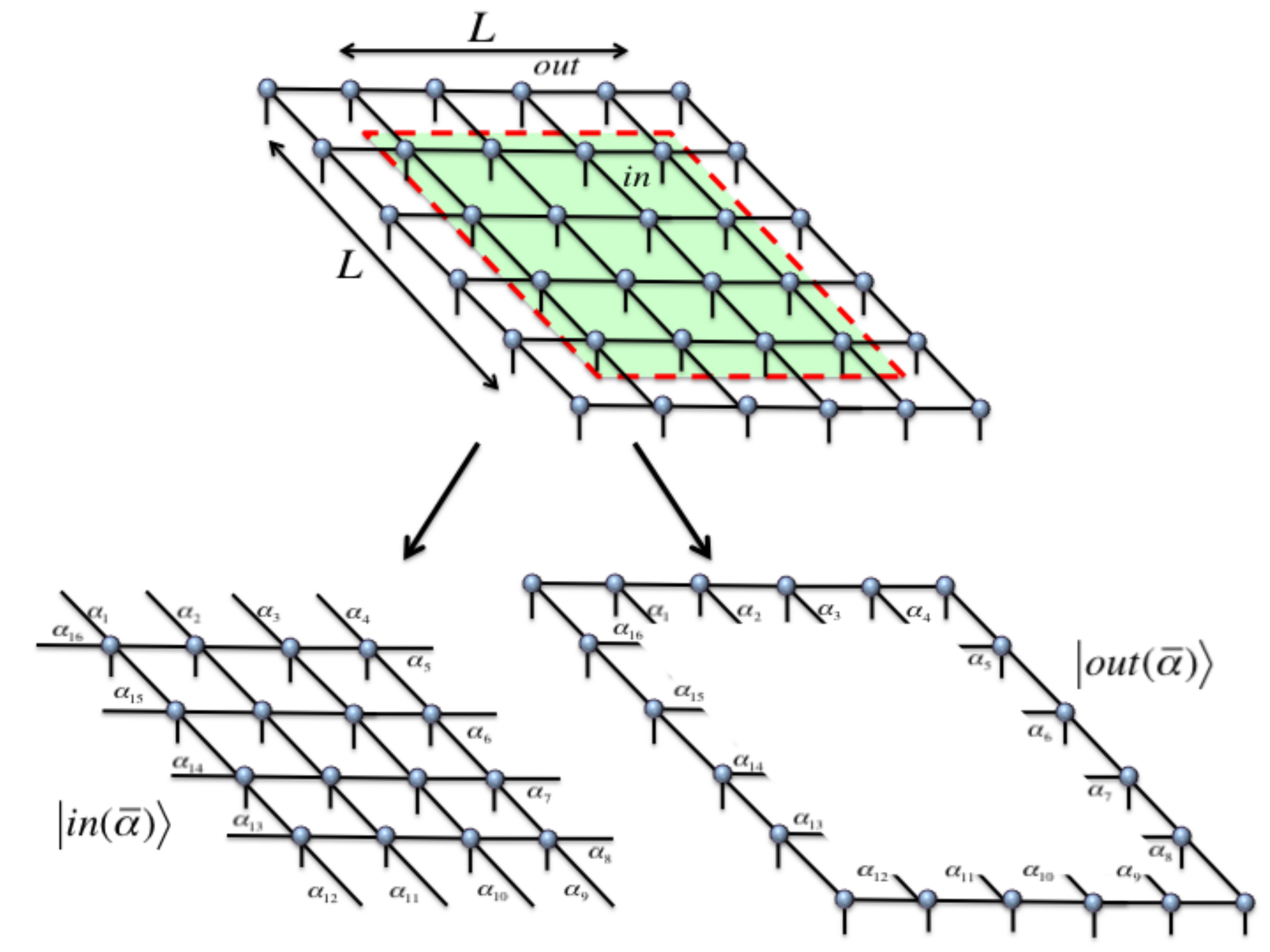} 
\par\end{centering}
\caption{(color online) States $\ket{in(\bar{\alpha})}$ and $\ket{out(\bar{\alpha})}$ for a $4 \times 4$ block of a $6 \times 6$ PEPS. \label{fig10}}
\end{figure}

To understand better how entanglement relates to the bond indices, let us give an example. Imagine that you are given a TN state with bond dimension $D$ for all the indices, and such as the one in Fig.(\ref{fig10}). This is an example of a TN state called \emph{Projected Entangled Pair State} (PEPS) \cite{PEPS}, which will also be further analyzed in the forthcoming sections.  Let us now estimate for this state the entanglement entropy of a block of linear length $L$ (see the figure). For this, we call $\bar{\alpha} = \{\alpha_1 \alpha_2 ... \alpha_{4L} \}$ the combined index of all the TN indices across the boundary of the block. Clearly, if the $\alpha$ indices can take up to $D$ values, then $\bar{\alpha}$ can take up to $D^{4L}$. We now write the state in terms of unnormalized kets for the inner and outer parts of the block (see also Fig.(\ref{fig10})) as
\beq
\ket{\Psi} = \sum_{\bar{\alpha}=1}^{D^{4L}} \ket{in(\bar{\alpha})} \otimes  \ket{out(\bar{\alpha})} \ .
\eeq
The reduced density matrix of e.g. the inner part is given by 
\beq
\rho_{in} = \sum_{\bar{\alpha}, \bar{\alpha'}} X_{\bar{\alpha} \bar{\alpha'}} \ket{in(\bar{\alpha})} \bra{in(\bar{\alpha'})} \ , 
\eeq
where $X_{\bar{\alpha} \bar{\alpha'}} \equiv   \braket{out(\bar{\alpha'})}{out(\bar{\alpha})}$. This reduced density matrix clearly has a rank that is, at most, $D^{4L}$. The same conclusions would apply if we considered the reduced density matrix of the outside of the block. Moreover, the entanglement entropy $S(L) = -{\rm tr} (\rho_{in} \log \rho_{in})$ of the block is upper bounded by the logarithm of the rank of $\rho_{in}$. So, in the end, we get
\beq
S(L) \le 4L \log D  \ ,
\eeq
which is nothing but an upper-bound version of the area-law for the entanglement entropy  \cite{arealaw}. In fact, we can also interpret this equation as \emph{every ``broken" bond index giving an entropy contribution of at most $\log D$}. 

Let us discuss the above result. First, if $D=1$ then the upper bound says that $S(L) = 0$ no matter the size of the block. That is, no entanglement is present in the wave function. This is a generic result for any TN: if the bond dimensions are trivial, then no entanglement is present in the wave function, and the TN state is just a product state. This is the type of ansatz that is used in e.g. mean field theory. Second, for any $D > 1$ we have that the ansatz can already handle an area-law for the entanglement entropy. Changing the bond dimension $D$ modifies only the multiplicative factor of the area-law. Therefore, in order to modify the scaling with $L$ one should change  the geometric pattern of the TN. This means that the entanglement in the TN is a consequence of both $D$ (the ``size" of the bond indices), and also the geometric pattern (the way these bond indices are connected). In fact, different families of TN states turn out to have very different entanglement properties, even for the same $D$. Third, notice that by limiting $D$ to a fixed value greater than one we can achieve TN representations of a quantum many-body state which are both computationally efficient (as in mean field theory) and quantumly correlated (as in exact diagonalization). In a way, by using TNs one gets the best of both worlds. 

TN states are also important because they have been proven to correspond to ground and thermal states of local, gapped Hamiltonians \cite{Hastings}. This means that TN states are, in fact, the states inside the relevant corner of the Hilbert space that was discussed in the previous section: they correspond to relevant states in Nature which obey the area-law and can, on top, be described efficiently using the tensor language. 

\section{MPS and PEPS: generalities}
\label{sec5}

Let us now present two families of well-known and useful TN states. These are Matrix Product States (MPS) and Projected Entangled Pair States (PEPS). Of course these two are not the only families of TN states, yet these will be the only two that we will consider in some detail here. For the interested reader, we briefly mention other families of TN states in Sec.\ref{sec8}. 

\subsection{Matrix Product States (MPS)}

The family of MPS \cite{MPS} is probably the most famous example of TN states. This is because it is behind some very powerful methods to simulate $1d$ quantum many-body systems, most prominently the Density Matrix Renormalization Group (DMRG) algorithm \cite{dmrg1, dmrg2, pbc1, dmrg4}. But it is also behind other well-known methods such as Time-Evolving Block Decimation (TEBD) \cite{tebd, itebd} and Power Wave Function Renormalization Group (PWFRG) \cite{pwfrg}. Before explaining any method, though, let us first describe what an MPS actually is, as well as some of its properties. 

\begin{figure}[h]
\begin{centering}
\includegraphics[width=10cm]{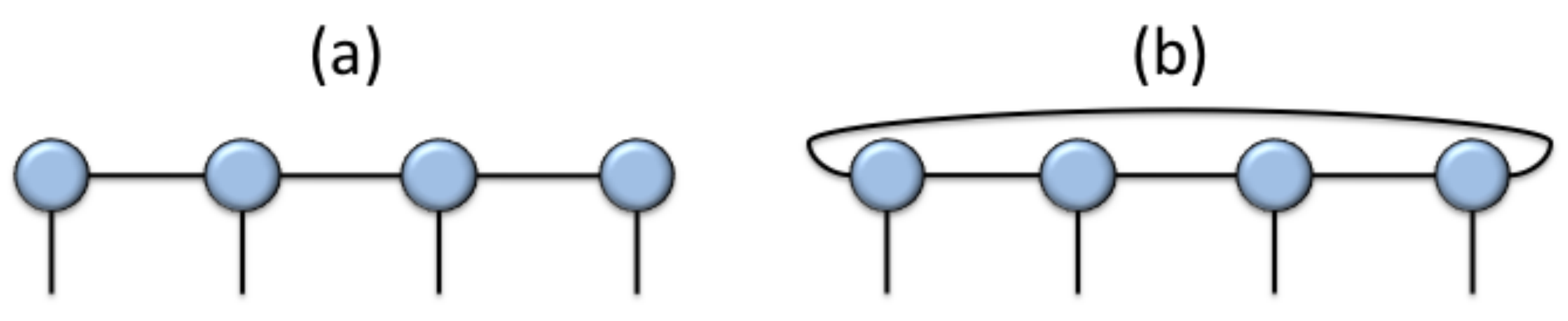} 
\par\end{centering}
\caption{(color online) (a) 4-site MPS with open boundary conditions; (b) 4-site MPS with periodic boundary conditions. \label{fig11}}
\end{figure}

MPS are TN states that correspond to a one-dimensional array of tensors, such as the ones in Fig.(\ref{fig11}). In a MPS there is one tensor per site in the many-body system. The connecting bond indices that glue the tensors together can take up to $D$ values, and the open indices correspond to the physical degrees of freedom of the local Hilbert spaces which can take up to $p$ values. In Fig.(\ref{fig11}) we can see two examples of MPS. The first one corresponds to a MPS with \emph{open} boundary conditions\footnote{Mathematicians sometimes call this the \emph{Tensor Train decomposition}\cite{lowrank}.}, and the second one  to a MPS with \emph{periodic} boundary conditions \cite{pbc1}. Both examples are for a finite system of 4 sites. 

\subsubsection{Some properties}

\begin{figure}
\begin{centering}
\includegraphics[width=9cm]{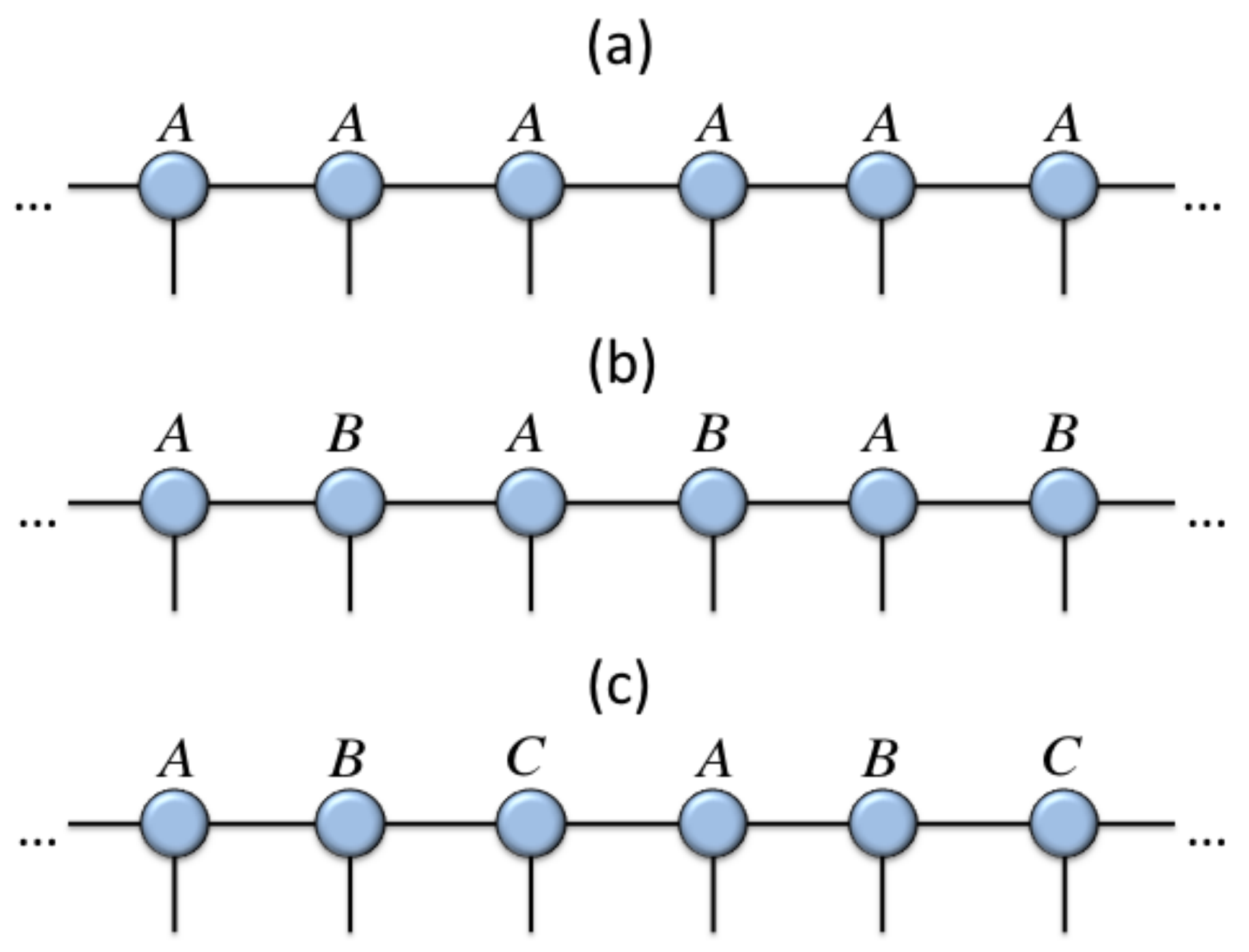} 
\par\end{centering}
\caption{(color online) infinite-MPS with (a) 1-site unit cell, (b) 2-site unit cell, and (3) 3-site unit cell. \label{fig12}}
\end{figure}

Let us now explain briefly some basic properties of MPS: 

\vspace{10pt}

{\bf \emph{1) $1d$ translational invariance and the thermodynamic limit.-}} In principle, all tensors in a finite-size MPS could be different, which means that the MPS itself is not translational invariant (TI). However, it is also possible to impose TI and take the thermodynamic limit of the MPS by choosing some fundamental unit cell of tensors that is repeated over the $1d$ lattice, infinitely-many times. This is represented in Fig.(\ref{fig12}). For instance, if the unit cell is made of one tensor, then the MPS will be TI over one-site shifts. For unit cells of two tensors, the MPS will be TI over two-site shifts. And so on. 
\begin{figure}
\begin{centering}
\includegraphics[width=9cm]{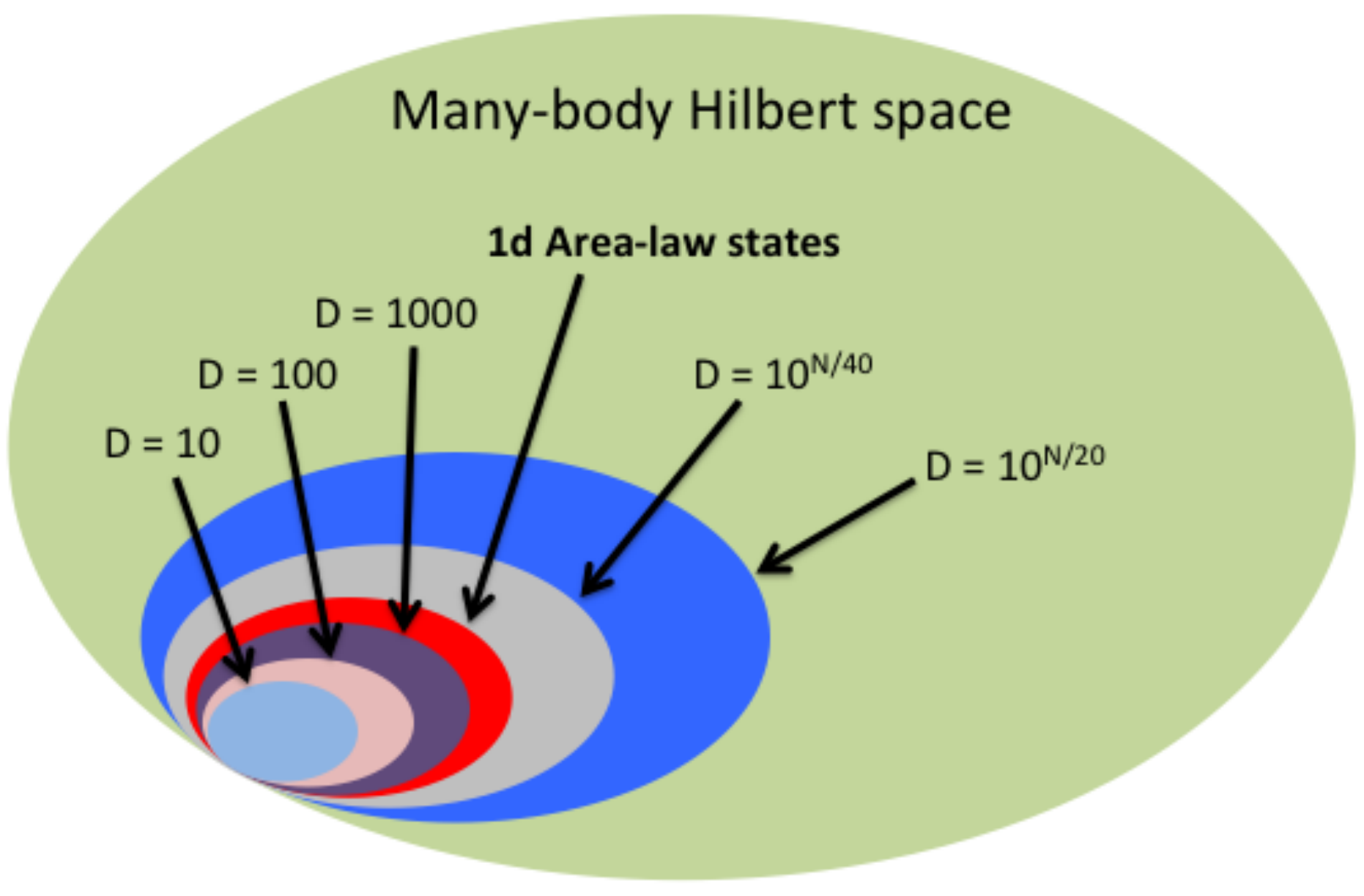} 
\par\end{centering}
\caption{(color online) Onion-like structure of the Hilbert space of a $1d$ quantum many-body system. MPS with finite bond dimension reproduce the properties of the corner of the Hilbert space satisfying the $1d$ area-law for the entanglement entropy. If the bond dimension increases, then the size of the manifold of accessible states also grows. For bond dimensions $D$ sufficiently large (i.e. exponential in the size $N$ of the system), MPS can actually reproduce states beyond the $1d$ area-law and, eventually, cover the whole Hilbert space. Compare this figure to Fig.(\ref{fig4}). \label{fig12b}}
\end{figure}

\vspace{10pt}

{\bf \emph{2) MPS are dense.-}} MPS can represent any quantum state of the many-body Hilbert space just by increasing sufficiently the value of $D$. To cover \emph{all} the states in the Hilbert space $D$ needs to be exponentially large in the system size. However, it is known that low energy states of gapped local Hamiltonians in $1d$ can be efficiently approximated with almost arbitrary accuracy by an MPS with a finite value of $D$ \cite{dmrg1}. For $1d$ critical systems, $D$ tends to diverge \emph{polynomially} in the size of the system \cite{arealaw}. These findings, in turn, explain the accuracy of some MPS-based methods for $1d$ systems such as DMRG. The main pictorial idea behind this property is represented in Fig.(\ref{fig12b}). 

\vspace{10pt}

{\bf \emph{3) One-dimensional area-law.-}} MPS satisfy the area-law scaling of the entanglement entropy adapted to $1d$ systems. This simply means that the entanglement entropy of a block of sites is bounded by a constant, more precisely $S(L) = -{\rm tr}(\rho_L \log \rho_L) = O(\log D)$,  with $\rho_L$ the reduced density matrix of the block. This is exactly the behavior that is usually observed in ground states of gapped $1d$ local Hamiltonians for large size $L$ of the block: precisely, $S(L) \sim  {\rm constant}$ for $L\gg1$ \cite{arealaw}. 

\vspace{10pt} 

\begin{figure}[h]
\begin{centering}
\includegraphics[width=14cm]{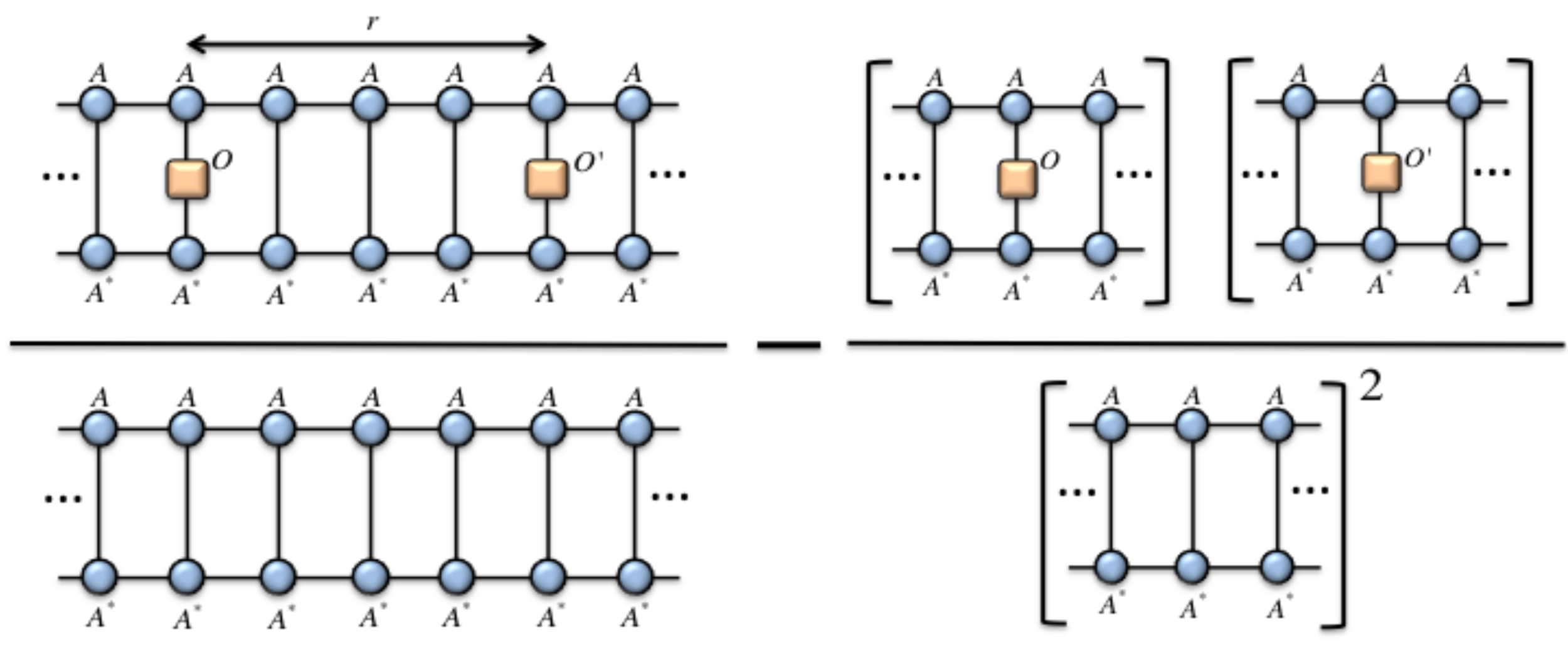} 
\par\end{centering}
\caption{(color online) Diagrams for the two-body correlator $C(r)$. \label{fig13}}
\end{figure}

{\bf \emph{4) MPS are finitely-correlated.-}} The correlation functions of an MPS decay always exponentially with the separation distance. This means that the correlation length of these states is always \emph{finite}, and therefore MPS can not reproduce the properties of critical or scale-invariant systems, where the correlation length is known to diverge \cite{mpsrev}. We can understand this easily with the following example: imagine that you are given a TI and infinite-size MPS defined in terms of one tensor $A$, as in Fig.(\ref{fig12}.a). The two-body correlator
\beq
C(r) \equiv \langle O_i O'_{i+r}  \rangle - \langle O_i \rangle \langle O'_{i+r} \rangle
\eeq
of one-body operators $O_i$ and $O'_{i+r}$ at sites $i$ and $i+r$ can be represented diagrammatically as in Fig.(\ref{fig13}). 

\begin{figure}[h]
\begin{centering}
\includegraphics[width=10cm]{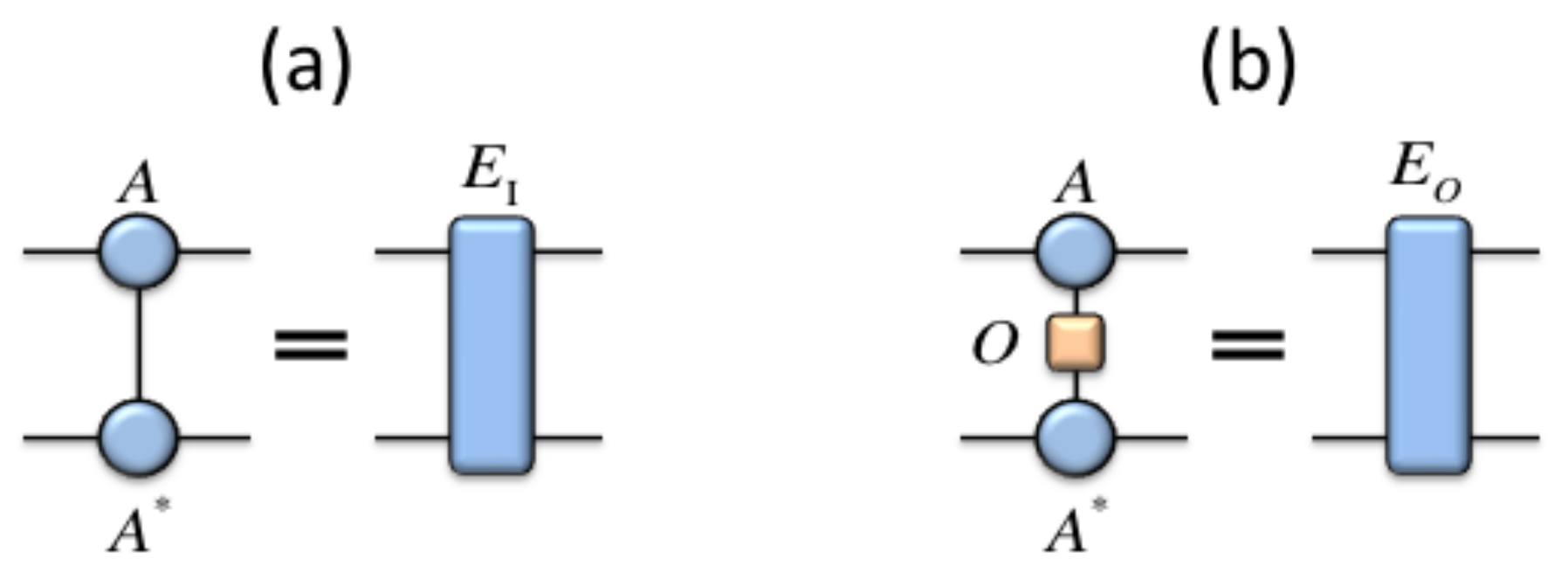} 
\par\end{centering}
\caption{(color online) (a) Transfer matrix $E_{{\mathbb I}}$; (b) matrix $E_{O}$. \label{fig14}}
\end{figure}

\begin{figure}
\begin{centering}
\includegraphics[width=13cm]{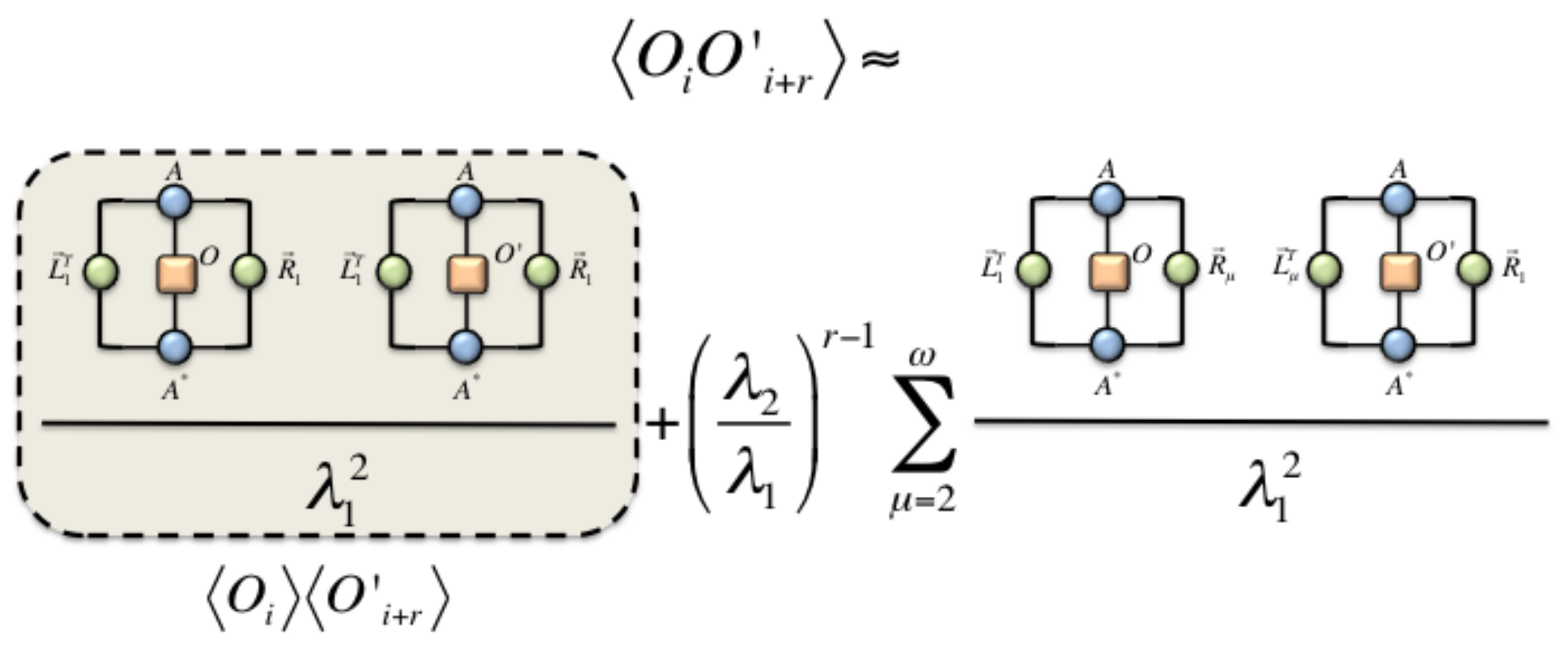} 
\par\end{centering}
\caption{(color online) Diagrams for $\langle O_i O'_{i+r}  \rangle$ for large separation distance $r$. The first part corresponds to $\langle O_i \rangle \langle O'_{i+r}  \rangle$. \label{fig15}}
\end{figure}

The zero-dimensional transfer matrix $E_{{\mathbb I}}$ in Fig.(\ref{fig14}.a) plays a key role in this calculation. In particular, we have that 
\beq
(E_{{\mathbb I}})^r = (\lambda_1)^r  \sum_{\mu = 1}^{D^2} \left(\frac{\lambda_i}{\lambda_1}\right)^r \vec{R}_i^T \vec{L}_i \ ,
\eeq
where $\lambda_i$ are the $i=1, 2, \ldots, D^2$ eigenvalues of $E_{{\mathbb I}}$ sorted in order of decreasing magnitude, and $\vec{R}_i, \vec{L}_i$ their associated right- and left-eigenvectors. Assuming that the largest magnitude eigenvalue $\lambda_1$ is non-degenerate, for $r \gg 1$ we have that 
\beq
(E_{{\mathbb I}})^r \sim (\lambda_1)^r  \left (\vec{R}_1^T \vec{L}_1 +  \left(\frac{\lambda_2}{\lambda_1}\right)^r \sum_{\mu = 2}^{\omega+1} \vec{R}_{\mu}^T \vec{L}_{\mu} \right) \ , 
\eeq
where $\omega$ is the degeneracy of $\lambda_2$. Defining the matrices $E_O$ and $E_{O'}$ as in Fig.(\ref{fig14}.b), and using the above equation, 
it is easy to see that 
\beq
\langle O_i O'_{i+r}  \rangle \sim \frac{(\vec{L}_1 E_O \vec{R}_1^T) (\vec{L}_1 E_{O'} \vec{R}_1^T)}{\lambda_1^2} +  \left(\frac{\lambda_2}{\lambda_1}\right)^{r-1} \sum_{\mu = 2}^{\omega +1} \frac{(\vec{L}_1 E_O \vec{R}_{\mu}^T) (\vec{L}_{\mu} E_{O'} \vec{R}_1^T)}{\lambda_1^2} \ ,
\eeq
which is expressed in terms of diagrams as in Fig.(\ref{fig15}). In this equation, the first term is nothing but $\langle O_i \rangle \langle O'_{i+r} \rangle$. Therefore, $C(r)$ is given for large $r$ by 
\beq
C(r) \sim \left(\frac{\lambda_2}{\lambda_1}\right)^{r-1} \sum_{\mu = 2}^{\omega + 1} \frac{(\vec{L}_1 E_O \vec{R}_{\mu}^T) (\vec{L}_{\mu} E_{O'} \vec{R}_1^T)}{\lambda_1^2}
\eeq
so that 
\beq
C(r) \sim f(r) a e^{-r/\xi} 
\eeq
with a proportionality constant $a = O(\omega)$, $f(r)$ a site-dependent phase $=\pm 1$ if $O$ and $O'$ are hermitian, and correlation length $\xi \equiv -1/\log{\left|\lambda_2 / \lambda_1 \right|}$. Importantly, this type of exponential decay of two-point correlation functions for large $r$ is the typical one in ground states of gapped non-critical $1d$ systems, which is just another indication that MPS are able to approximate well this type of states. 

\vspace{10pt}

\begin{figure}
\begin{centering}
\includegraphics[width=10cm]{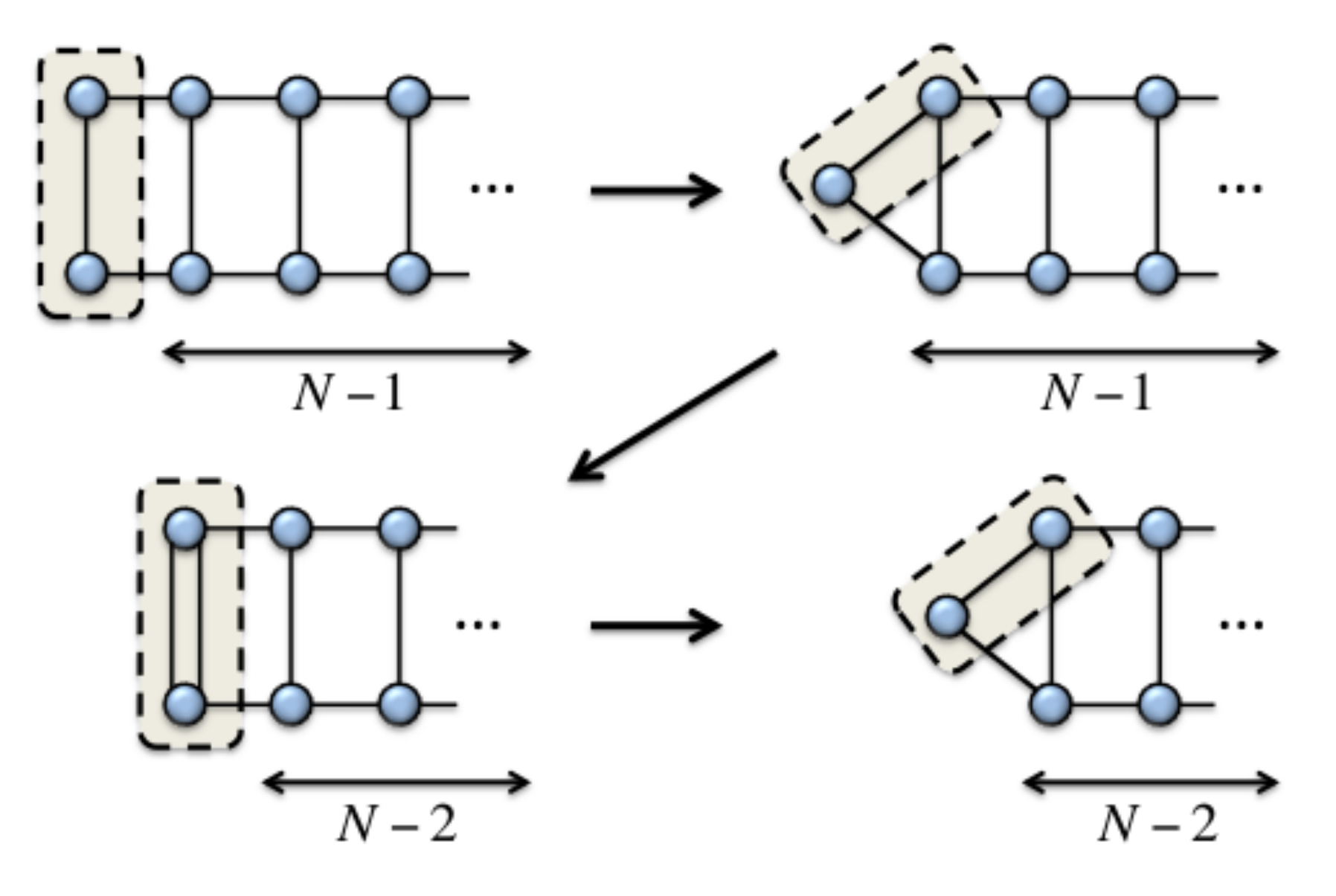} 
\par\end{centering}
\caption{(color online) Order of contractions for the scalar product of a finite MPS, starting from the left. The same strategy could be used for an infinite MPS, starting with some boundary condition at infinity and iterating until convergence. 
\label{fig16}}\end{figure}

{\bf \emph{5) Exact calculation of expectation values.-}} The exact calculation of the scalar product between two MPS for $N$ sites can always be done exactly in a time $O(N p D^3)$. We explain the basic idea for this calculation in Fig.(\ref{fig16}). For an infinite system, the calculation can be done in  $O(p D^3)$ using similar techniques as the ones explained above for the calculation of the two-point correlator $C(r)$ (namely, finding the dominant eigenvalue and dominant left/right eigenvectors of the transfer matrix $E_{{\mathbb I}}$). In general, expectation values of local observables such as correlation functions, energies, and local order parameters, can also be computed using the same kind of tensor manipulations. 

\vspace{10pt}

{\bf \emph{6) Canonical form and the Schmidt decomposition.-} } Given a quantum state $\ket{\Psi}$ in terms of an MPS with open boundary conditions, there is a choice of tensors called \emph{canonical form of the MPS} \cite{tebd, itebd} which is extremely convenient. This is defined as follows: for a given MPS with open boundary conditions (either for a finite or infinite system), we say that it is in its \emph{canonical form} \cite{itebd} if, for each bond index $\alpha$, the index corresponds to the labeling of Schmidt vectors in the Schmidt decomposition of $\ket{\Psi}$ across that index, i.e:
\beq
\ket{\Psi} = \sum_{\alpha = 1}^D \lambda_{\alpha} \ket{\Phi_{\alpha}^L} \otimes \ket{\Phi_{\alpha}^R} \ .
\eeq
In the above equation, $\lambda_{\alpha}$ are Schmidt coefficients ordered into decreasing order ($\lambda_1 \ge \lambda_2 \ge \cdots \ge 0$), and the Schmidt vectors form orthonormal sets, that is, $\bra{\Phi_{\alpha}^L}\Phi_{\alpha'}^L \rangle = \bra{\Phi_{\alpha}^R}\Phi_{\alpha'}^R \rangle = \delta_{\alpha \alpha'}$. 

\begin{figure}[h]
\begin{centering}
\includegraphics[width=11cm]{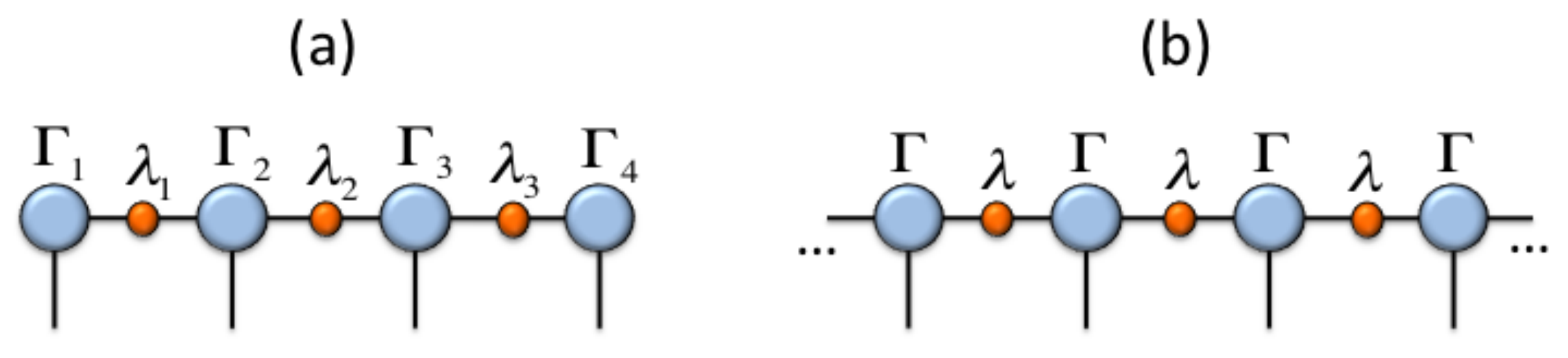} 
\par\end{centering}
\caption{(color online) (a) 4-site MPS in canonical form; (b) infinite MPS with 1-site unit cell in canonical form. \label{fig17}}
\end{figure}

For a finite system of $N$ sites \cite{tebd}, the above condition corresponds to having the decomposition for the coefficient of the wave-function
\beq
C_{i_1 i_2 \ldots i_N} = \Gamma^{[1] i_1}_{\alpha_1} \lambda^{[1]}_{\alpha_1} \Gamma^{[2] i_2}_{\alpha_1 \alpha_2} \lambda^{[2]}_{\alpha_2} \Gamma^{[3] i_3}_{\alpha_2 \alpha_3} \lambda^{[3]}_{\alpha_3} \cdots  \lambda^{[N-1]}_{\alpha_{N-1}} \Gamma^{[N] i_N}_{\alpha_{N-1}} \ ,  
\label{canonimps}
\eeq
where the $\Gamma$ tensors correspond to changes of basis between the different Schmidt basis and the computational (spin) basis, and the vectors $\lambda$ correspond to the Schmidt coefficients. In the case of an infinite MPS with one-site traslation invariance \cite{itebd}, the canonical form corresponds to having just one tensor $\Gamma$ and one vector $\lambda$ describing the whole state. Regarding the Schmidt coefficients as the entries of a diagonal matrix, the TN diagram for both the finite and infinite MPS in canonical form are shown in Fig.(\ref{fig17}). 

Let us now show a way to obtain the canonical form of an MPS $\ket{\Psi}$ for a finite system from successive Schmidt decompositions \cite{tebd}. If we perform the Schmidt decomposition between the site $1$ and the remaining $N-1$, we can write the state as
\begin{equation}
|\Psi \rangle = \sum_{\alpha_1=1}^{{\rm min}(p,D)} \lambda^{[1]}_{\alpha_1} |\tau^{[1]}_{\alpha_1}\rangle \otimes |\tau^{[2 \cdots N]}_{\alpha_1}\rangle \ , 
\label{smdec}
\end{equation}
where $\lambda^{[1]}_{\alpha_1}$ are the Schmidt coefficients, and $|\tau^{[1]}_{\alpha_1}\rangle$, $|\tau^{[2 \cdots N]}_{\alpha_1}\rangle $ are the corresponding left and right Schmidt vectors. If we rewrite the left Schmidt vector in terms of the local basis $\ket{i_1}$ for site $1$,  the state $\ket{\Psi}$ can then be written as
\begin{equation}
|\Psi \rangle = \sum_{i_1=1}^p \sum_{\alpha_1=1}^{{\rm min}(p,D)} \Gamma^{[1] i_1}_{\alpha_1} \lambda^{[1]}_{\alpha_1} |i_1 \rangle \otimes |\tau^{[2 \cdots N]}_{\alpha_1}\rangle \ ,
\label{gamaequation}
\end{equation}
where $\Gamma^{[1] i_1}_{\alpha_1}$ correspond to the change of basis $|\tau^{[1]}_{\alpha_1}\rangle = \sum_{i_1}  \Gamma^{[1]i_1}_{\alpha_1} |i_1 \rangle $. Next, we expand each Schmidt vector $|\tau^{[2 \cdots N]}_{\alpha_1}\rangle$ as 
\begin{equation}
 |\tau^{[2 \cdots n]}_{\alpha_1}\rangle = \sum_{i_2=1}^p |i_2 \rangle \otimes |\omega^{[3 \cdots N]}_{\alpha_1 i_2} \rangle \ . 
 \label{cosita}
 \end{equation}
We now write the unnormalised quantum state  $|\omega^{[3 \cdots N]}_{\alpha_1 i_2} \rangle$ in terms of the at most $p^2$ eigenvectors of the reduced density matrix for systems $[3, \ldots, N]$, that is, in terms of the right Schmidt vectors $|\tau^{[3 \cdots n]}_{\alpha_2} \rangle $ of the bipartition between subsystems $[1, 2]$ and the rest, together with the corresponding Schmidt coefficients $\lambda^{[2]}_{\alpha_2}$: 
\begin{equation}
|\omega^{[3 \cdots N]}_{\alpha_1 i_2} \rangle = \sum_{\alpha_2=1}^{{\rm min}(p^2,D)} \Gamma^{[2] i_2}_{\alpha_1 \alpha_2} \lambda^{[2]}_{\alpha_2} |\tau^{[3 \cdots N]}_{\alpha_2} \rangle \  . 
\label{ecc}
\end{equation}
Replacing the last two expressions into Eq.(\ref{gamaequation}) we get 
\begin{equation}
|\Psi \rangle = \sum_{i_1, i_2 = 1}^p \sum_{\alpha_1=1}^{{\rm min}(p,D)} \sum_{\alpha_2=1}^{{\rm min}(p^2,D)} \left( \Gamma^{[1] i_1}_{\alpha_1} \lambda^{[1]}_{\alpha_1} \Gamma^{[2] i_2}_{\alpha_1 \alpha_2} \lambda^{[2]}_{\alpha_2} \right)  |i_1\rangle \otimes |i_2 \rangle \otimes |\tau^{[3 \cdots N]}_{\alpha_2} \rangle \ . 
\label{iuid}
\end{equation}
Iterating the above procedure for all subsystems, we finally get 
\begin{equation}
\label{gammalambda}
 |\Psi \rangle =
\sum_{\{i\}}\sum_{\{\alpha \}}\left( \Gamma^{[1]i_1}_{\alpha_1} \lambda^{[1]}_{\alpha_1}
\Gamma^{[2]i_2}_{\alpha_1\alpha_2} \lambda^{[2]}_{\alpha_2}
\dots \lambda^{[N-1]}_{\alpha_{N-1}}\Gamma^{[N] i_N}_{\alpha_{N-1}} \right) \ket{i_1} \otimes \ket{i_2} \otimes \cdots \otimes \ket{i_N}
 \ , 
 \end{equation} 
where the sum over each index in $\{s\}$ and $\{\alpha\}$ runs up to their respective allowed values. And this is nothing but the representation that we mentioned in Eq.(\ref{canonimps}). 
 
For an infinite MPS one can also compute the canonical form \cite{itebd}. In this case one just needs to notice that, in the canonical form, the bond indices of the MPS always correspond to orthonormal vectors to the left and right. Thus, finding the canonical form of an MPS is commonly referred to also as \emph{orthonormalizing the indices} of the MPS.  For an infinite system defined by a single tensor $A$ this canonical form can be found following the procedure indicated in the diagrams in Fig.(\ref{fig18}). This can be summarized in three main steps: 

\begin{figure}[h]
\begin{centering}
\includegraphics[width=12cm]{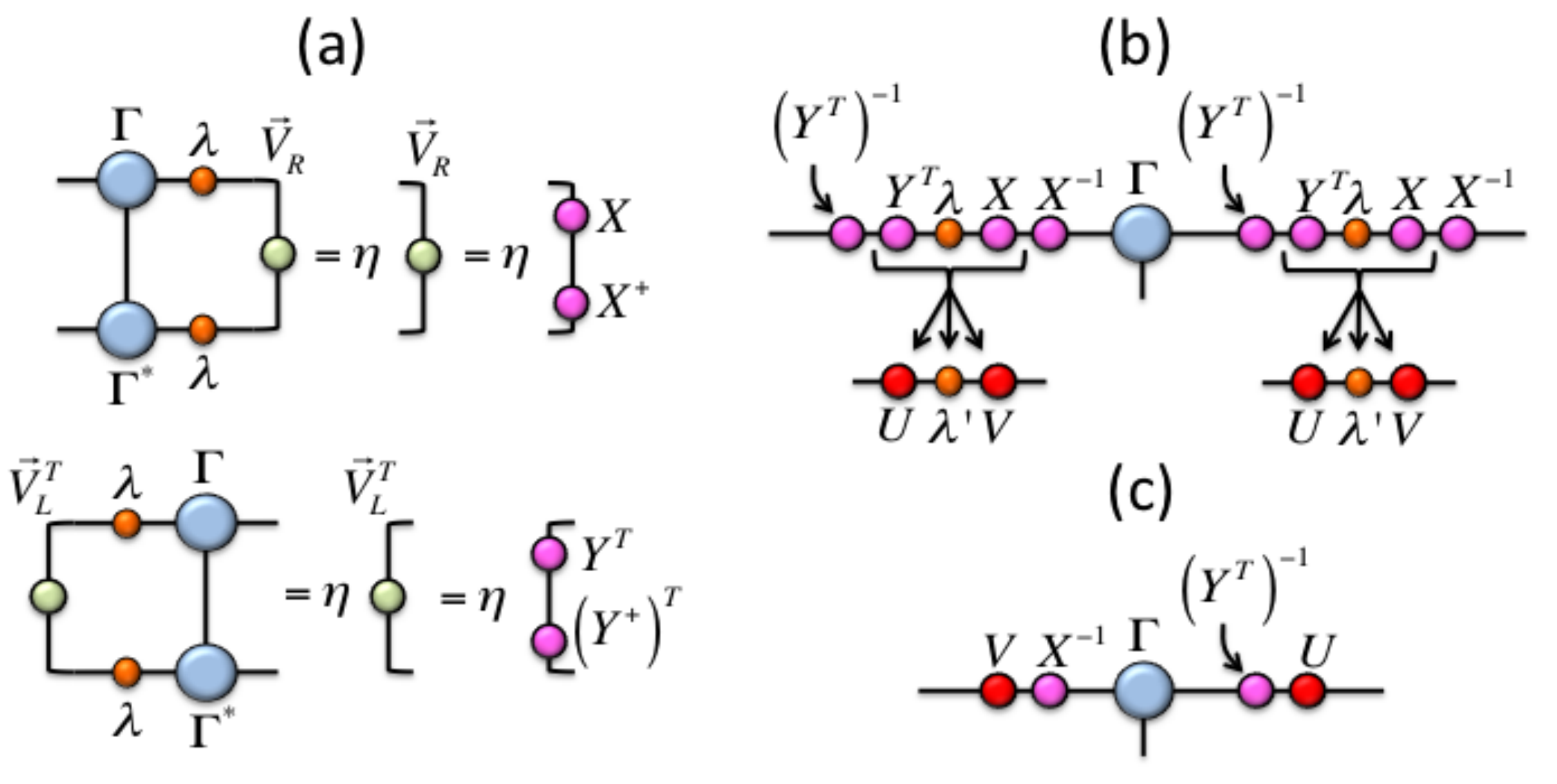} 
\par\end{centering}
\caption{(color online) Canonicalization of an infinite MPS with 1-site unit cell (see text).\label{fig18}}
\end{figure}

\vspace{10pt} 

(i) Find the dominant right eigenvector $\vec{V}_R$ and the dominant left eigenvector $\vec{V}_L$ of the transfer matrices defined in Fig.(\ref{fig18}.a). Regarding the bra/ket indices, $\vec{V}_R$ and $\vec{V}_L$ can also be understood as hermitian and positive matrices. Decompose these matrices as squares, $V_R = XX^{\dagger}$ and $V_L = Y^{\dagger} Y$, as shown in the figure.

\vspace{10pt}

(ii) Introduce $\mathbb{I} = (Y^T)^{-1} Y^T$ and $\mathbb{I} = X X^{-1}$ in the bond indices of the MPS as shown in Fig.(\ref{fig18}.b). Next, calculate the singular value decomposition of the matrix product $Y^T \lambda X = U \lambda' V$, where $U$ and $V$ are unitary matrices and $\lambda'$ are the singular values. It is easy to show that these singular values correspond to the Schmidt coefficients of the Schmidt decomposition of the MPS. 

\vspace{10pt}

(iii) Arrange the remaining tensors into a new tensor $\Gamma'$, as shown in Fig.(\ref{fig18}.c). The MPS is now defined in terms of $\lambda'$ and $\Gamma'$.

The above procedure produces an infinite MPS such that all its bond indices correspond to orthonormal Schmidt basis, and is therefore in canonical form by construction\footnote{Two comments are in order: first, here we do not consider the case in which the dominant eigenvalues of the transfer matrices are degenerate. Second, this canonical form can also be achieved by running the iTEBD algorithm on an MPS with an identity time evolution until convergence \cite{itebd}. Quite probably, the same result can be obtained by running iDMRG \cite{iDMRG} with identity Hamiltonian and alternating left and right sweeps until convergence.}.

\begin{figure}[h]
\begin{centering}
\includegraphics[width=10cm]{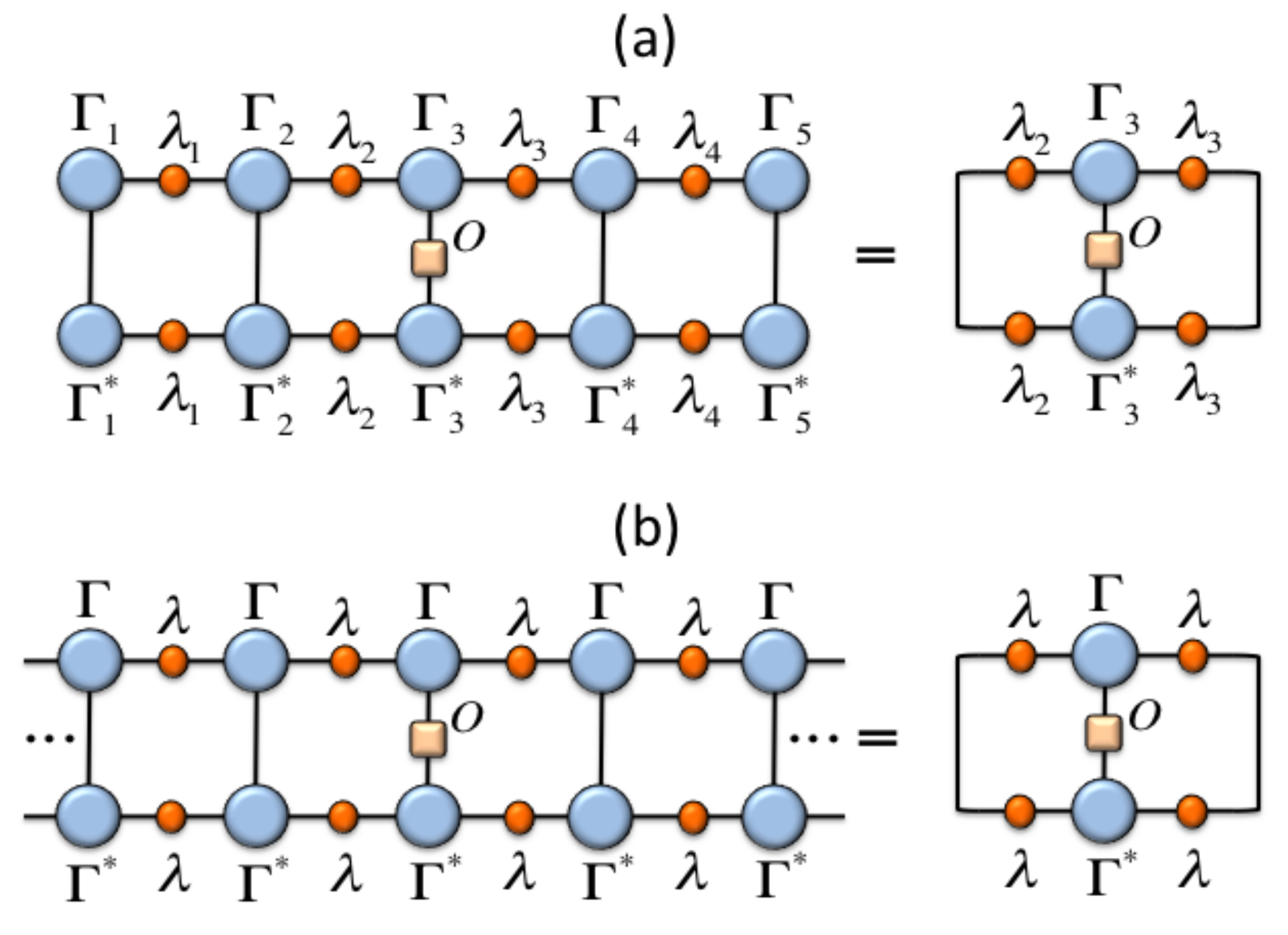} 
\par\end{centering}
\caption{(color online) Expectation value of a 1-site observable for an MPS in canonical form: (a) 5-site MPS and (b) infinite MPS with 1-site unit cell.  \label{fig19}}
\end{figure}

The canonical form of an MPS has a number of properties that make it very useful for MPS calculations. First, the eigenvalues of the reduced density matrices of different ``left vs right" bipartitions are just the square of the Schmidt coefficients, which is very useful for calculations of e.g. entanglement spectrums and entanglement entropies. Moreover, the calculation of expectation values of local operators simplifies a lot, see the diagrams in Fig.(\ref{fig19}). 
 
But most importantly, the canonical form provides a prescription for the truncation of the bond indices of an MPS in numerical simulations: just keep the largest $D$ Schmidt coefficients at every bond at each simulation step. This truncation procedure is optimal for a finite system as long as we keep the locality of the truncation (i.e. only the tensors involved in the truncated index are modified), see e.g. Ref.\cite{tebd}. This prescription for truncating the bond indices turns out to be really useful, and is at the basis of the TEBD method and related algorithms for $1d$ systems. 

\subsubsection{Some examples}

\begin{figure}[h]
\begin{centering}
\includegraphics[width=8cm]{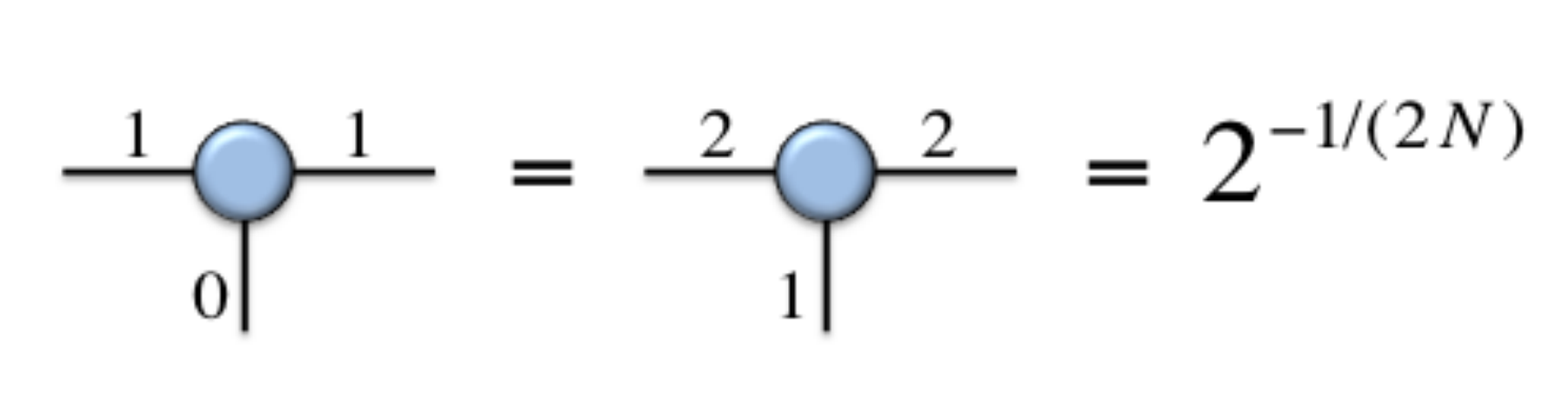} 
\par\end{centering}
\caption{(color online) Non-zero components for the MPS tensors of the GHZ state. \label{fig20}}
\end{figure}

Let us now give four specific examples of non-trivial states that can be represented exactly by MPS: 

\vspace{10pt} 

{\bf \emph{1) GHZ state.-}} The GHZ state of $N$ spins-$1/2$ is given by 
\beq
\ket{GHZ} = \frac{1}{\sqrt{2}} \left( \ket{0}^{\otimes N} + \ket{1}^{\otimes N} \right) \ ,
\eeq
where $\ket{0}$ and $\ket{1}$ are e.g. the eigenstates of the Pauli $\sigma_z$ operator (spin ``up" and ``down") \cite{ghz}. This is a highly entangled quantum state of the $N$ spins, which has some non-trivial entanglement properties (e.g. it violates certain $N$-partite Bell inequalities). Still, this state can be represented exactly by a MPS with bond dimension $D = 2$ and periodic boundary conditions. The non-zero coefficients in the tensor are shown in the diagram of Fig.(\ref{fig20}). 

\vspace{10pt}

\begin{figure}
\begin{centering}
\includegraphics[width=10cm]{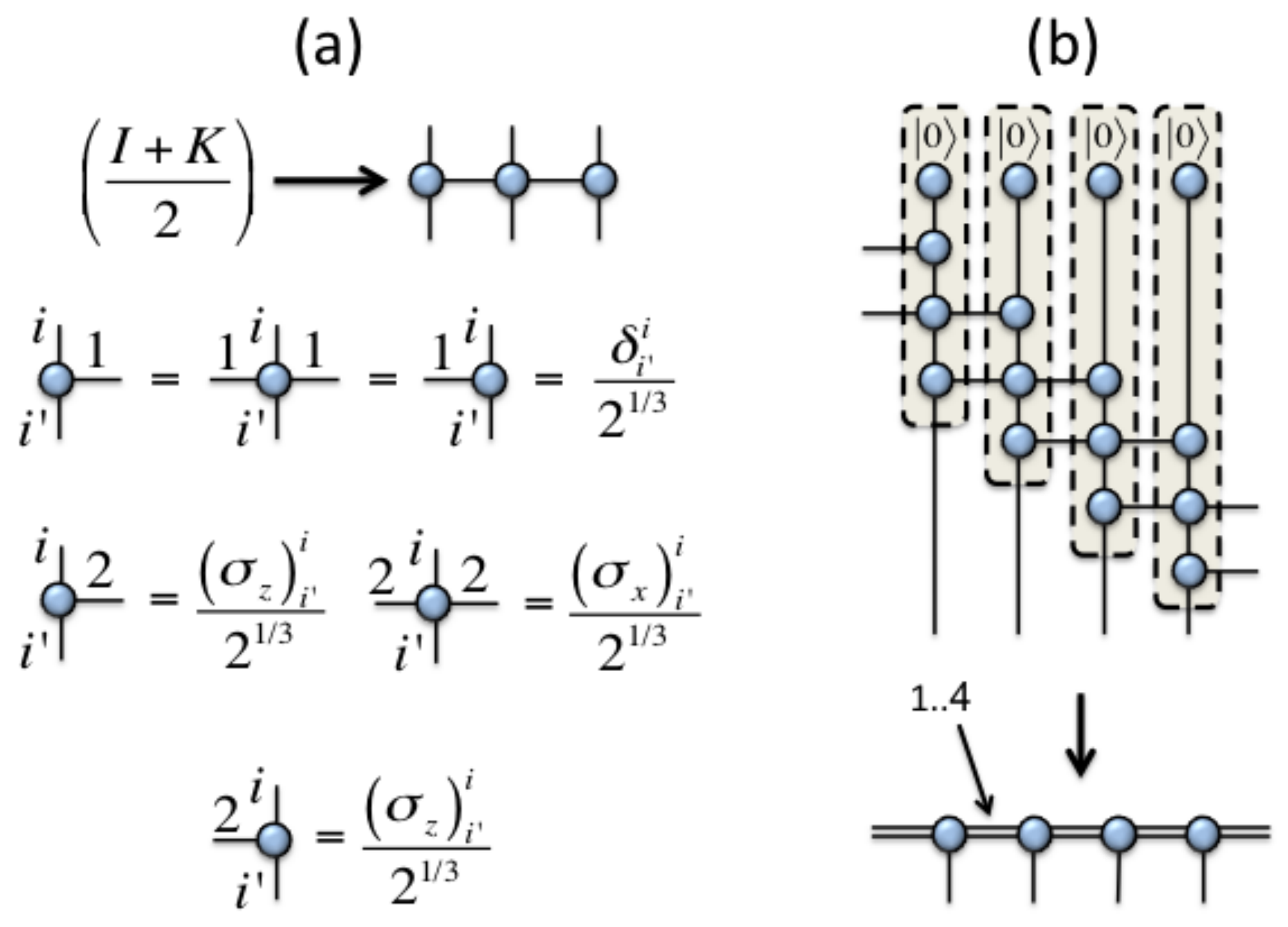} 
\par\end{centering}
\caption{(color online) MPS for the $1d$ cluster state: (a) tensor network decomposition of the operator $\left( \mathbb{I} + K^{[i]}\right)/2$ and non-zero coefficients of the tensors; (b) construction of the infinite MPS with 1-site unit cell. \label{fig21}}
\end{figure}

{\bf \emph{2) $1d$ cluster state.-}} Introduced by Raussendorf and Briegel \cite{cluster}, the cluster state in a $1d$ chain can be seen as the $+1$ eigenstate of a set of mutually commuting stabilizer operators $\{ K^{[i]} \}$ defined as
\beq
K^{[i]} \equiv \sigma_z^{i-1}\sigma_x^{i} \sigma_z^{i+1} \ ,
\eeq
where the $\sigma_{\alpha}^{i}$ are the usual spin-$1/2$ Pauli matrices with $\alpha \in \{ x,y,z \}$ at lattice site $i$. Since $(K^{i})^2 = \mathbb{I}$, for an infinite system this quantum state can be written (up to an overall normalization constant) as
\beq
\ket{\Psi_{1dCL}} =  \prod_{i} \frac{\left( \mathbb{I} + K^{[i]}\right)}{2}\ket{0}^{\otimes N \rightarrow \infty} \ .  
\eeq
Each one of the terms $\left( \mathbb{I} + K^{[i]}\right)/2$ is a projector that admits a TN representation with bond dimension $2$ as in Fig.(\ref{fig21}.a). From here, it is easy to obtain an MPS description with bond dimension $D=4$ for the $1d$ cluster state $\ket{\Psi_{1dCL}}$, as shown in Fig.(\ref{fig21}.b). The non-zero coefficients of the MPS tensors follows easily from the corresponding TN contractions in the diagram.  
\begin{figure}
\begin{centering}
\includegraphics[width=12cm]{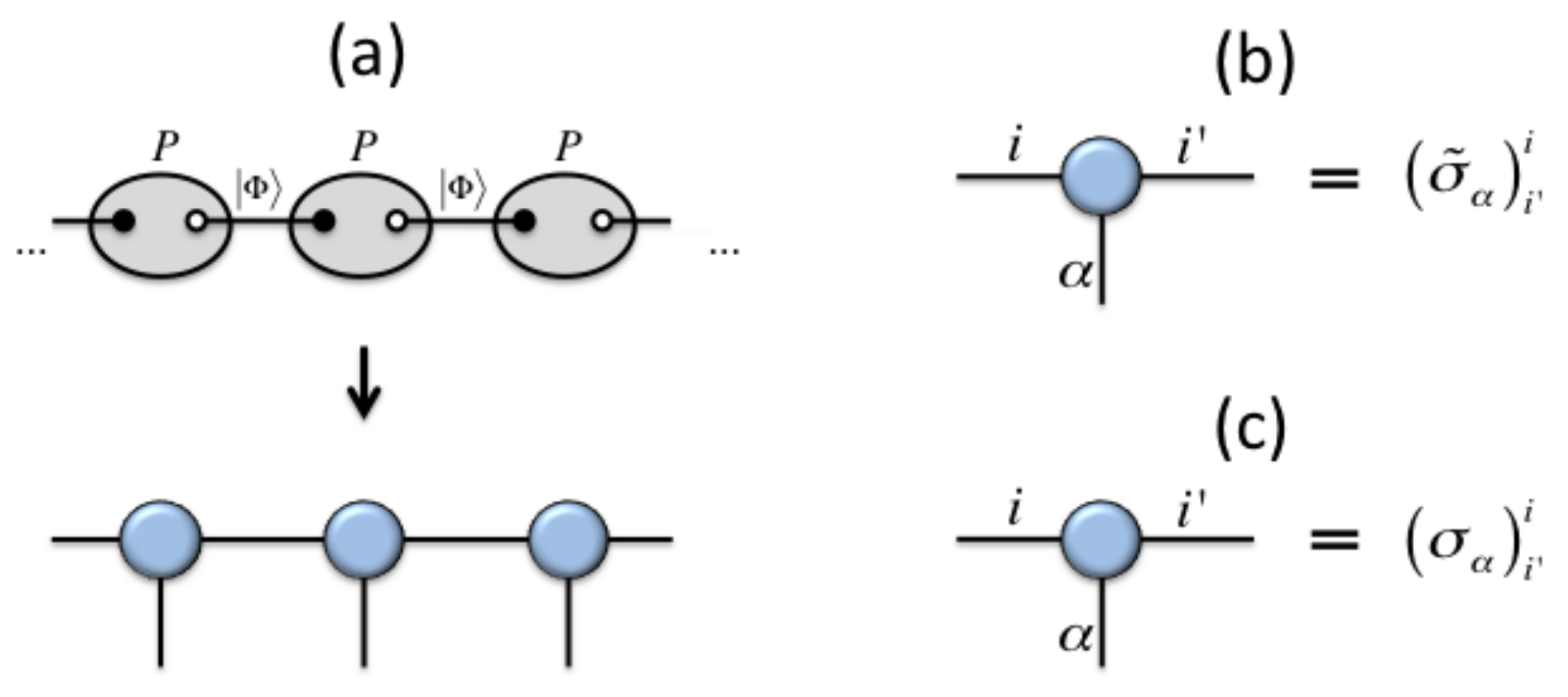} 
\par\end{centering}
\caption{(color online) MPS for the AKLT state: (a) spin-$1/2$ particles arranged in singlets $\ket{\Phi} = 2^{-1/2} (\ket{0}\otimes \ket{1} - \ket{1} \otimes \ket{0})$, and projected by pairs into the spin-1 subspace by projector $P$; (b-c) non-zero components of the tensors for the infinite MPS with 1-site unit cell: (b) in terms of $ \widetilde{\sigma}_1 = \sqrt{2} \sigma_+, \widetilde{\sigma}_2 = -\sqrt{2} \sigma_-$ and $\widetilde{\sigma}_3 = \sigma_z$, with $\sigma_{\pm} = (\sigma_x \pm \sigma_y)/2$; (c) There is a gauge in which these coefficients are given by the three spin-1/2 Pauli matrices ${\sigma}_1 = \sigma_x,  {\sigma}_2 = \sigma_y$ and ${\sigma}_3 = \sigma_z$. \label{fig22}}
\end{figure}

\vspace{10pt}

{\bf \emph{3) $1d$ AKLT model.-}} The state we consider now is the ground state of the $1d$ AKLT model \cite{aklt}. This is a quantum spin chain of spin-$1$, that is given by the Hamiltonian 
\beq
H = \sum_i \left( \vec{S}^{[i]} \vec{S}^{[i+1]} + \frac{1}{3}(\vec{S}^{[i]} \vec{S}^{[i+1]})^2\right) \ , 
\label{1daklt}
\eeq
where $\vec{S}^{[i]}$ is the vector of spin-$1$ operators at site $i$, and where again we assumed an infinite-size system. This model was introduced by Affleck, Kennedy, Lieb and Tasaki in Ref.\cite{aklt}, and it was the first analytical example of a quantum spin chain supporting the so-called Haldane's conjecture: it is a local spin-$1$ Hamiltonian with Heisenberg-like interactions and a non-vanishing spin gap in the thermodynamic limit. What is also remarkable about this model is that its ground state is given exactly, and by construction, in terms of a MPS with bond dimension $D = 2$. This can be understood in terms of a collection of spin-$1/2$ singlets, whose spins are paired and projected into spin-$1$ subspaces as indicated in Fig.(\ref{fig22}.a). This, by construction, is an MPS with $D=2$. Interestingly, there is a choice of tensors for the MPS (i.e. a \emph{gauge}) such that these are given by the three spin-$1/2$ Pauli matrices, which are the generators of the irreducible representation of SU(2) with $2 \times 2$ matrices (see Fig.(\ref{fig22}.c)). We will not enter into details of why this representation for the MPS tensors is possible. For the curious reader, let us simply mention that this is a consequence of the SU(2) symmetry of the Hamiltonian, which is inherited by the ground state, and which is also reflected at the level of the individual tensors of the MPS.  

\vspace{10pt}

\begin{figure}[h]
\begin{centering}
\includegraphics[width=13cm]{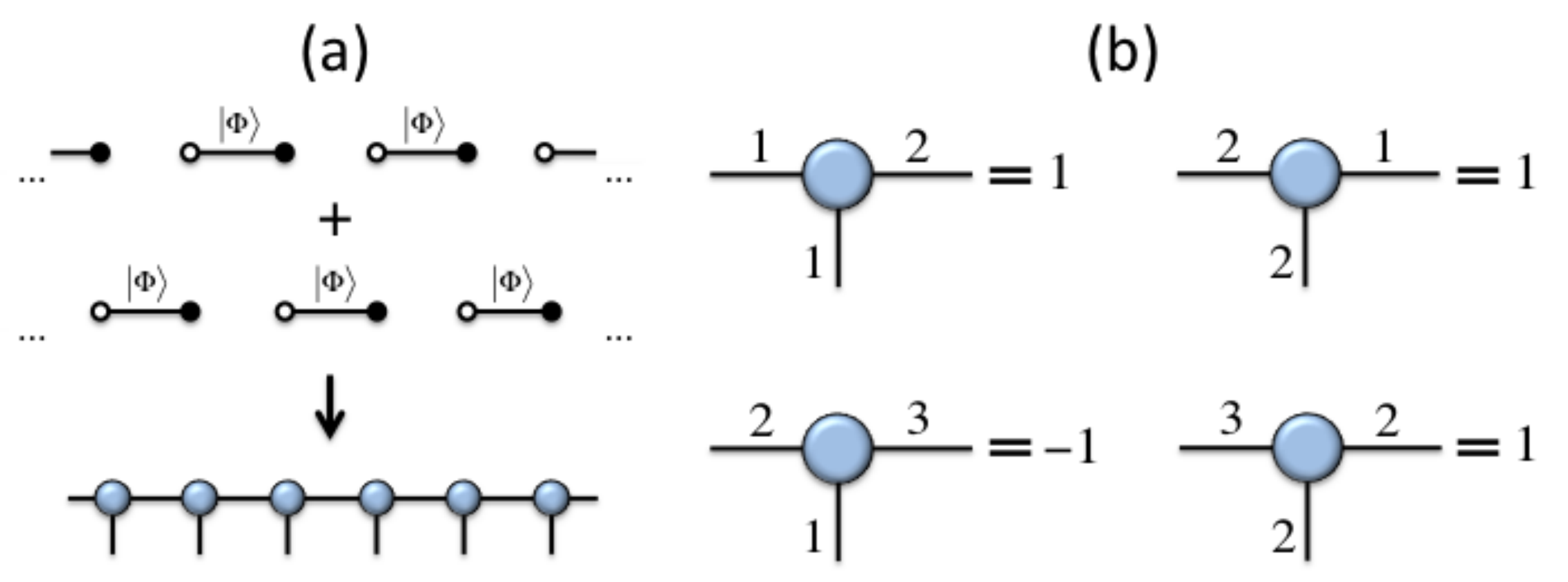} 
\par\end{centering}
\caption{(color online) MPS for the Majumdar-Gosh state: (a) the superposition of two dimerized states of singlets $\ket{\Phi}$ in (a) can be written in terms of an infinite MPS with 1-site unit cell, with non-zero coefficients as in (b). \label{fig23}}
\end{figure}

{\bf \emph{4) Majumdar-Gosh model.-} } We now consider the ground state of the Majumdar-Gosh model \cite{majgosh}, which is a frustrated $1d$ spin chain defined by the Hamiltonian
\beq
H = \sum_i \left( \vec{S}^{[i]} \vec{S}^{[i+1]} + \frac{1}{2} \vec{S}^{[i]} \vec{S}^{[i+2]} \right) \ ,
\eeq
where $\vec{S}^{[i]}$ is the vector of spin-$1/2$ operators at site $i$. The ground state of this model is given by singlets between nearest-neighbor spins, as shown in Fig.(\ref{fig23}). Nevertheless, to impose translational invariance we need to consider the superposition between this state and its traslation by one lattice site. The resultant state can be written in compact notation with an MPS of bond dimension $D=3$, also as shown in Fig.(\ref{fig23}). 

\vspace{10pt}

\subsection{Projected Entangled Pair States (PEPS)}

The family of PEPS \cite{PEPS} is just the natural generalization of MPS to higher spatial dimensions. Here we shall only consider the $2d$ case. $2d$ PEPS are at the basis of several methods to simulate $2d$ quantum lattice systems, e.g. PEPS \cite{PEPS} and infinite-PEPS \cite{iPEPS} algorithms, as well as Tensor Renormalization Group (TRG) \cite{TRG}, Second Renormalization Group (SRG) \cite{SRG}, Higher-Order Tensor Renormalization Group (HOTRG) \cite{HOTRG}, and methods based on Corner Transfer Matrices (CTM) and Corner Tensors \cite{dctm,ctmrg,ctens}. In Secs.\ref{sec6}-\ref{sec7} of these notes we will describe basic aspects of some of these methods.   

\begin{figure}[h]
\begin{centering}
\includegraphics[width=12cm]{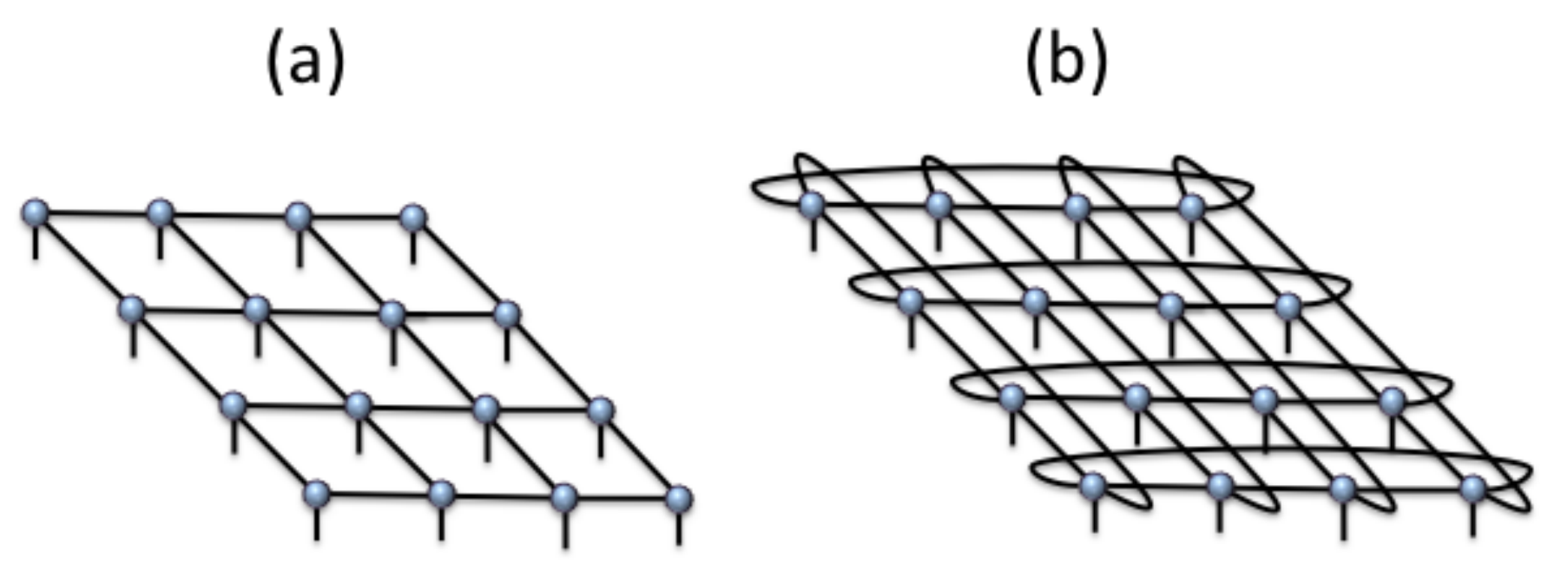} 
\par\end{centering}
\caption{(color online) $4 \times 4$ PEPS: (a) open boundary conditions, and (b) periodic boundary conditions. \label{fig24}}
\end{figure}

PEPS are TNs that correspond to a $2d$ array of tensors. For instance, for a $4 \times 4$ square lattice, we show the corresponding PEPS in Fig.(\ref{fig24}), both with open and periodic boundary conditions. As such this generalization may look quite straightforward, yet we will see that the properties of PEPS are remarkably different from those of MPS. Of course, one can also define PEPS for other types of $2d$ lattices e.g. honeycomb, triangular, kagome...  yet, in these notes we mainly consider the square lattice case. Moreover, and as expected, there are also two types of indices in a PEPS: physical indices, of dimension $p$, and bond indices, of dimension $D$. 

\subsubsection{Some properties}

\begin{figure}[h]
\begin{centering}
\includegraphics[width=14cm]{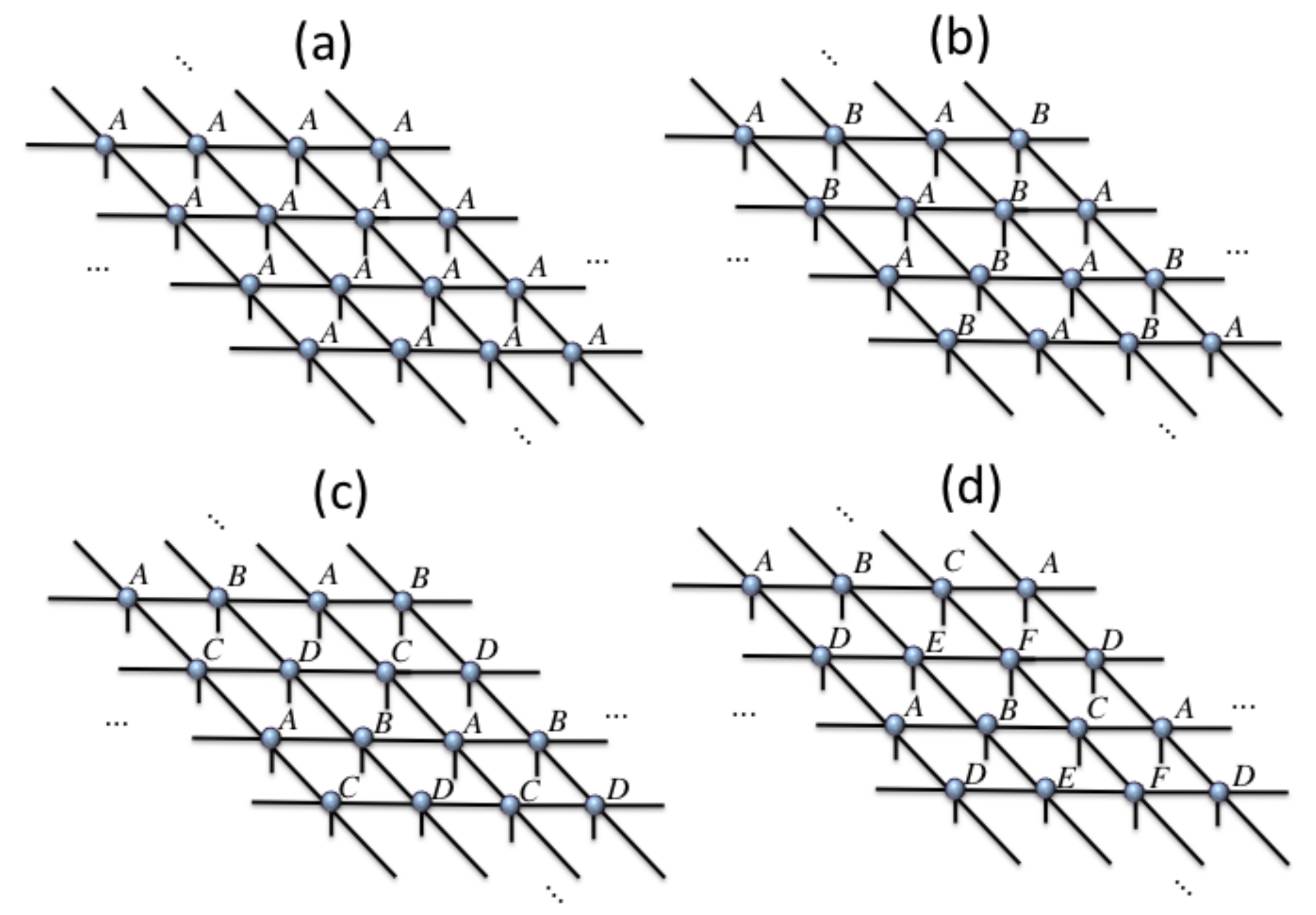} 
\par\end{centering}
\caption{(color online) Infinite PEPS with (a) $1 \times 1$ unit cell, (b) $2 \times 2$ unit cell with 2 tensors, (c) $2 \times 2$ unit cell with 4 tensors, and (d) $2 \times 3$ unit cell with 6 tensors. \label{fig25}}
\end{figure}

We now sketch some of the basic properties of PEPS: 

\vspace{10pt} 

{\bf \emph{1) $2d$ Translational invariance and the thermodynamic limit.-}} As in the case of MPS, one is free to choose all the tensors in the PEPS to be different, which leads to a PEPS that is not TI. Yet, it is again possible to impose TI and take the thermodynamic limit by choosing a fundamental unit cell that is repeated all over the (infinite) $2d$ lattice, see e.g. Fig.(\ref{fig25}). As expected for higher-dimensional systems, TI needs to be imposed in all the spatial directions of the lattice. 

\vspace{10pt}

{\bf \emph{2) PEPS are dense.-}} As was the case of MPS, PEPS is a dense family of states, meaning that they can represent any quantum state of the many-body Hilbert space just by increasing the value of the bond dimension $D$. As happens for MPS, the bond dimension $D$ of a PEPS needs to be exponentially large in the size of the system in order to cover the whole Hilbert space. Nevertheless, one expects $D$ to be reasonably small and finite for low-energy states of interesting $2d$ quantum models. In practice this is observed numerically \cite{iPEPS}, but there are also theoretical arguments in favor of this property. For instance, it is well known that $D=2$ is sufficient to handle polynomially-decaying correlation functions (and hence critical states) \cite{critPEPS}, and that PEPS can also approximate with arbitrary accuracy thermal states of $2d$ local Hamiltonians \cite{Hastings}. In this case, a similar picture to the one in Fig.(\ref{fig12b}) would also apply adapted to the case of PEPS and the $2d$ area-law. 

\vspace{10pt} 

{\bf \emph{3) $2d$ Area-law.-}} PEPS satisfy also the area-law scaling of the entanglement entropy. This was shown already in the example of Fig.(\ref{fig10}), but the validity is general. In practice, the entanglement entropy of a block of boundary $L$ of a PEPS with bond dimension $D$ is always $S(L) = O(L \log D)$. As discussed before, this property is satisfied by many interesting quantum states of quantum many-body systems, such as some ground states and low-energy excited states of local Hamiltonians.

\vspace{10pt} 

{\bf \emph{4) PEPS can handle polynomially-decaying correlations.-}} A remarkable difference between PEPS and MPS is that PEPS can handle two-point correlation functions that decay polynomially with the separation distance \cite{critPEPS}. And this happens already for the smallest non-trivial bond dimension $D=2$. This property is important, since correlation functions that decay polynomially (as opposed to exponentially) are characteristic of critical points, where the correlation length is infinite and the system is scale invariant. Hence, the class of PEPS is suitable to describe, in principle, gapped phases as well as critical states of matter. 

\begin{figure}[h]
\begin{centering}
\includegraphics[width=12cm]{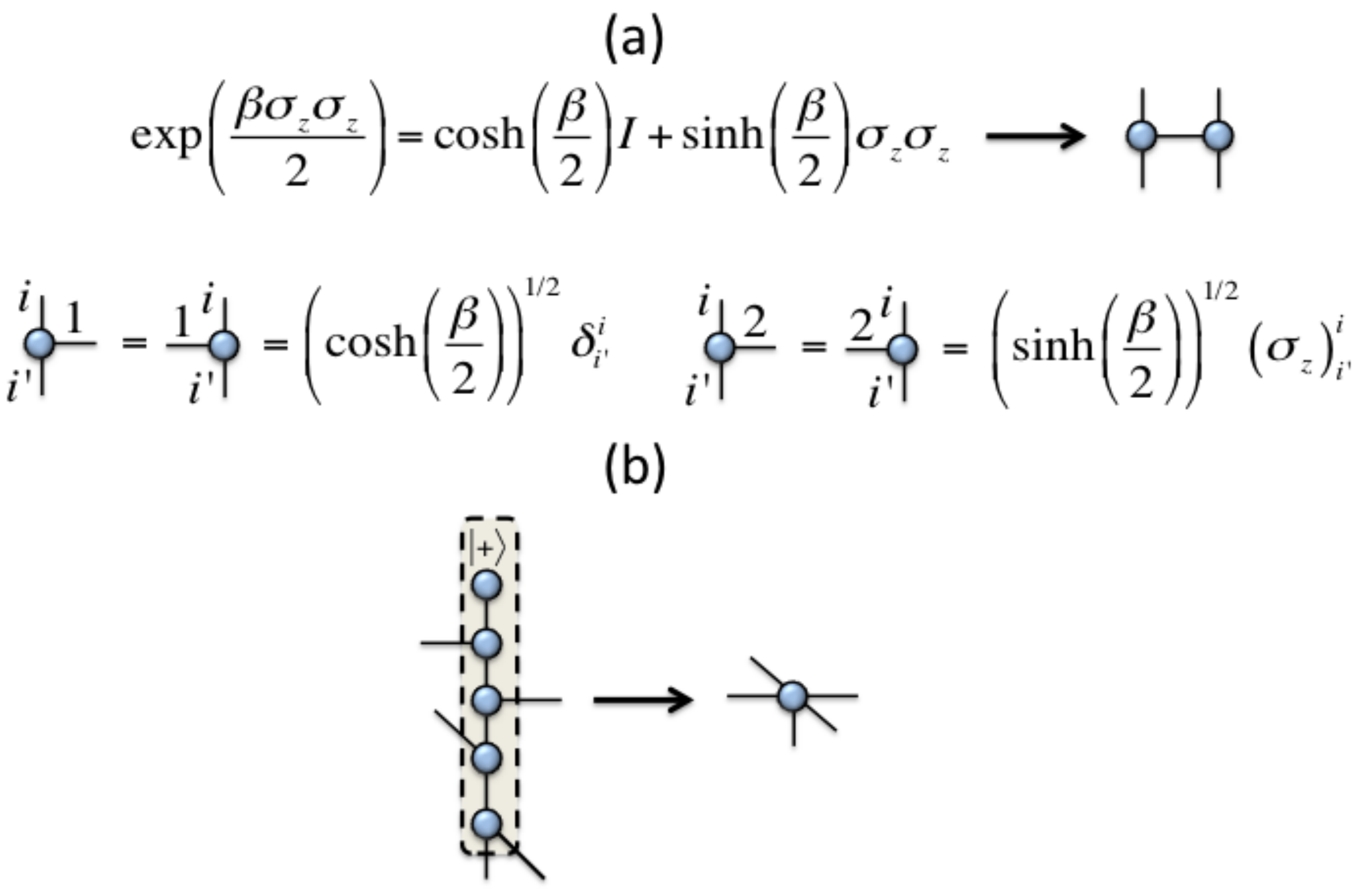} 
\par\end{centering}
\caption{(color online) PEPS for a (classical) thermal state: (a) tensor network decomposition of the operator $\exp(-\beta \sigma_z^{[\vec{r}]} \sigma_z^{[\vec{r}']} /2)$ and non-zero coefficients of the tensors; (b) construction of the infinite PEPS with 1-site unit cell. \label{fig26}}
\end{figure}

This property can be seen with the following example \cite{critPEPS}: consider the unnormalized state
\beq
\ket{\Psi(\beta)} = e^{-\beta H/2} \ket{+}^{\otimes N \rightarrow \infty} \ ,
\eeq
with $\ket{+} = 2^{-1/2}(\ket{0} + \ket{1})$, and $H$ given by 
\beq
H =  - \sum_{ \langle \vec{r}, \vec{r}^{\prime} \rangle }\sigma_z^{[ \vec{r} ]} \sigma_z^{[ \vec{r}^{\prime}]} \ . 
\eeq
After some simple algebra it is easy to see that the norm of this quantum state is proportional to the partition function of the $2d$ classical Ising model on a square lattice at inverse temperature $\beta$, i.e.:
\beq
\braket{\Psi(\beta)}{\Psi(\beta)} \propto Z(\beta) = \sum_{ \{ s \} } e^{-\beta K( \{ s \})} \ , 
\eeq
with $K( \{ s \})$ the classical Ising Hamiltonian, 
\beq
 K( \{ s \}) = - \sum_{ \langle \vec{r}, \vec{r}^{\prime} \rangle }s^{[ \vec{r} ]} s^{[ \vec{r}^{\prime}]} \ , 
 \eeq
where $s^{[ \vec{r} ]} = \pm 1$ is a classical spin variable at lattice site $\vec{r}$, and $\{ s \}$ is some configuration of all the classical spins. It is also easy to see that the expectation values of local operators in $\ket{\Psi(\beta)}$ correspond to classical expectation values of local observables in the classical model. For instance, the expectation value of $\sigma_z^{[\vec{r}]}$ corresponds to the classical magnetization at site $\vec{r}$ at inverse temperature $\beta$, 
\beq
\langle s^{[\vec{r}]}\rangle_{\beta} = \frac{\bra{\Psi (\beta)} \sigma_z^{[\vec{r}]} \ket{\Psi(\beta)}}{\bra{\Psi(\beta)} \Psi (\beta) \rangle} = \frac{1}{Z(\beta)} \sum_{ \{ s \} } s^{[\vec{r}]} e^{-\beta K( \{ s \})} \ . 
\eeq
Also, the two-point correlation functions in the quantum state correspond to classical correlation functions of the classical model. For instance, the correlation function for operators $\sigma_z^{[\vec{r}]}$ and $\sigma_z^{[\vec{r}^\prime]}$ at sites $\vec{r}$ and $\vec{r}'$ corresponds to the usual correlation function of the classical Ising variables, 
\beq
\langle s^{[\vec{r}]} s^{[\vec{r}^\prime]} \rangle_{\beta} = \frac{\bra{\Psi (\beta)} \sigma_z^{[\vec{r}]}  \sigma_z^{[\vec{r}^\prime]} \ket{\Psi(\beta)}}{\bra{\Psi(\beta)} \Psi (\beta) \rangle} = \frac{1}{Z(\beta)} \sum_{ \{ s \} } s^{[\vec{r}]} s^{[\vec{r}^\prime]} e^{-\beta K( \{ s \})} \ . 
\eeq
At the critical inverse temperature $\beta_c = (\log (1 + \sqrt{2} ))/2$ it is well known that the correlation length of the system diverges, and the above correlation function decays polynomially for long separation distances as
\beq
\langle s^{[\vec{r}]} s^{[\vec{r}^\prime]} \rangle_{\beta_c} \approx \frac{a}{|\vec{r} - \vec{r}'|^{1/4}} \ , 
\eeq
for some constant $a = O(1)$ and $|\vec{r} - \vec{r}'| \gg 1$. The next step is to realize that, actually, the quantum state $\ket{\Psi(\beta)}$ is a $2d$ PEPS with bond dimension $D=2$. This is shown in the tensor network diagrams in Fig.(\ref{fig26}). Therefore, at the critical value $\beta = \beta_c$, the resultant quantum state $\ket{\Psi (\beta_c)}$ is an example of a $2d$ PEPS with finite bond dimension $D=2$ and with polynomially decaying correlation functions (and hence infinite correlation length).This should be considered as a proof of principle about the possibility of having some criticality in a PEPS with finite bond dimension. As a remark, notice that this is totally different to the case of $1d$ MPS, where we saw before that two-point correlation functions always decay exponentially fast with the separation distance. 

Notice, though, that the criticality obtained in this PEPS is essentially classical, since we just codified the partition function of the $2d$ classical Ising model as the squared norm of a $2d$ PEPS. Recalling the quantum-classical correspondence that a quantum  $d$-dimensional  lattice model is equivalent to some classical $d+1$-dimensional lattice model \cite{sachdev}, one realizes that this criticality is actually the one from the $1d$ quantum Ising model, but ``hidden" into a $2d$ PEPS with small $D$. By generalizing this idea, we could actually think as well of e.g. ``codifying" the quantum criticality of a $2d$ quantum model into a $3d$ PEPS with small $D$. Yet, it is still unclear under which conditions a $2d$ PEPS could handle this true $2d$ quantum criticality for small $D$ as well. 

\vspace{10pt}

{\bf \emph{5) Exact contraction is $\sharp$P-Hard:}} The exact calculation of the scalar product between two PEPS is an exponentially hard problem. This means that for two arbitrary PEPS of $N$ sites, it will always take a time $O(\exp(N))$, no matter the order in which we try to contract the different tensors. This statement can be done mathematically precise. From the point of view of computational complexity, the calculation of the scalar product of two PEPS is a problem in the complexity class $\sharp$P-Hard \cite{sharp}. We shall not enter into detailed definitions of complexity classes here, yet let us explain in plain terms what this means. The class $\sharp$P-Hard is the class of problems related to counting the number of solutions to NP-Complete problems. Also, the class NP-Complete is commonly understood as a class of very difficult problems in computational complexity\footnote{The interested reader can take a look at e.g. Ref.\cite{papa}.}, and it is widely believed that there is no classical (and possibly quantum) algorithm that can solve the problems in this class using polynomial resources in the size of the input. A similar statement is also true for the class $\sharp$P-Hard. Therefore, unlike for $1d$ MPS, computing exact scalar products of arbitrary $2d$ PEPS is, in principle, inefficient. 

However, and as we will see in Sec.\ref{sec6}, it is possible in practice to approximate these expectation values using clever numerical methods. Moreover, recent promising results in the study of entanglement spectra seem to indicate that these approximate calculations can be done with a very large accuracy (possibly even exponential), at least for $2d$ PEPS corresponding to ground states of gapped, local $2d$ Hamiltonians \cite{entspec}. 

\vspace{10pt} 

\begin{figure}[h]
\begin{centering}
\includegraphics[width=11cm]{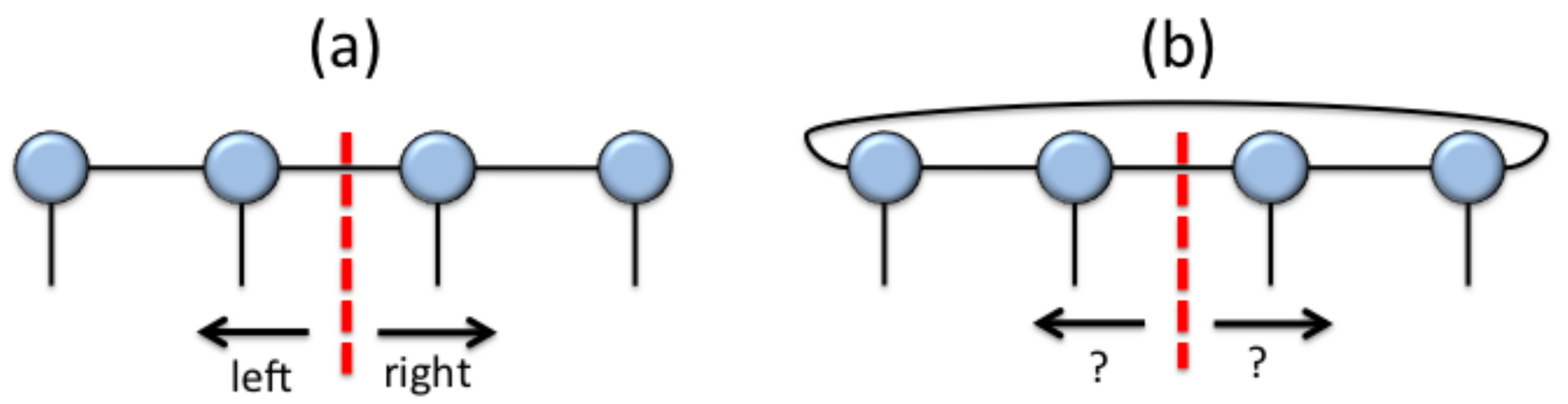} 
\par\end{centering}
\caption{(color online) By ``cutting" a link in a TN, one can define left and right pieces if there are no other connecting indices between the two pieces (a), whereas this is not possible if other connecting indices exist (b). \label{fig27}}
\end{figure}

{\bf \emph{6) No exact canonical form.-}} Unlike for MPS with open boundary conditions, there is no canonical form of a PEPS, in the sense that it is not possible to choose orthonornal basis simultaneously for all the bond indices. In fact, this happens already for MPS with periodic boundary conditions or, more generally, \emph{as long as we have a loop in the TN}. Loosely speaking, a loop in the TN means that we can not formally split the network into left and right pieces by just cutting one index, so that a Schmidt decomposition between left and right does not make sense. In practice this means that we can not define orthonormal basis (i.e. Schmidt basis) to the left and right of a given index, and hence we can not define a canonical form in this sense (see Fig.(\ref{fig27})). 

Nevertheless, it is observed numerically that for non-critical PEPS it is usually possible to find a \emph{quasi-canonical form}, which leads to approximate numerical methods for finding ground states (a variation of the so-called ``simple update" approach \cite{simple}). We refer the interested reader to Ref.\cite{clpeps} for more details about this. 

\vspace{10pt} 

\subsubsection{Some examples}

\begin{figure}[h]
\begin{centering}
\includegraphics[width=13cm]{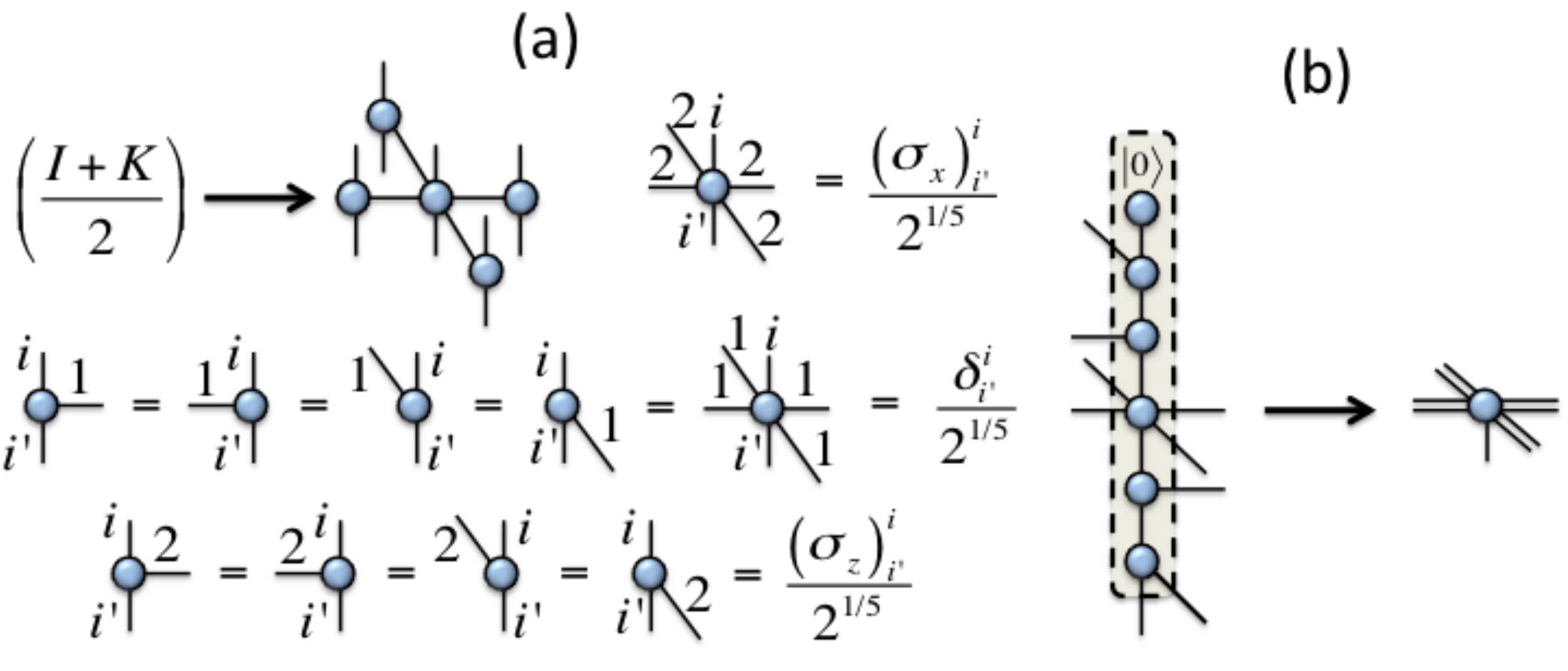} 
\par\end{centering}
\caption{(color online) PEPS for the $2d$ cluster state: (a) tensor network decomposition of the operator $\left( \mathbb{I} + K^{[\vec{r}]}\right)/2$ and non-zero coefficients of the tensors; (b) construction of the infinite PEPS with 1-site unit cell. \label{fig28}}
\end{figure}

In what follows we provide some examples of interesting quantum states for $2d$ lattices that can be expressed exactly using the PEPS formalism. These are the following: 

\vspace{10pt} 

{\bf \emph{1) $2d$ Cluster State.-}} The cluster state in a $2d$ square lattice \cite{cluster} is highly-entangled quantum state that can be used as a resource for performing universal measurement-based quantum computation. This quantum state is  the $+1$ eigenstate of a set of mutually commuting stabilizer operators $\{ K^{[\vec{r}]} \}$ defined as
\beq
K^{[\vec{r}]} \equiv \sigma_x^{[\vec{r}]} \bigotimes_{\vec{p} \in \Gamma(\vec{r})} \sigma_z^{[\vec{p}]} \ ,
\eeq
 where $\Gamma(\vec{r})$ denotes the four nearest-neighbor spins of lattice site $\vec{r}$ and the $\sigma_{\alpha}^{[\vec{r}]}$ are the usual spin-$1/2$ Pauli matrices with $\alpha \in \{ x,y,z \}$ at lattice site $\vec{r}$. For the infinite square lattice, these stabilizers are five-body operators. Noticing that $(K^{[\vec{r}]})^2 = \mathbb{I}$, this quantum state can be written (up to an overall normalization constant) as
\beq
\ket{\Psi_{2dCL}} =  \prod_{\vec{r}} \frac{\left( \mathbb{I} + K^{[\vec{r}]}\right)}{2}\ket{0}^{\otimes N \rightarrow \infty} \ .  
\eeq
Each one of the terms $\left( \mathbb{I} + K^{[\vec{r}]}\right)/2$ is a projector that admits a TN representation as in Fig.(\ref{fig28}). From here, it is easy to obtain a TN description for the $2d$ cluster state $\ket{\Psi_{2dCL}} $ in terms of a $2d$ PEPS with bond dimension $D=4$, as shown in Fig.(\ref{fig28}). In the figure we also give the value of the non-zero coefficients of the PEPS tensors. 

\vspace{10pt} 

{\bf \emph{2) Toric Code model.-}} The Toric Code, introduced by Kitaev \cite{toric}, is a model of spins-1/2 on the links of a $2d$ square lattice. The Hamiltonian can be written as
\beq
H = -J_a \sum_s A_s - J_b \sum_p B_p \ , 
\eeq
where $A_s$ and $B_p$ are star and plaquette operators such that 
\beq
A_s = \prod_{\vec{r} \in s} \sigma_x^{[\vec{r}]}   \ , \ \ \ \ \ \  B_p = \prod_{\vec{r} \in p} \sigma_z^{[\vec{r}]} \ . 
\eeq
In other words, $A_s$ is the product of $\sigma_x$ operators for the spins around a star, and $B_p$ is the product of $\sigma_z$ operators for the spins around a plaquette. Here we will be considering the case of an infinite $2d$ square lattice.

The Toric Code is of relevance for a variety of reasons. First, it can be seen as the Hamiltonian of a $\mathbb{Z}_2$ lattice gauge theory with a ``soft" gauge constraint \cite{kogut} (i.e. if we send either $J_a$ or $J_b$ to infinity, then we recover the low-energy sector of the lattice gauge theory). But also, it is important because it is the simplest known model such that its ground state displays the so-called ``topological order", which is a kind of order in many-body wave-functions related to a pattern of long-range entanglement that prevades over the whole quantum state (here we shall not discuss topological order in detail; the interested reader can have a look at the vast literature on this topic, e.g. Ref.\cite{topo}). Furthermore, the Toric Code is important in the field of quantum computation, since one could use its degenerate ground state subspace on a torus geometry to define a topologically protected qubit that is inherently robust to local noise \cite{toric}. 

\begin{figure}[h]
\begin{centering}
\includegraphics[width=12cm]{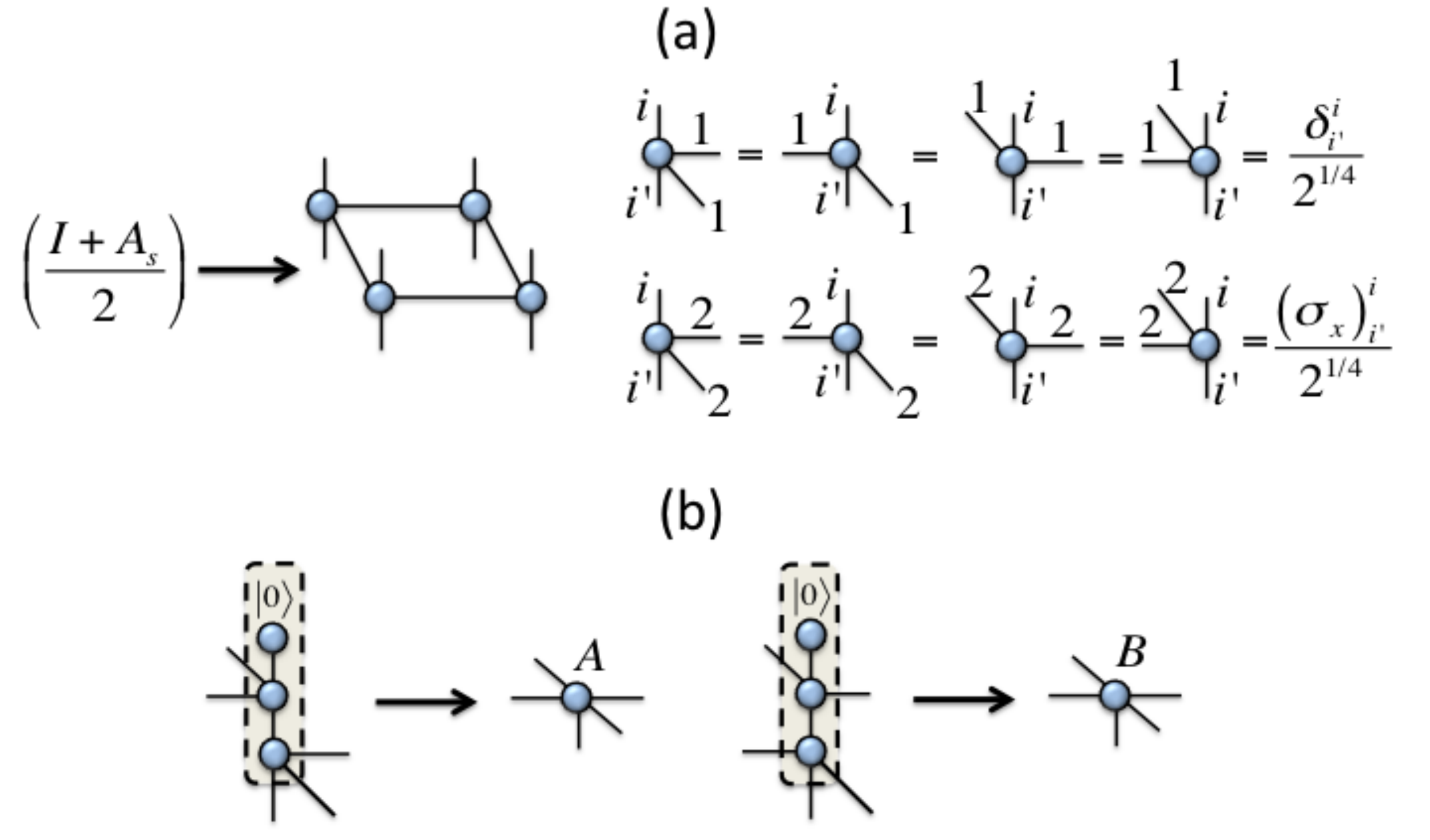} 
\par\end{centering}
\caption{(color online) PEPS for the Toric Code: (a) tensor network decomposition of operator $\left( \mathbb{I} + A_s\right)/2$ and non-zero coefficients of the tensors; (b) construction of the infinite PEPS with $2 \times 2$ unit cell and 2 tensors. \label{fig29}}
\end{figure}

An interesting feature about the Toric Code is that, again, it can be understood as a sum of mutually commuting stabilizer operators. This time the stabilizers are the set of star and plaquette operators $\{ A_s \}$ and $\{ B_p \}$. As for the cluster state, it is easy to check that the square of the stabilizer operators equals the identity ($A_s^2 = B_p^2 = \mathbb{I} \ \forall s,p$). The ground state of the system is the $+1$ eigenstate of all these stabilizer operators. For an infinite $2d$ square lattice this ground state is unique, and can be written (up to a normalization constant) as
\beq
\ket{\Psi_{TC}} =  \prod_{s} \frac{\left( \mathbb{I} + A_s\right)}{2} \prod_{p} \frac{\left( \mathbb{I} + B_p\right)}{2}\ket{0}^{\otimes N \rightarrow \infty} =  \prod_{s} \frac{\left( \mathbb{I} + A_s\right)}{2} \ket{0}^{\otimes N \rightarrow \infty} \ ,
\eeq
where the last equality follows from the fact that the state $\ket{0}^{\otimes N \rightarrow \infty}$ is already a $+1$ eigenstate of the operators $B_p$ for any plaquette $p$. The above state can be written easily as a PEPS with bond dimension $D=2$ \cite{critPEPS}. For this, notice that each one of the terms $(\mathbb{I} + A_s)/2$ for any star $s$ admits the TN representation from Fig.(\ref{fig29}.a). From here, a PEPS representation with a 2-site unit cell as in Fig.(\ref{fig25}.b) follows easily, see Fig.(\ref{fig29}.b). As we can see with this example, a PEPS with the smallest non-trivial bond dimension $D=2$ can already handle topologically ordered states of matter. 

\vspace{10pt}

\begin{figure}[h]
\begin{centering}
\includegraphics[width=13cm]{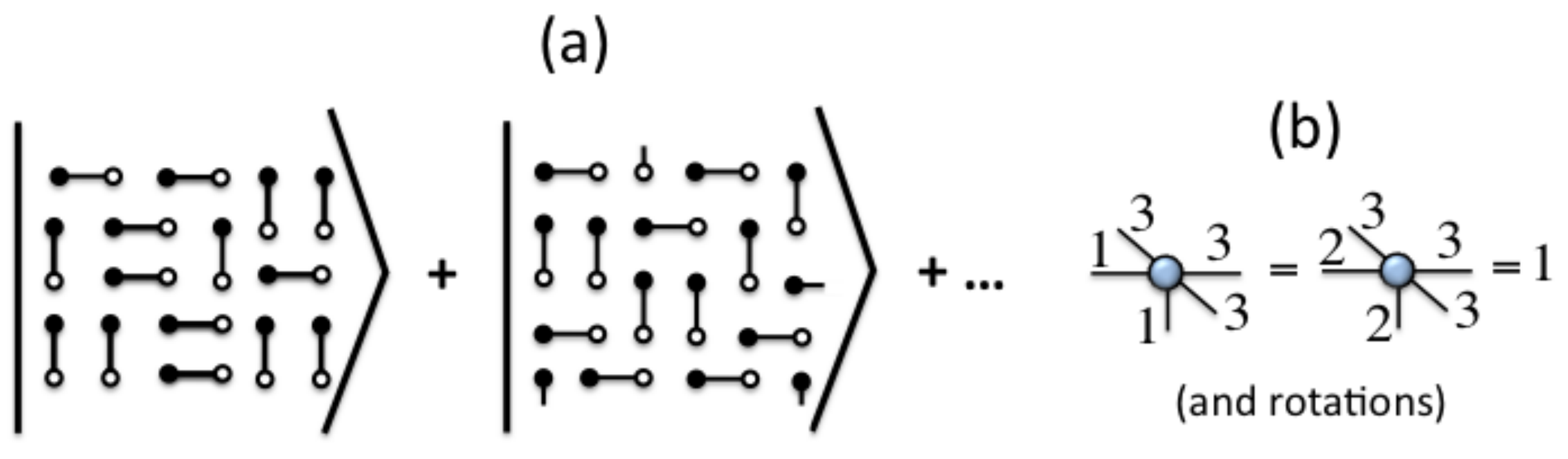} 
\par\end{centering}
\caption{(color online) A $2d$ Resonating Valence Bond state built from nearest-neighbor singlets (a) can be written as an infinite PEPS with 1-site unit cell, with non-zero coefficients of the tensors as in (b). \label{fig30}}
\end{figure}

{\bf \emph{3) $2d$ Resonating Valence Bond State.-}} The $2d$ Resonating Valence Bond (RVB) state \cite{rvb}, is a quantum state proposed by Anderson in 1987 in the context of trying to explain the mechanisms behind high-$T_c$ superconductivity. For our purposes this state corresponds to the equal superposition of all possible nearest-neighbor dimer coverings of a lattice, where each dimer is a SU(2) singlet, 
\beq
\ket{\Phi} = \frac{1}{\sqrt{2}}\left( \ket{0} \otimes \ket{1} - \ket{1} \otimes \ket{0} \right) \ ,
\eeq
which is a maximally-entangled state of two qubits (also known as EPR pair, or Bell state). For the $2d$ square lattice this is represented in Fig.(\ref{fig30}.a). This state is also important since it is the arquetypical example of a quantum spin liquid: a quantum state of matter that does not break any symmetry (neither translational symmetry, nor SU(2)). Importantly, this RVB state can also be written as a PEPS with bond dimension $D=3$. The non-zero coefficients of the tensors are given in Fig.(\ref{fig30}.b). 

 \vspace{10pt}

\begin{figure}[h]
\begin{centering}
\includegraphics[width=10cm]{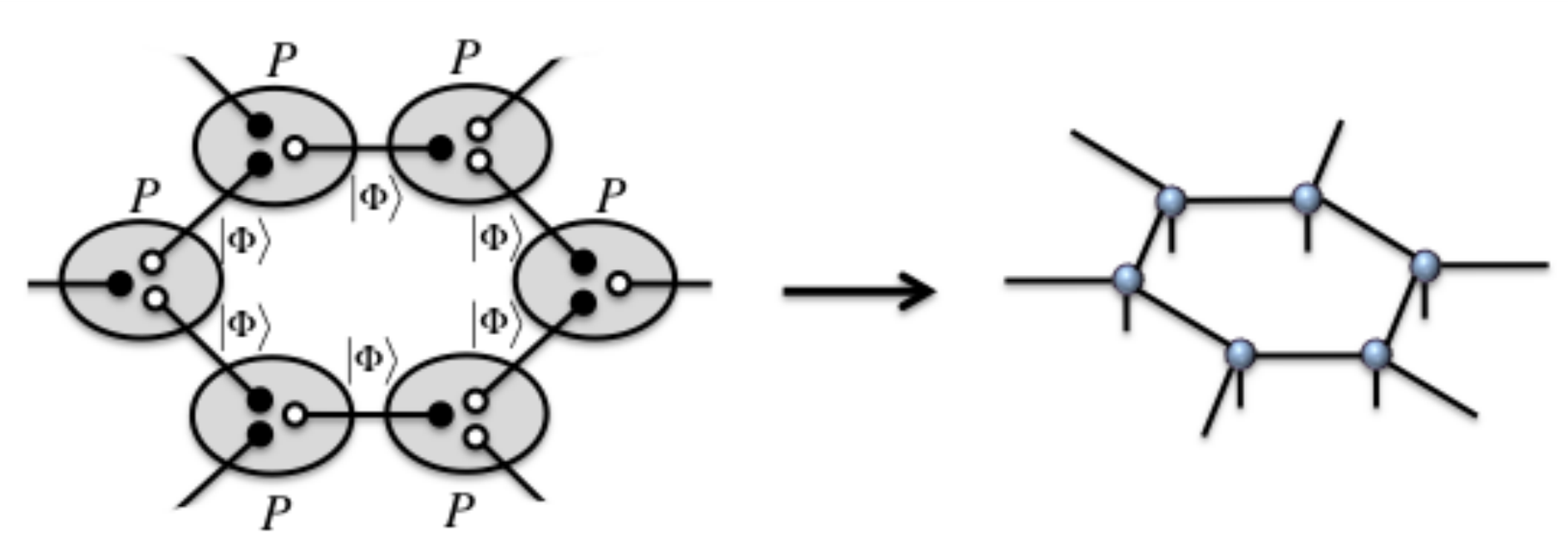} 
\par\end{centering}
\caption{(color online) PEPS for the $2d$ AKLT state on the honeycomb lattice. Spin-$1/2$ particles arranged in singlets $\ket{\Phi} = 2^{-1/2} (\ket{0}\otimes \ket{1} - \ket{1}\otimes \ket{0})$, and projected in trios using projector $P = (\ket{1}\bra{0}\otimes \bra{0}\otimes \bra{0} + \ket{2}\bra{1}\otimes \bra{1}\otimes \bra{1} + \ket{3}\bra{W} + \ket{4}\bra{\bar{W}})$, with $\ket{W} = 3^{-1/2} (\ket{0}\otimes\ket{0}\otimes\ket{1} + \ket{0}\otimes\ket{1}\otimes\ket{0} + \ket{1}\otimes\ket{0}\otimes\ket{0})$ and $\ket{\bar{W}} = 3^{-1/2} (\ket{1}\otimes\ket{1}\otimes\ket{0} + \ket{1}\otimes\ket{0}\otimes\ket{1} + \ket{0}\otimes\ket{1}\otimes\ket{1})$. \label{fig31}}
\end{figure}

{\bf \emph{4) $2d$ AKLT model.-}} The $2d$ AKLT model on a honeycomb lattice \cite{aklt, 2daklt} is given by the Hamiltonian
\beq
H = \sum_{ \langle \vec{r}, \vec{r}'  \rangle } \left( \vec{S}^{[\vec{r}]} \vec{S}^{[\vec{r}^\prime]} + \frac{116}{243} \left(  \vec{S}^{[\vec{r}]} \vec{S}^{[\vec{r}^\prime]} \right)^2 + \frac{16}{243} \left(  \vec{S}^{[\vec{r}]} \vec{S}^{[\vec{r}^\prime]} \right)^3 \right) \ , 
\eeq
where $\vec{S}^{[\vec{r}]}$ is the vector of spin-$3/2$ operators at site $\vec{r}$, and the sum is over nearest-neighbor spins. As explained in Ref.\cite{aklt, 2daklt}, this model can be seen as a generalization of the $1d$ AKLT model in Eq.(\ref{1daklt}) to two dimensions. The ground state can be understood in terms of a set of spin-$1/2$ singlets, which are brought together into groups of 3 at every vertex of the honeycomb lattice, and projected into their symmetric subspace (i.e. spin-$3/2$), see Fig.(\ref{fig31}). By construction, this is a $2d$ PEPS with bond dimension $D=2$. 

\section{Extracting information: computing expectation values}
\label{sec6}

An important problem for TN is how to extract information from them. This is usually achieved by computing expectation values of local observables. It turns out that such expectation values can be computed efficiently, either exactly (for MPS) or approximately (for PEPS). This is very important, since otherwise it would not make any sense to have an efficient representation of a quantum state: we also need to be able to extract information from it! 
 
In what follows we explain how expectation values can be extracted efficiently from MPS and PEPS, both for finite and infinite systems. In fact, in the case of MPS with open boundary conditions a lot of the essential information was already introduced in Sec.\ref{sec4}, when talking about the exponential decay of two-point correlation functions and the exact calculation of norms. 

We will see that the calculation of expectation values follows many times a \emph{dimensional reduction} strategy. More precisely, the $1d$ problem for an MPS is reducible to a $0d$ problem, and this can be solved exactly. Also, the $2d$ problem for a PEPS is reducible to a $1d$ problem that can be solved approximately using MPS techniques. This MPS problem, in turn, is itself reducible to a $0d$ problem that is again exactly solvable. Such a dimensional reduction strategy is nothing but an implementation, in terms of TN calculations, of the ideas of the holographic principle \cite{holo}. 

\subsection{Expectation values from MPS}

Expectation values of local operators can be computed exactly for an MPS without the need for further approximations. In the case of open boundary conditions, this is achieved for finite and infinite systems using the techniques that we already introduced in Fig.(\ref{fig16}). As explained in Sec.\ref{sec5}, all those manipulations can be done easily in $O(NpD^3)$ time for finite systems, and $O(pD^3)$ for infinite \cite{mpsrev}. The trick for infinite systems is to contract from the left and from the right as indicated in the figure, assuming some boundary condition placed at infinity. In practice, such a contraction is equivalent to finding the dominant left and right eigenvectors of the $0d$ transfer matrix from Fig.(\ref{fig14}). Let us remind that these calculations are very much simplified if the MPS is in canonical form, see Fig.(\ref{fig19}). 

\begin{figure}[h]
\begin{centering}
\includegraphics[width=10cm]{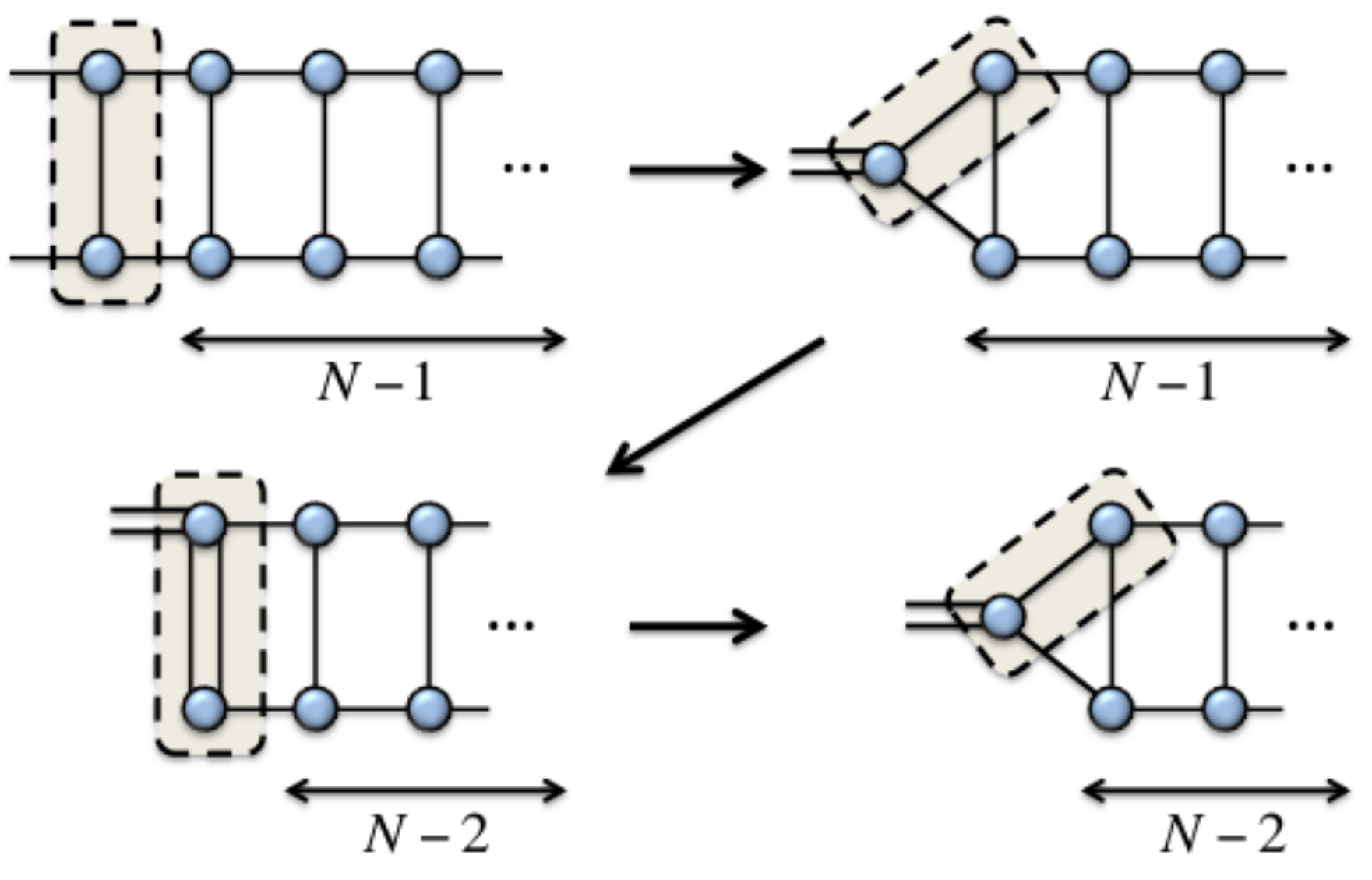} 
\par\end{centering}
\caption{(color online) order of contractions for an MPS with periodic boundary conditions \label{fig32}}
\end{figure}

If the MPS has periodic boundary conditions \cite{pbc1}, then the TN contractions for expectation values are similar to the ones in Fig.(\ref{fig32}). This time the calculation can be done in $O(NpD^5)$ time. As expected, the calculation for periodic boundary conditions is less efficient than the one for open boundary conditions, since one needs to carry more tensor indices at each calculation step. 

\subsection{Expectation values from PEPS}

Unlike for MPS, expectation values for PEPS need to be computed using approximations. Here we explain some techniques to do this for the case of finite systems with open boundary conditions, as well as for infinite systems. The methods explained here are by far not the only ones (see e.g. Ref.\cite{ctmrg} and Ref.\cite{HOTRG} for some alternatives), yet they are quite representative, and constitute also the first step towards understanding many of the calculations that are involved in the numerical algorithms for finite and infinite-PEPS. 

\begin{figure}
\begin{centering}
\includegraphics[width=10cm]{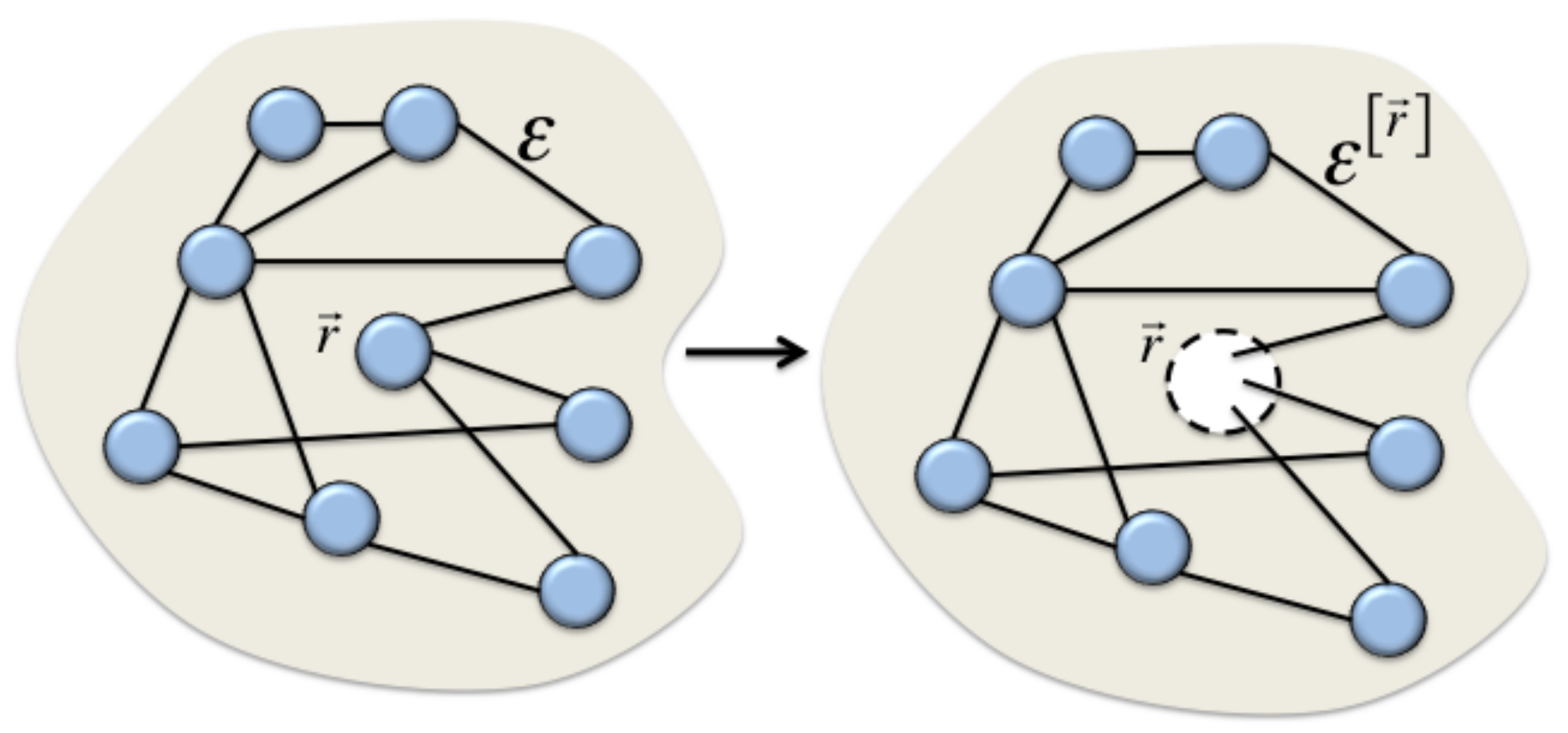} 
\par\end{centering}
\caption{(color online) The environment of a site corresponds to the contraction of the whole tensor network except for the tensors at that site. \label{fig33}}
\end{figure}

Let us introduce a couple of definitions before proceeding any further. We call the \emph{environment} $\mathcal{E}^{[\vec{r}]}$ of a site $\vec{r}$ the TN consisting of all the tensors in the original TN, except those at site $\vec{r}$, see Fig.(\ref{fig33}). Unlike for MPS, environments for sites of a PEPS can not be computed exactly. Therefore, we call \emph{effective environment} $\mathcal{G}^{[\vec{r}]}$ of a site $\vec{r}$ the approximation of the contraction of the exact environment of site $\vec{r}$ by using some ``truncation" criteria in the relevant degrees of freedom. In what follows we shall be more explicit with what we mean by this ``truncation".  

\subsubsection{Finite systems}
\begin{figure}[h]
\begin{centering}
\includegraphics[width=11cm]{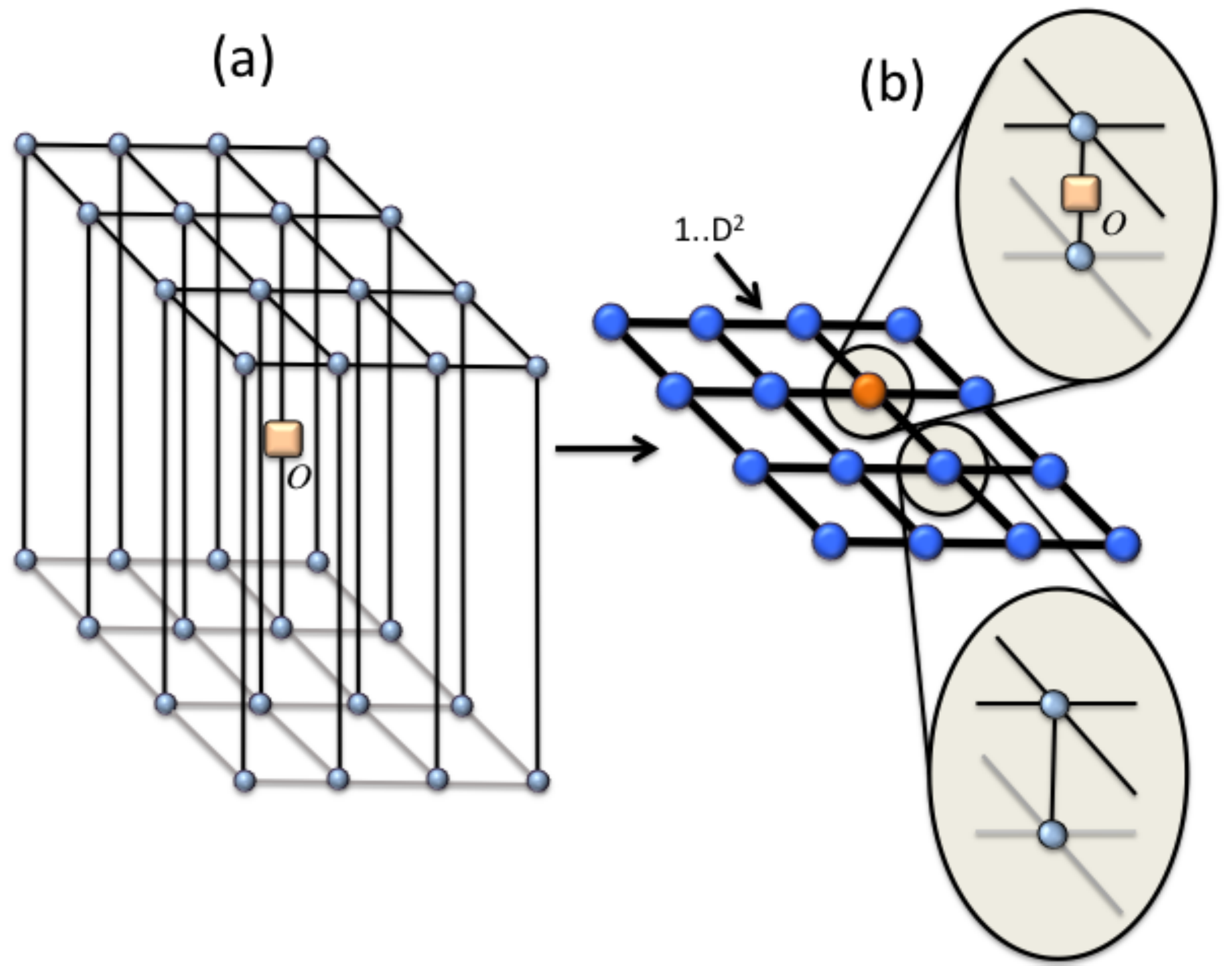} 
\par\end{centering}
\caption{(color online) (a) Expectation value of a 1-site observable for a $4 \times 4$ PEPS; (b) Contraction of the $4 \times 4$ lattice of reduced tensors. We use thick lines to indicate ``double" indices ranging from 1 to $D^2$.\label{fig34}}
\end{figure}

Consider the expectation value of a local one-site observable for a finite PEPS \cite{PEPS}. The TN that needs to be contracted corresponds to the one in the diagram in Fig.(\ref{fig34}.a) which can be understood in terms of a $2d$ lattice of \emph{reduced tensors} as in Fig.(\ref{fig34}.b). As explained earlier, the exact contraction of such a TN is a $\sharp$P-Hard problem, and therefore must be approximated. 

\begin{figure}
\begin{centering}
\includegraphics[width=13cm]{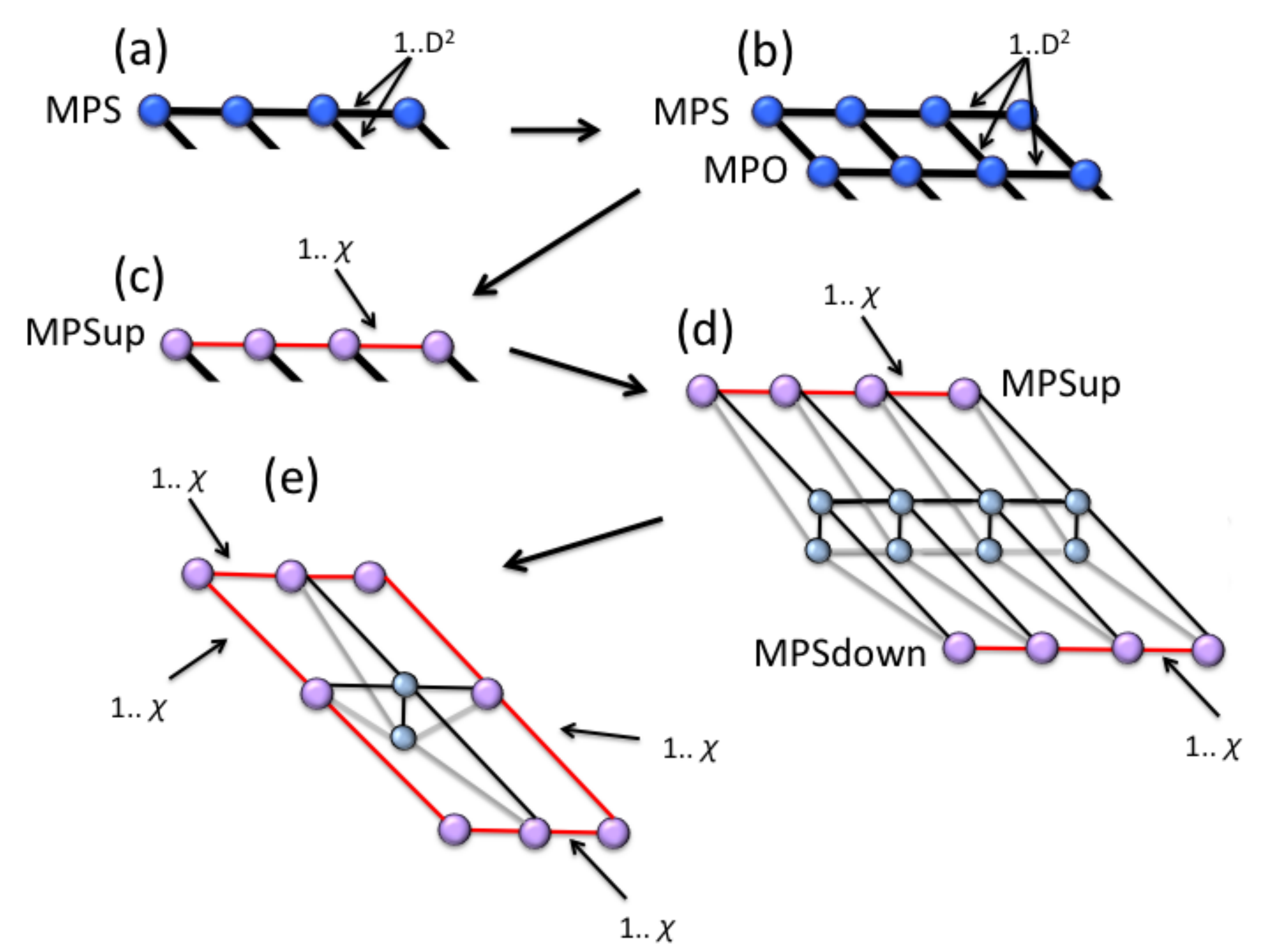} 
\par\end{centering}
\caption{(color online) Strategy to contract the lattice of reduced tensors for a finite PEPS. \label{fig35}}
\end{figure}

The way to approximate this calculation is by reducing the original $2d$ problem to a series of $1d$ problems that we can solve using MPS methods such as DMRG or TEBD. Let us do this as follows: first, consider the upper-row boundary of the system (in this case we have a square lattice). The tensors within this row can be understood as forming an MPS if the physical indices are contracted at each site, see Fig.(\ref{fig35}.a). Next, the contraction of this MPS with the next row of tensors is equivalent to the action of a Matrix Product Operator (MPO) on the MPS, see Fig.(\ref{fig35}.b). This is, by construction, an MPS of higher bond dimension than the original (actually of bond dimension $D^4$). Therefore, in order to maintain the bond dimension under control, we approximate the resultant MPS by another MPS of lower bond dimension which we call $\chi$, see Fig.(\ref{fig35}.c). This approximation can be done using e.g. DMRG or TEBD methods. We can proceed in this way in order to add each one of the rows of the $2d$ lattice, and of course we can do the same but starting from the lower boundary. Proceeding in this way, both from up and down, we end up eventually in the contraction of the $1d$ TN shown in Fig.(\ref{fig35}.d). But remarkably, what is now left is a $1d$ problem! Therefore the remaining contractions can be computed exactly by reducing everything to a $0d$ problem as explained in the previous section, i.e. we start contracting the tensors from the left, and then from the right, until we end up in the TN of Fig.(\ref{fig35}.e). The final expectation value just follows from this contraction, dividing by the corresponding norm of the state (same TN but without the observable operator). Overall, this procedure has a computational cost of $O(Np^2 \chi^2 D^6)$ in time.

In case you did not notice, what we just computed in the above procedure was an effective environment for the site where we placed the observable. This is not exclusive of this calculation: all the methods to compute expectation values of $2d$ systems are based on techniques to compute effective environments of different sites. And not only that: some of the minimization techniques for $2d$ PEPS that will be discussed in the next section rely also heavily on the accurate calculation of effective environments. 

\subsubsection{Infinite systems} 

For infinite systems, the calculation of expectation values reduces again to the contraction of a lattice of reduced tensors (this time infinite), and also, to the calculation of effective environments. Here we explain briefly three families of methods available to achieve this calculation: boundary-MPS \cite{iPEPS}, Corner Transfer Matrix (CTM) \cite{ctmrg, dctm}, and tensor coarse-graining methods \cite{TRG, SRG, HOTRG}. 

\vspace{10pt} 

\begin{figure}
\begin{centering}
\includegraphics[width=13cm]{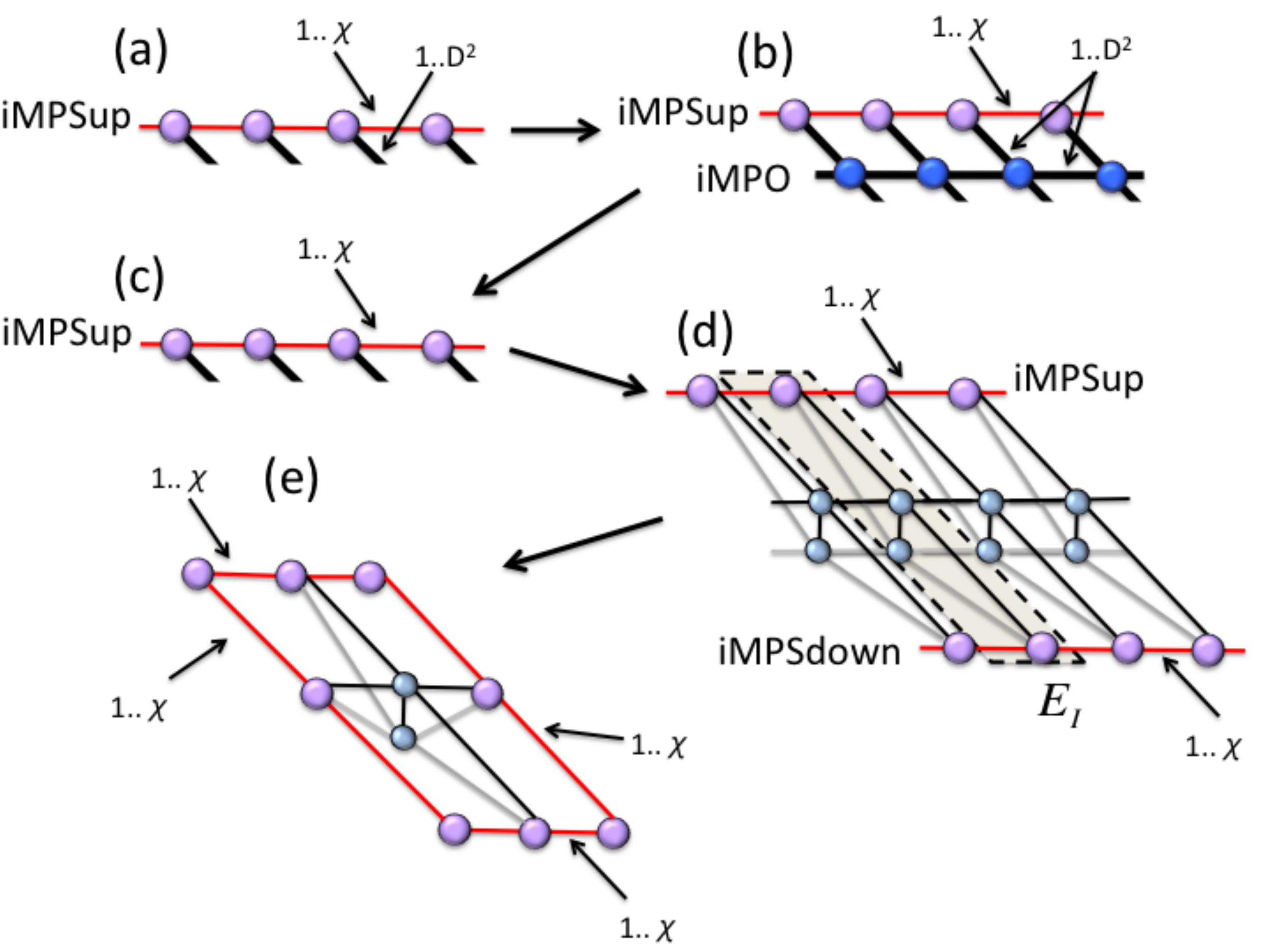} 
\par\end{centering}
\caption{(color online) Strategy to contract the lattice of reduced tensors for an infinite PEPS. \label{fig36}}
\end{figure}

{\bf \emph{1) Boundary MPS methods.-}} This method corresponds to the generalization of the calculation of effective environments for finite systems that we explained before using MPS, to the case of infinite systems. For this, notice that the contraction of an infinite $2d$ lattice of reduced tensors can be understood, at least theoretically, as (i) placing an infinite \emph{boundary MPS} with bond dimension $\chi$ at the boundary of the $2d$ lattice at infinity (Fig.(\ref{fig36}.a)), (ii) applying infinite MPOs to this infinite MPS (Fig.(\ref{fig36}.b)) and producing new infinite MPSs of bond dimension $\chi$ using iDMRG or iTEBD (Fig.(\ref{fig36}.c)), and (iii) repeating the process infinitely-many times. In practice, repeating infinitely-many times is equivalent to repeating until the resultant boundary MPS has converged for a given $\chi$ with a given precision. 

As such, this procedure, is nothing but a power method to approximate the dominant left/right eigenvector of the infinite MPO in Fig.(\ref{fig36}.b) by an infinite MPS (that is, the eigenvector of the infinite MPO with the largest magnitude eigenvalue). Hence, we have reduced the problem of contracting the infinite $2d$ lattice to that of approximating the dominant eigenvector of a $1d$ transfer matrix given by an infinite MPO. 

Again, the steps above can be done for two parallel boundaries (say ``up" and ``down"), being thus equivalent to finding the dominant left and right eigenvectors of the MPO.  So, after convergence of the boundary MPSs we end up with the contraction of an infinite $1d$ stripe of tensors. This remaining contraction can be done exactly, just by reducing the problem to that of finding the dominant left and right eigenvectors of the $0d$ transfer matrix $E_\mathbb{I}$ in Fig.(\ref{fig36}.d). The final contraction produces the effective environment given in Fig.(\ref{fig36}.e). 

\begin{figure}[h]
\begin{centering}
\includegraphics[width=12cm]{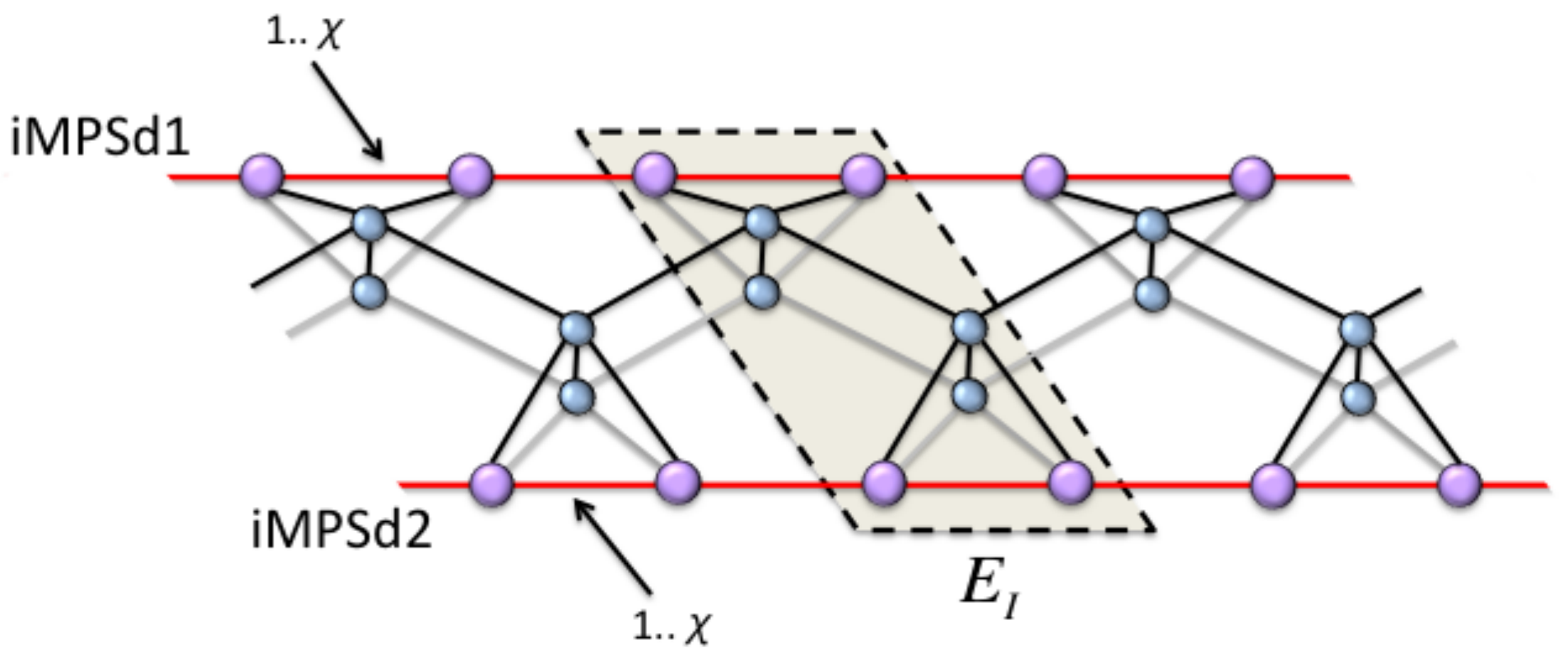} 
\par\end{centering}
\caption{(color online) 1-dimensional problem obtained when contracting the infinite $2d$ lattice of reduced tensors using diagonal boundary MPS. \label{fig37}}
\end{figure}

A comment is now in order. Notice that in this procedure we decided to place our infinite boundary MPS at the ``up" and ``down" boundaries, hence using horizontal MPSs. But this choice is not unique. For instance,  we could have chosen equally well the ``left" and ``right" boundaries, hence using vertical MPSs. What is more, we could have chosen to work with diagonal boundary MPS such as the ones in Fig.(\ref{fig37}). In such a case, the infinite MPO as well as the relevant $0d$ transfer matrix $E_\mathbb{I}$ have a different construction (see the figure). While the final result does not depend too much on how we choose to place the boundary MPS, the computational cost does. In particular, the horizontal (or vertical) MPS method has a computational cost of $O(\chi^3 D^6 + p \chi^2 D^8)$, whereas the diagonal MPS method has a slightly lower computational cost of $O(\chi^3 D^4 + p \chi^2 D^6)$ \cite{iPEPS}. 

\vspace{10pt} 

\begin{figure}[h]
\begin{centering}
\includegraphics[width=8.5cm]{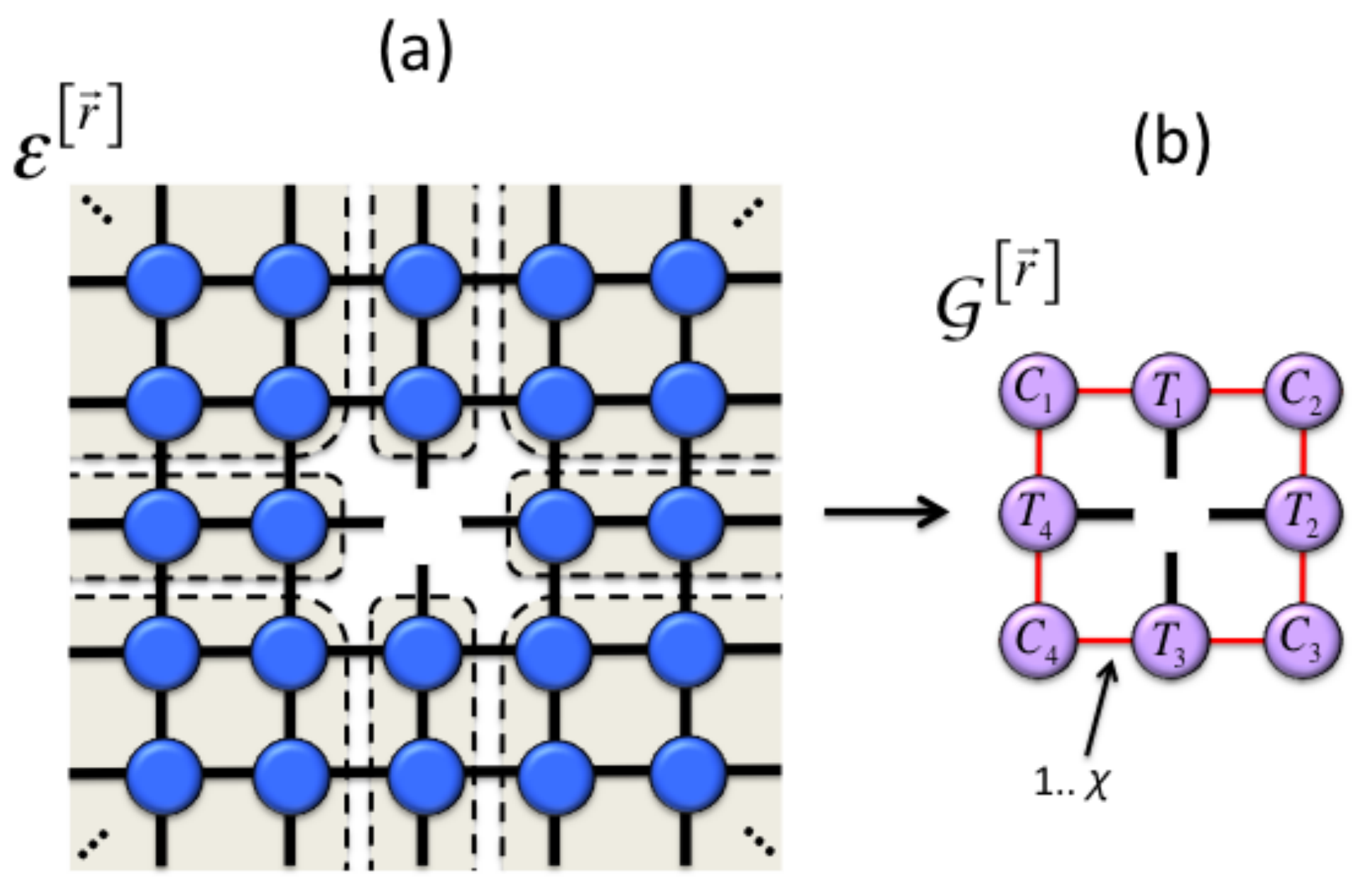} 
\par\end{centering}
\caption{(color online) General strategy of Corner Transfer Matrix methods. The problem is reduced to contracting four CTMs, two half-row and  two half-column transfer matrices. \label{fig38}}
\end{figure}

{\bf \emph{2) CTM methods.-}} CTM methods are a different way of computing effective environments for $2d$ infinite PEPS. To understand this, let us consider the environment $\mathcal{E}^{[\vec{r}]}$ of lattice site $\vec{r}$. Another way of understanding this environment is to think that it comes from the contractions of the four corners of tensors, together with two half-row and two half-column $1d$ transfer matrices, as shown in Fig.(\ref{fig38}).  

The contraction of the TN corresponding to all the tensors in a corner is a well-known object called \emph{corner transfer matrix}, and is known to have very nice spectral properties (see e.g. the works by Baxter in Ref.\cite{baxter}). The aim of CTM contraction methods is to find an approximate, renormalized version of the exact CTMs, half-row and half-column transfer matrices. This can be done numerically in several ways. One possibility is to use the Corner Transfer Matrix Renormalization Group (CTMRG) developed by Nishino \emph{et al} \cite{ctmrg}. Here we explain a variation of this method, called directional CTM approach \cite{dctm}, which is more efficient and essentially as accurate (in most cases) as the original CTMRG method. 

\begin{figure}
\begin{centering}
\includegraphics[width=10cm]{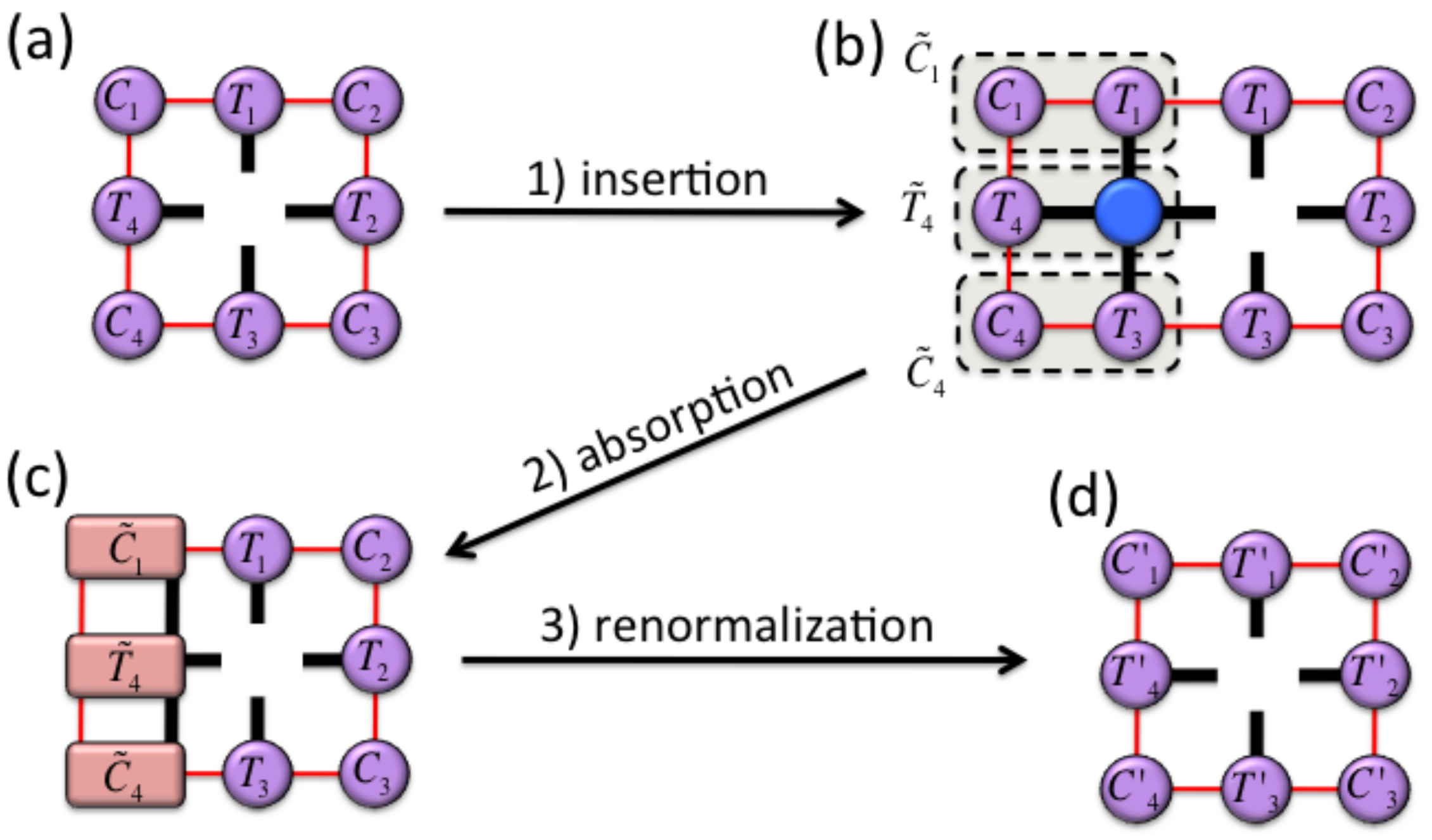} 
\par\end{centering}
\caption{(color online) Different steps in the directional CTM approach (see text).}
\label{fig39}
\end{figure}
\begin{figure}
\begin{centering}
\includegraphics[width=10cm]{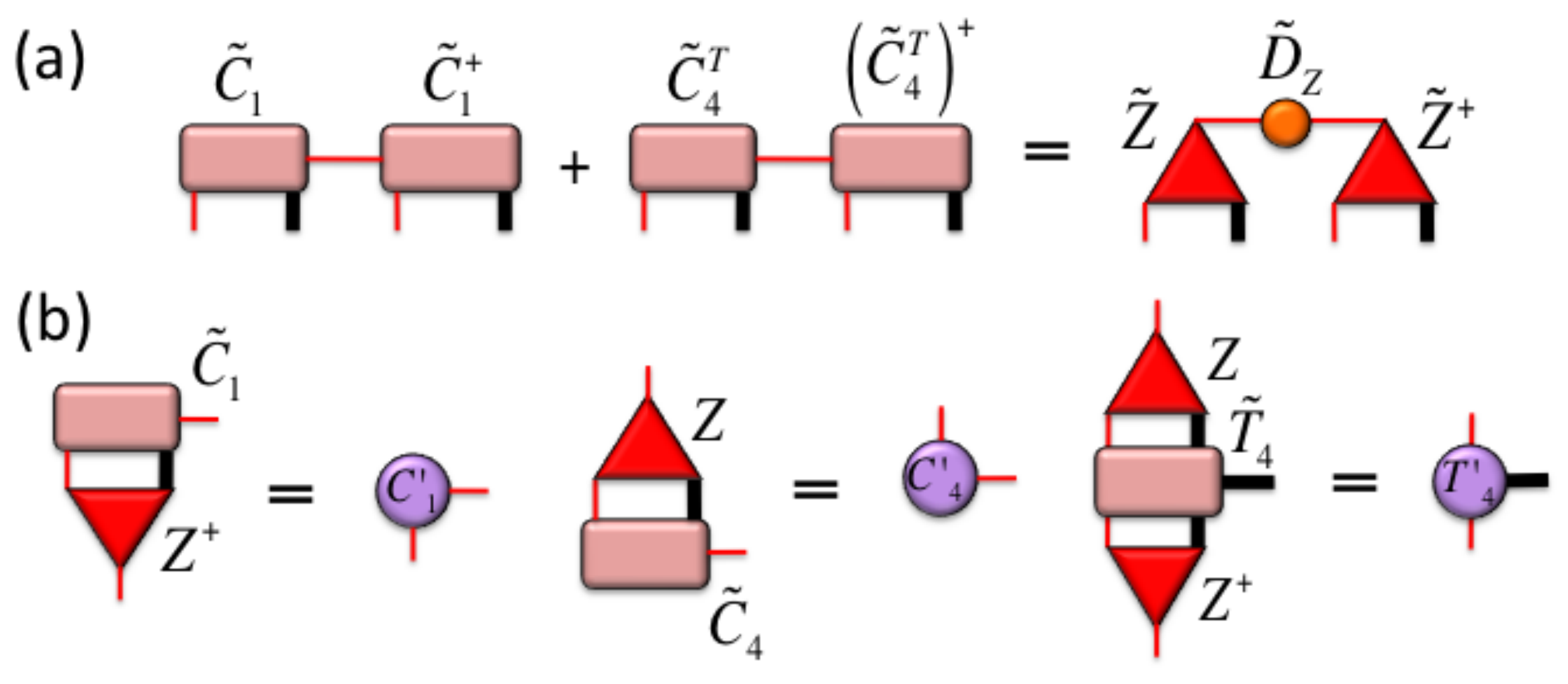} 
\par\end{centering}
\caption{(color online) (a) Calculation of isommetries in the directional CTM approach; (b) renormalization of tensors with the truncated isommetries (see text).}
\label{fig40}
\end{figure}

In the directional CTM approach, the tensors are updated according to four \emph{directional moves}, to the left, right, up and down, which are iterated until the environment converges. To be more specific, we start from an effective environment $\mathcal{G}^{[\vec{r}]} = \{C_1, T_1, C_2, T_2, C_3, T_3, C_4, T_4 \}$. Then, a move to e.g. the left is implemented by the following three steps: 
\vspace{10pt} 

\underline{(1) \emph{Insertion}}: starting from the effective environment $\mathcal{G}^{[\vec{r}]}$ in Fig.(\ref{fig39}.a), insert a new column as in Fig.(\ref{fig39}.b).
\vspace{10pt} 

\underline{(2) \emph{Absorption}}: contract tensors $C_1$ and $T_1$, tensors $C_4$ and $T_3$, and also tensors $T_4$ and the reduced PEPS tensor as indicated in Fig.(\ref{fig39}.b). As a result, obtain two CTMs $\widetilde{C}_1$ and $\widetilde{C}_4$, and a half-row transfer matrix $\widetilde{T}_4$, see Fig.(\ref{fig39}.c). 
\vspace{10pt} 

\underline{(3) \emph{Renormalization}}: compute isommetry $\widetilde{Z}$ as in Fig.(\ref{fig40}.a). Truncate it by keeping at most $\chi$ values of the right index corresponding to the largest singular values $\widetilde{D}_Z$ in the singular value decomposition, thus getting the truncated isommetry of $Z$. Finally, find two new CTMs $C'_1 = Z^{\dagger} \widetilde{C}_1$, $C'_4 = \widetilde{C}_4 Z$ and a new half-row transfer matrix $T'_4$ as indicated in Fig.(\ref{fig40}.b).
\vspace{10pt} 

After these steps the new effective environment $\mathcal{G}'^{[\vec{r}]}$ of a site in the lattice is given by the tensors $\mathcal{G}'^{[\vec{r}]} = \{C'_1, T_1, C_2, T_2, C_3, T_3, C'_4, T'_4 \}$, see Fig.(\ref{fig39}.d). Also, the effective size of the system increases by one column. The whole method continues by iterating this procedure in the different directions until convergence (of e.g. the singular values of the different CTMs). On the whole, the asymptotic computational cost in time of the directional CTM approach is $O(\chi^3 D^6)$. Several other remarks are provided e.g. in Ref.\cite{dctm}. The computation of one-body local observables and two-point functions follows easily from this. 

\vspace{10pt} 

\begin{figure}[h]
\begin{centering}
\includegraphics[width=9cm]{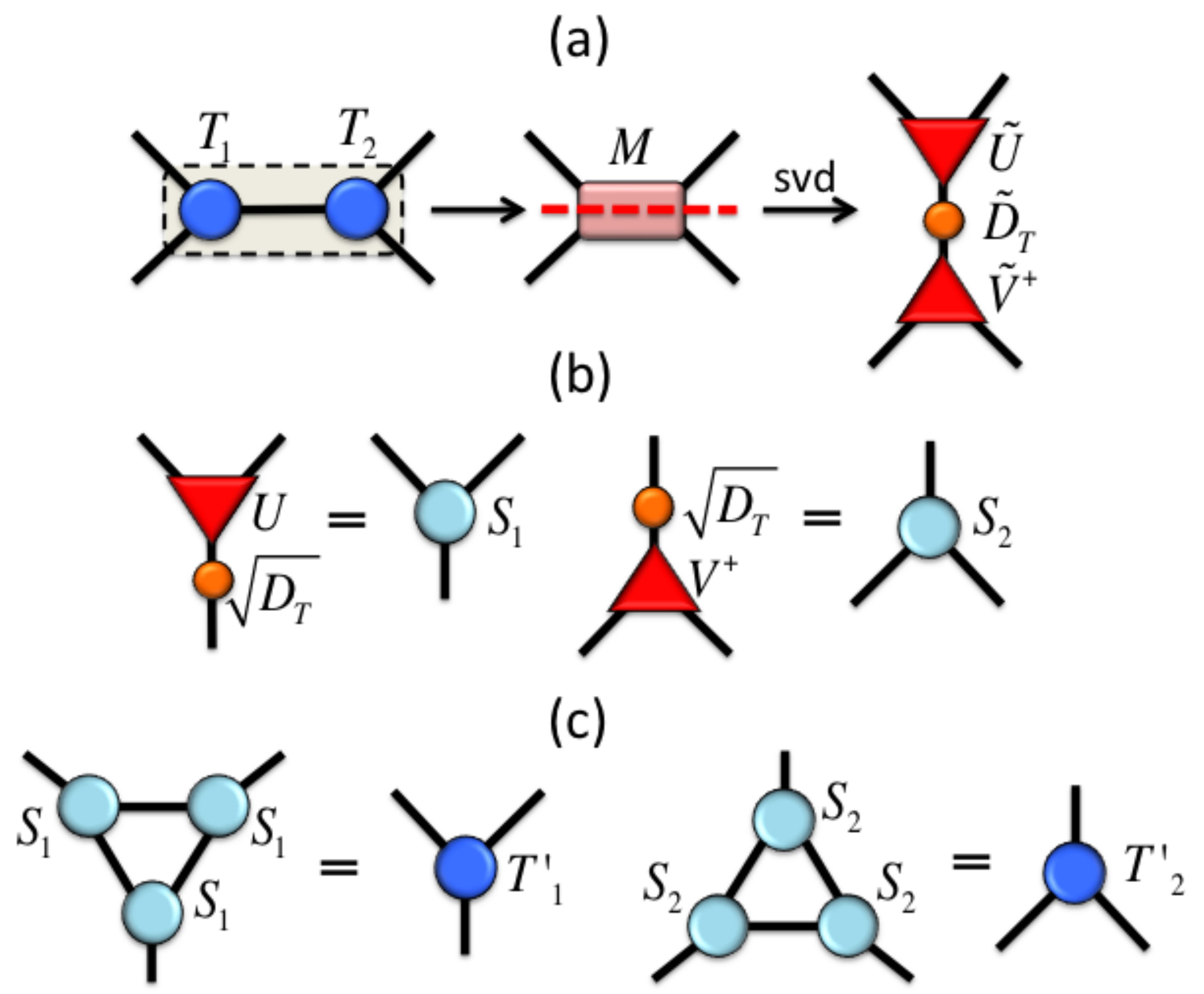} 
\par\end{centering}
\caption{(color online) Different steps in the TRG method: (a) rewiring and SVD, (b) calculation of $S_1$ and $S_2$ and (c) decimation.}
\label{fig41}
\end{figure}

{\bf \emph{3) Tensor coarse-graining.-}} The main idea of tensor coarse-graining methods is, as the name indicates, to perform a coarse-graining of the tensors in the $2d$ TN until some fixed-point tensor is reached, in a way similar to the Kadanoff blocking in classical statistical mechanical models. Some methods belonging to this family are e.g. Tensor Renormalization Group (TRG) \cite{TRG}, Second Renormalization Group (SRG) \cite{SRG}, Higher-Order TRG (HOTRG) \cite{HOTRG}, and Higher-Order SRG (HOSRG). Here we shall just sketch the basics of the simplest of these methods, namely TRG. This goes as follows: 

Consider a honeycomb lattice of reduced tensors $T_1$ and $T_2$ with two-site traslation invariance. The reason why we choose a honeycomb lattice with this unit cell structure is because it makes the explanation simpler, but everything can also be adapted to other types of lattices. The TRG method follows as in Fig.(\ref{fig40}) from three main steps: rewiring, truncation and decimation. Let us explain these briefly: 

\vspace{10pt}

\underline{(1) \emph{Rewiring}}: picking up two neighboring tensors in the lattice, say $T_1$ and $T_2$, we contract them along one direction as in Fig.(\ref{fig41}.a) and perform the SVD of the resulting tensor $M$ along the perpendicular direction, as in Fig.(\ref{fig41}.b), so that $M = \widetilde{U} \widetilde{D}_T \widetilde{V}^{\dagger}$. The resulting singular values $\widetilde{D}_T$ measure the entanglement between the two new tensors $\widetilde{U}$ and $\widetilde{V}^{\dagger}$ obtained from the SVD.
\vspace{10pt}

\underline{(2) \emph{Truncation}}: truncate the singular values so that you keep only the $\chi$ largest ones, obtaining the truncated isommetries $U$ and $V^\dagger$ as well as the truncated matrix of singular values $D_T$. 
\vspace{10pt}

\underline{(3) \emph{Decimation}}: define tensors $S_1 = U \sqrt{D_T}, S_2 = \sqrt{D_T}V^{\dagger}$, and contract them as in Fig.(\ref{fig41}.c) to define new tensors $T_1^{\prime}$ and $T_2^{\prime}$. 
\vspace{10pt}

After these steps, the resultant $2d$ lattice is again a honeycomb lattice but this time made of renormalized (coarse-grained) tensors. The method just proceeds by iterating these steps until convergence is reached, e.g. in the singular values that are kept. From here, the calculation of expectation values of physical observables follows also easily.

The accuracy of TRG can be largely improved by including in the truncation step the effect of the environment of tensors $U$ and $V^{\dagger}$. This technique is called the SRG method. Moreover, it is possible to improve the accuracy of both TRG and SRG by using the so-called Higher-Order SVD (or Tucker decomposition), which gives rise to the HOTRG and HOSRG methods. These shall not be explained here, and the interested reader is referred to Refs.\cite{TRG, SRG, HOTRG} for more information. 

\section{Determining the tensors: finding ground states}
\label{sec7}

Independently of their analytical properties, MPS and PEPS are very useful because they can be used as numerical ansatzs for approximating interesting quantum many-body wave-functions, such as ground states and low-energy excitations of local Hamiltonians. As we explained before, this is a very good idea since TNs are the states that target the relevant corner of the many-body Hilbert space. To put it in other words: MPS and PEPS have in their built-in structure the area-law for the entanglement entropy, and therefore the correct entanglement behavior for many ground states of gapped local systems (albeit with some exceptions at criticality \cite{fermion, spinbose, heisenberg}). Generally speaking, it is also known that MPS and PEPS approximate well the finite-temperature properties of gapped local Hamiltonians \cite{Hastings}. It is not surprising, thus, that MPS and PEPS turn out to be very valuable \emph{ans\"atze} when it comes to numerics. 

However, there is still a basic question that needs to be addressed: how do we fill in the coefficients of the tensors of MPS and PEPS? The answer is not unique: this can be done in different ways depending on the type of system (finite or infinite, $1d$ or $2d$...) and the type of state that is targeted (ground state, excitation, time-evolved state...). Explaining in detail all these different scenarios is far beyond the reach of this paper. Instead, here we explain the basic idea behind the two most widely-used families of methods to find ground states with TNs, namely: variational optimization, and imaginary time evolution. This should be a good starting point for the reader interested in learning the details of other techniques.  

\subsection{Variational optimization}
Given a Hamiltonian $H$, the variational principle states that for a given quantum state $\ket{\Psi}$ it will always be the case that 
\beq
\frac{\bra{\Psi} H \ket{\Psi}}{\braket{\Psi}{\Psi}} \ge E_0 \ , 
\eeq
with $E_0$ the lowest eigenvalue of $H$ (i.e. the ground state energy). Therefore, if $\ket{\Psi}$ is a TN belonging to some family of states like MPS or PEPS with a fixed bond dimension $D$, we can always approach the ground state energy from above by minimizing this expectation value over the relevant family, i.e. 
\beq
\min_{\ket{\Psi} \in {\rm family}} \left( \bra{\Psi} H \ket{\Psi} - \lambda \braket{\Psi}{\Psi} \right) \ , 
\eeq
where by ``family" we mean either MPS, or PEPS, or any other family of TN state with bond dimension $D$, and where $\lambda$ is a Lagrange multiplier that introduces the constraint that $\ket{\Psi}$ must have norm one.  

Ideally, the above minimization should be done simultaneously over all the free parameters of the TN state, hence over all the coefficients of all the tensors for all sites. However this is normally quite difficult to implement and not particularly efficient. Instead of this, the strategy is to proceed \emph{tensor by tensor}. This is, to minimize with respect to one tensor while keeping the others fixed, then move to another tensor and repeat the minimization, and so on. In practice one sweeps over all the tensors several times until the desired convergence in expectation values is attained. 

The way to minimize with respect to one tensor works as follows: imagine that we fix all the tensors in the TN except one of them which we call $A$. The coefficients of $A$ are now our variational parameters. The minimization is thus written as 
\beq
\min_{A} \left( \bra{\Psi} H \ket{\Psi} - \lambda \braket{\Psi}{\Psi} \right) =  \min_{A}  \left( \vec{A}^{\dagger} \mathcal{H}_{eff} \vec{A} - \lambda \vec{A}^{\dagger} \mathcal{N} \vec{A} \right) \ . 
\label{vareq}
\eeq
\begin{figure}
\begin{centering}
\includegraphics[width=8.5cm]{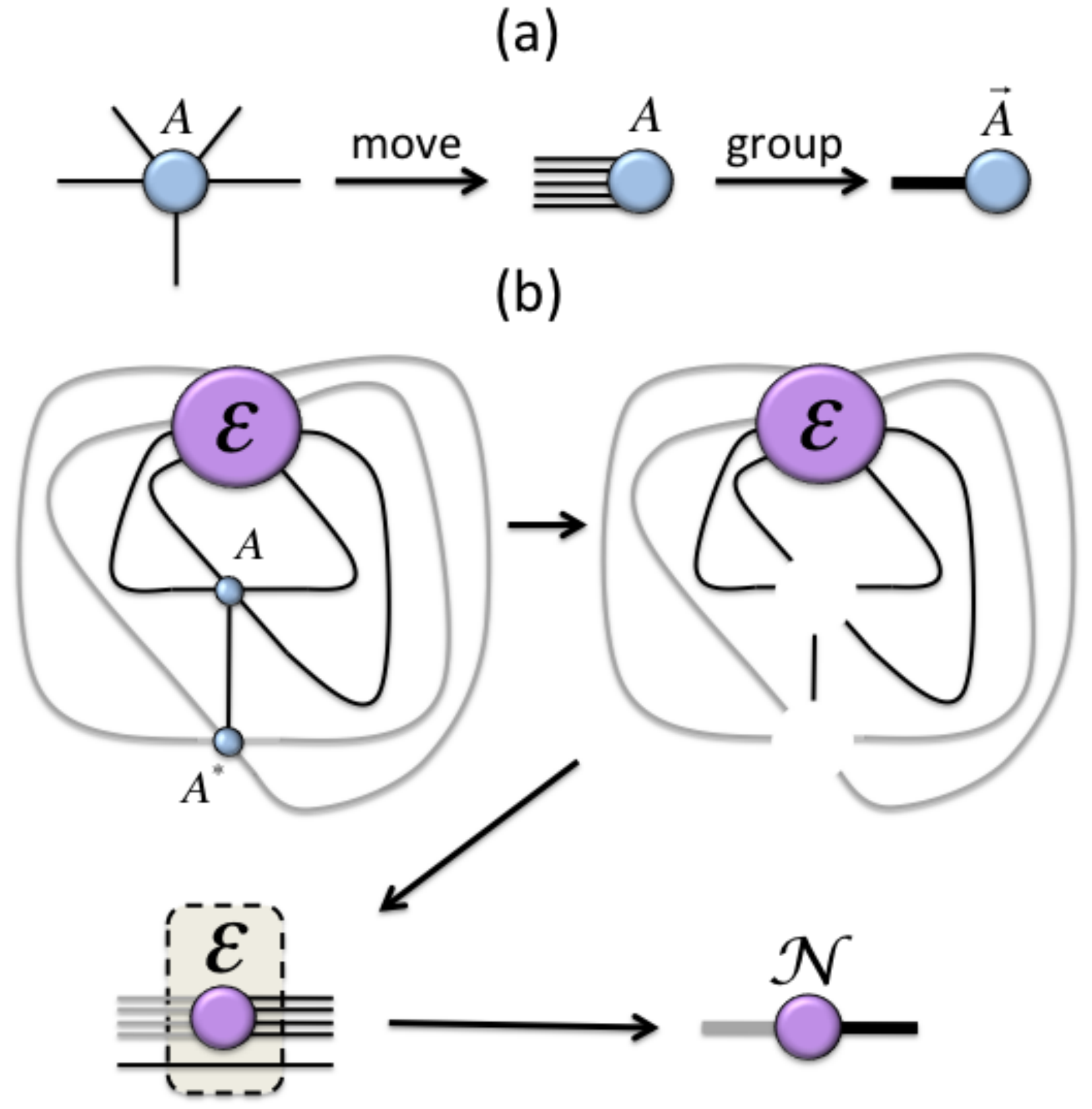} 
\par\end{centering}
\caption{(color online) (a) A tensor can be transformed into a vector by moving and grouping its indices; (b) procedure to get the normalization matrix $\mathcal{N}$.}
\label{fig42}
\end{figure}
In the above equation, $\vec{A}$ is just the tensor $A$ but with all its coefficients arranged as a vector (see Fig.(\ref{fig42}.a)), $\mathcal{H}_{eff}$ is an effective Hamiltonian, and $\mathcal{N}$ is a normalization matrix. In practice, $\mathcal{H}_{eff}$ and $\mathcal{N}$ correspond respectively to the TNs for $\bra{\Psi} H \ket{\Psi}$ and $\braket{\Psi}{\Psi}$ but without tensors $A$ and $A^*$. That is, the \emph{environment} of tensors $A$ and $A^*$ in the two TNs, but written in matrix form (see Fig.(\ref{fig42}.b)). Now, this minimization can be done as 
\beq
\frac{\partial}{\partial \vec{A}^{\dagger}} \left( \vec{A}^{\dagger} \mathcal{H}_{eff} \vec{A} - \lambda \vec{A}^{\dagger} \mathcal{N} \vec{A} \right) = 0 \ , 
\eeq
which leads to the generalized eigenvalue problem 
\beq
\mathcal{H}_{eff} \vec{A} = \lambda  \mathcal{N} \vec{A} \ . 
\eeq
Once $\mathcal{H}_{eff}$ and $\mathcal{N}$ are known this generalized eigenvalue problem can be solved numerically by using standard linear algebra packages. And to calculate both $\mathcal{H}_{eff}$ and $\mathcal{N}$ one must use the tools explained in the previous section to compute expectation values of physical observables and (effective) environments in a TN. In practice, this means that both  $\mathcal{H}_{eff}$ and $\mathcal{N}$ can be computed efficiently and exactly for an MPS, whereas for a PEPS this calculation is also efficient but approximate. 

A few remarks are in order. First, if we perform this optimization for an MPS with open boundary conditions, then what we recover is nothing but the famous Density Matrix Renormalization Group algorithm (DMRG) in the language of MPS \cite{dmrg1, dmrg2, mpsrev}. This is particularly appealing, since using this MPS picture is also now quite clear how to extend the method to periodic boundary conditions \cite{pbc1}. Second, this variational technique has also been applied successfully for infinite MPS \cite{iDMRG}. Not only that, but the method has also been applied successfully for e.g. finite PEPS \cite{PEPS}, as well as other TNs such as TTNs \cite{TTN} and the MERA \cite{ER}. Third, there are stability issues in the algorithm depending on the behavior of the normalization matrix $\mathcal{N}$. We postpone the discussion about this point to Sec.(\ref{secN}). 

\subsection{Imaginary time evolution} The imaginary time evolution technique proceeds by evolving some initial state in imaginary time with the Hamiltonian of interest. For very long times the resultant quantum state tends to be in the ground state $\ket{E_0}$, namely
\beq
\ket{E_0} = \lim_{\tau \rightarrow \infty} \frac{e^{-\tau H} \ket{\Psi(0)}}{\sqrt{|\braket{\Psi(\tau)}{\Psi(\tau)}|}}  ~~~~~~~~~~~~ \ket{\Psi(\tau)} = e^{-\tau H} \ket{\Psi(0)} \ , 
\eeq
for some initial quantum state $\ket{\Psi(0)}$ that has non-zero overlap with the ground state, and where $\tau$ is the imaginary time.   

\begin{figure}[h]
\begin{centering}
\includegraphics[width=12cm]{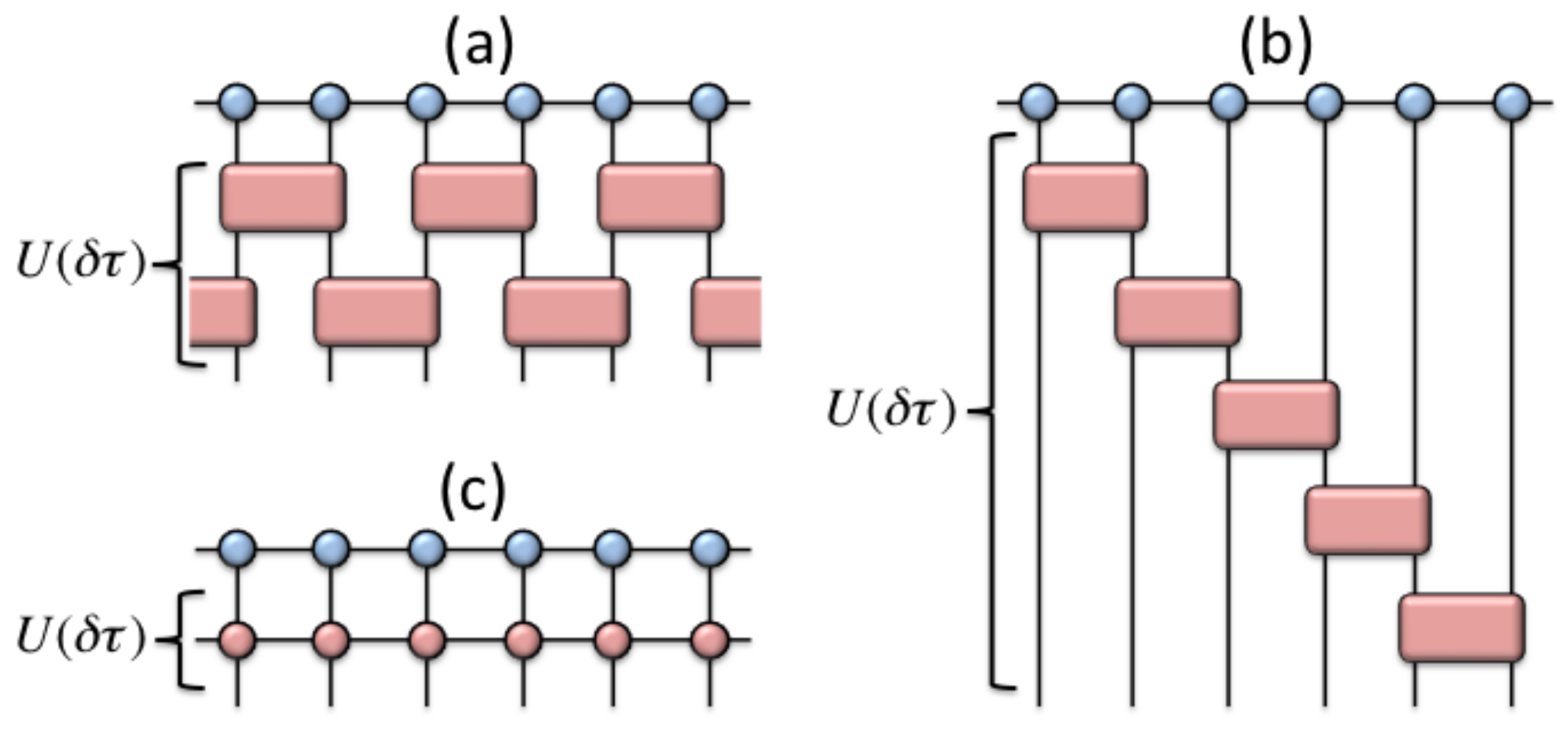} 
\par\end{centering}
\caption{(color online) Different ways to apply the time evolution operator $U(\delta \tau)$ to an MPS: (a) parallel, (b) sequential, and (c) Matrix Product Operator.}
\label{fig43}
\end{figure}

The idea now is to find a way to implement such an evolution on the family of relevant TN states, say e.g. MPS or PEPS, for a given quantum Hamiltonian. A way to achieve this is to split the time-evolution operator into small imaginary-time steps, 
\beq
e^{-\tau H} = \left(e^{-\delta \tau H} \right)^{m} \ ,
\eeq
where $m = \tau/\delta \tau \gg 1$. Next, let us assume for the sake of simplicity that the Hamiltonian is the sum of two-body nearest-neighbor terms, i.e. 
\beq
H = \sum_{\langle ij \rangle} h_{i,j} \ , 
\eeq
where sites $i$ and $j$ are nearest neighbors in the lattice\footnote{Other types of interactions could also be considered easily.}. With this in mind, we can make the approximation to first order in $\delta \tau$
\beq
e^{-\delta \tau H} = \prod_{\langle i, j \rangle} e^{-\delta \tau h_{ij}} + O(\delta \tau^2) \ . 
\eeq
The above is the so-called \emph{first-order Suzuki-Trotter expansion} \cite{Trotter}. Further orders in this approximation can also be considered if necessary if one wishes to reduce the associated error. Each of the terms in the multiplication above is also called a \emph{two-body gate}, in a notation that is reminiscent from that used in the language of quantum circuits. This is, the two-body gate $g_{ij}$ between sites $i$ and $j$ is given by 
\beq
g_{ij} \equiv e^{-\delta \tau h_{ij}} \ .
\eeq
As such, all the above equations mean that the imaginary-time evolution operator can be approximated by $m \gg 1$ repetitions of the operator $U(\delta \tau) \equiv \prod_{\langle i,j \rangle} g_{ij}$. This simplifies a lot the original problem, but still, one needs to know how to apply this operator on a given TN state. Luckily enough there are many ways of achieving this, see e.g. the TN diagrams for MPSs from Fig.(\ref{fig43}). 

Once the above points are clear, one must proceed according to the following strategy: 
\vspace{5pt}

\underline{(1) \emph{Infinitesimal evolution}}: apply operator $U(\delta \tau)$ on a given TN state $\ket{\Psi}$ (e.g. MPS or PEPS) with bond dimension $D$, obtaining a new TN state $\ket{\widetilde{\Psi}} = U(\delta \tau) \ket{\Psi}$ with bond dimension $D' \ge D$.
 
\underline{(2) \emph{Truncation}}: approximate $\ket{\widetilde{\Psi}}$ by a new TN state $\ket{\Psi '}$ back with bond dimension $D$. 

\begin{figure}
\begin{centering}
\includegraphics[width=10.5cm]{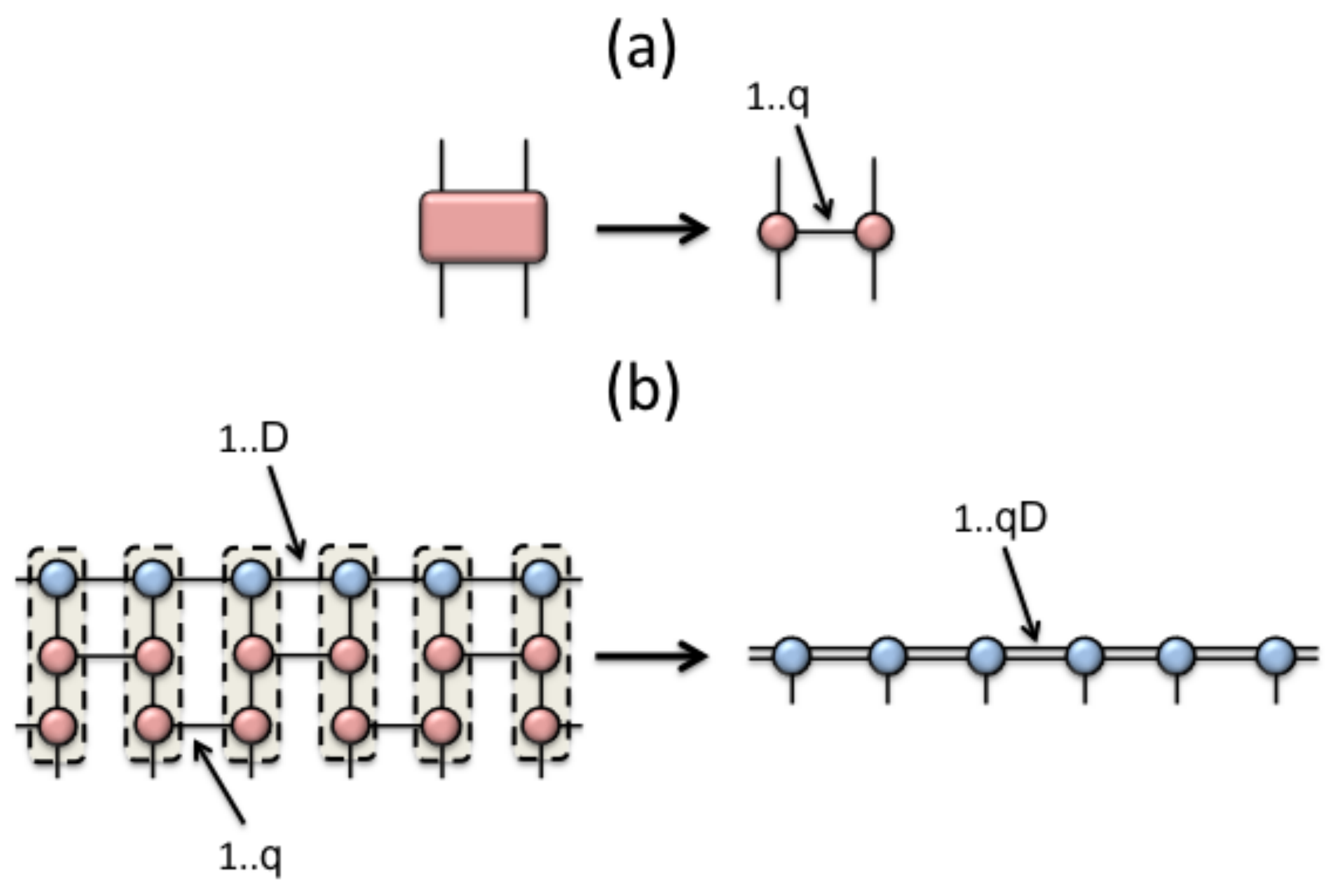} 
\par\end{centering}
\caption{(color online) (a) Tensor decomposition of the time evolution gate into two tensors. This can be done by using e.g. singular value or eigenvalue decompositions, as well as other ($q$ is the number of values of the corresponding bond index); (b) by applying in parallel the evolution gates to some MPS one obtains a new MPS with larger bond dimension.}
\label{fig44}
\end{figure}

\vspace{5pt}
The above two steps are very general, and there are \emph{many} ways of implementing them depending on (i) the dimensionality of the system, and (ii) the accuracy and efficiency that is desired. As expected, more accurate algorithms have a larger computational cost thus being less efficient, while less accurate algorithms, while being more efficient, may produce incorrect results if the entanglement in the system is large\footnote{This is specially true in the vicinity of a quantum critical point.}. Just as an example, step (1) above can be implemented for $1d$ systems as in Fig.(\ref{fig44}) (the extension to $2d$ is straightforward). 
Furthermore, step (2) can be implemented by e.g. minimizing the squared distance error 
\beq
err = \left|\left|\ket{\Psi'} - \ket{\widetilde{\Psi}} \right|\right|^2 \ .
\eeq
This minimization is done again using a tensor-by-tensor strategy, i.e. we fix all the tensors in $\ket{\Psi '}$ except, say, tensor $A$. The minimization over this tensor then reads
\beq
\min_A  \left|\left|\ket{\Psi'} - \ket{\widetilde{\Psi}} \right|\right|^2 = \min_A \left( \vec{A}^{\dagger} \mathcal{N} \vec{A} - \vec{A}^{\dagger} \vec{\mathcal{B}} - \vec{\mathcal{B}}^{\dagger} \vec{A} + \mathcal{C} \right) \ , 
\eeq
where $\vec{A}$ are the components of tensor $A$ arranged as a vector, $\mathcal{N}$ is again the normalization matrix that we encountered in the variational minimization in Eq.(\ref{vareq}), $\vec{\mathcal{B}}$ is a vector corresponding to the environment of tensor $A$ in $\braket{\Psi '}{\widetilde{\Psi}}$, and $\mathcal{C} = \braket{\widetilde{\Psi}}{\widetilde{\Psi}}$ is a scalar. Again, $\mathcal{N}, \vec{\mathcal{B}}$ and $\mathcal{C}$ can be computed using the strategies explained before in Sec.\ref{sec6}. The minimization then proceeds as
\beq
\frac{\partial}{\partial \vec{A}^{\dagger}} \left( \vec{A}^{\dagger} \mathcal{N} \vec{A} - \vec{A}^{\dagger} \vec{\mathcal{B}} - \vec{\mathcal{B}}^{\dagger} \vec{A} + \mathcal{C} \right) = 0 \ , 
\eeq
so that we end up with 
\beq
\mathcal{N} \vec{A} =   \vec{\mathcal{B}} \ , 
\eeq
which is nothing but a linear system of equations for the components of tensor $A$, and which can be formally solved as $\vec{A} = \mathcal{N}^{-1} \vec{\mathcal{B}}$. So in the end, once matrix $\mathcal{N}$ and vector $\vec{\mathcal{B}}$ are known, one just needs to solve this linear system of equations. 

Some comments are in order: first, the system of linear equations that we just described can be solved easily for small $D$. However, for large $D$ the complexity grows and one needs to use  specialized numerical packages. Second, as for the variational optimization there are stability issues in the algorithm depending on the behavior of the normalization matrix $\mathcal{N}$, which we shall discuss in the next section. 

This strategy to implement step (2) is normally called \emph{full update}, in the sense that the full environment of tensor $A$ has been taken into account in order to update its components. Despite having a relatively large computational cost, this approach is also the most accurate one. Nevertheless, its large cost leads to demanding (in terms of resources) numerical algorithms, very specially for $2d$ systems. Thus, alternative and more efficient approaches have been proposed at the cost of sacrificing part of the accuracy: these are the so-called \emph{simple update} methods. In a broad sense, these methods avoid the full calculation of the environment of tensor $A$ during its update, which usually is the bottleneck of the calculation. Instead, they use some ``simple" environment for which almost no calculation is needed. Such a ``simple" environment implements a huge oversimplification of the effect of quantum correlations in the system, and in fact most of the time follows from some type of mean-field approximation (or Bethe lattice approximation \cite{bethe}). This strategy turns out to be really good and almost exact for $1d$ systems as long as the infinitesimal time step $\delta \tau$ is sufficiently small, and leads to the so-called Time-Evolving Block Decimation (TEBD) method \cite{tebd}. However, for $2d$ systems its  ``accuracy vs efficiency" tradeoff is still a matter of debate: results for gapped phases are normally acceptable and can give a good idea of the physics of the system at hand with a fast algorithm that allows to reach large values of $D$. But however, the accuracy of the results drops down substantially when approaching a quantum critical point, i.e. when the correlation length is large and the full effect of all the surrounding environment of a tensor is important. We shall not explain here more details about the simple update method, and refer the reader to Ref.\cite{simple} for more details. 

\subsection{Stability and the normalization matrix $\mathcal{N}$.}
\label{secN}

A delicate issue in the numerics of the above two approaches is the conditioning of matrix $\mathcal{N}$. Notice that, by definition, this matrix is hermitian and positive since it can always be written as $\mathcal{N} = X^{\dagger}X$ for some matrix $X$ (this is just a consequence of the ket-bra structure of the corresponding TN). Therefore, all its eigenvalues are positive. Taking this into account, one can identify two possible problems: 

\begin{enumerate}
\item{The matrix, even if computed exactly (as in MPS), has eigenvalues very close to zero.}
\item{The matrix can only be approximated (as in PEPS), and therefore it may have some negative eigenvalues.}  
\end{enumerate} 

In either of the two cases, the corresponding solution to the generalized eigenvalue problem or linear system of equations may produce wrong results. If not somehow corrected, the error usually amplifies over the iterations in the algorithm, hence killing the convergence and producing instabilities. Therefore, it is really important to have, at every step, matrices $\mathcal{N}$ as accurate and as well conditioned as possible. 
 
Depending on the situation there are several ways of getting around the above two problems. Some options are: 

\begin{enumerate}
\item{If the TN has no loops (e.g. MPS with open boundary conditions), then it is always possible to choose an appropriate gauge for the tensors in which $\mathcal{N} = \mathbb{I}$. This is actually achieved by the canonical form in Fig.(\ref{fig17}) for MPS with open boundary conditions. The algorithm is thus more stable and the problems to be solved are far more simple, yet the price to be paid is that one needs to compute this canonical form at every step.}
\item{If the TN has loops (e.g. PEPS), then one can think of adding/subtracting to $\mathcal{N}$ some infinitesimal constant term proportional to the identity with the aim to kill eigenvalues close to zero or slightly negative, i.e. $\mathcal{N} + \epsilon \mathbb{I}$ with $\epsilon \ll 1$. One can also think of making ``as hermitian as possible" the matrix at every step by e.g. using $(\mathcal{N} + \mathcal{N}^{\dagger})/2$ instead of $\mathcal{N}$ as long as the introduced difference is small enough. Another option is to choose a gauge for the tensors in such a way that $\mathcal{N}$ is as close as possible to the identity (yet not exactly equal to it), and which can normally be achieved by using a quasi-canonical form for the TN, see e.g. Ref.\cite{clpeps}. Finally, it is also possible to think of approximation schemes for $\mathcal{N}$ which preserve the hermitian and positive structure at every step, yet at the price of getting lower accuracy, see e.g. Ref.\cite{singlelayer}.}

\end{enumerate} 
 
These are just some examples of ``stabilizing tricks". Of course, in a realistic situation one tries to combine all of them in order to make the algorithm as stable as possible.

\section{Final remarks}
\label{sec8}

Hopefully we offered here a plausible introduction to some key aspects of TN methods for quantum many-body systems for the non-experienced reader. A number of topics were addressed, including some important ideas to deal numerically with $1d$ systems using MPS and $2d$ systems using PEPS. Yet, what we explained here is only the top of the iceberg. 

There are many things which were not covered in this introduction. For instance, we did not enter into the detailed implementation of most algorithms (such as TEBD \cite{tebd}, iPEPS \cite{iPEPS}, and many more). We also did not talk about the very interesting properties of the MERA and its related numerical methods \cite{ER, branching}. Other important topics were not addressed, such as the implementation of symmetries in TNs \cite{sym}, how to deal with fermionic systems in $2d$ \cite{ferm}, and contraction/update schemes based on Montecarlo sampling \cite{StringBond1, EntangPlaq, MCMPS}. We did not talk either about the continuum limit of a TN, which defines e.g. the family of continuous MPS \cite{CMPS}, and which is of relevance as variational ansatz to simulate quantum field theories in the continuum. Also, we did not discuss $2d$ implementations of DMRG \cite{frus}, as well as TN methods for thermal states and dissipative systems \cite{thermal}. From a more theoretical point of view, we did not explain the deep relation between TNs and the entanglement spectrum \cite{entspec}, neither the relation to the holographic principle and the AdS/CFT correspondence \cite{holographic} and other potential applications to quantum gravity \cite{geo}. Other important topics that we did not touch here are parent and uncle Hamiltonians for $2d$ PEPS \cite{uncle}, as well as mathematical proofs for the connection between thermal states of gapped local Hamiltonians and TNs \cite{Hastings}. 

The above are just some examples of topics that we believe to be important, but which were not discussed here because its explanation lies beyond our current purpose. In fact, new developments in the field appear almost on a daily basis, and sometimes in totally unforeseen and surprising directions. It is thus our hope that this paper helps the newcomers to digest better the big amount of literature and information already available on the exciting topic of Tensor Networks. 

\noindent {{\bf Acknowledgements}}
The author acknowledges financial support from the Johannes-Gutenberg Universit\"at, as well as the participants of the \emph{DMRG101 Winter School 2012} in Taipei for interesting discussions which motivated this paper. Insightful comments by T. Nishino, G. Vidal, M. Rizzi, F. Verstraete, J. I. Cirac, M. C. Ba\~nuls and T. Xiang are also acknowledged. 

{}

\end{document}